\documentclass[a4paper,fleqn,usenatbib]{mnras}

\usepackage{newtxtext,newtxmath}
\usepackage[T1]{fontenc}
\usepackage{ae,aecompl}

\usepackage{graphicx}	
\usepackage{amsmath}	
\usepackage{cleveref}
\usepackage{longtable}

\usepackage{subfig}
\usepackage[utf8]{inputenc}		
\usepackage{pifont}
\usepackage{romannum}

\def\gs{\mathrel{\raise0.35ex\hbox{$\scriptstyle >$}\kern-0.6em
\lower0.40ex\hbox{{$\scriptstyle \sim$}}}}
\def\ls{\mathrel{\raise0.35ex\hbox{$\scriptstyle <$}\kern-0.6em
\lower0.40ex\hbox{{$\scriptstyle \sim$}}}}

\title{The  kinematics of massive high-redshift dusty star-forming galaxies}
\author[A.\ Amvrosiadis et al.]
{A.\ Amvrosiadis$^{1}$\thanks{E-mail: aristeidis.amvrosiadis@durham.ac.uk},
J.\,L.\ Wardlow$^{2}$,
J.\,E.\ Birkin$^{3,4}$,
I.\ Smail$^{3}$,
A.\,M.\ Swinbank$^{3}$, \newauthor 
J.\ Nightingale$^{1}$, 
F.\ Bertoldi$^{5}$, 
W.\,N.\ Brandt$^{6, 7, 8}$,
C.\,M.\ Casey$^{9}$,
S.\,C.\ Chapman$^{10}$, \newauthor
C.\--C.\ Chen$^{11}$,  
P.\ Cox$^{12}$,
E.\ da Cunha$^{13, 14, 15}$, 
H.\ Dannerbauer$^{16, 17}$, 
U.\ Dudzevi\v{c}i\={u}t\.{e}$^{3, 18}$, \newauthor
B.\ Gullberg$^{19}$,
J.\,A.\ Hodge$^{20}$,
K.\,K.\ Knudsen$^{19}$, 
K.\ Menten$^{21}$, 
F.\ Walter$^{18}$,
P.\ van der Werf$^{20}$
\vspace{4mm}\\
$^{1}$ Institute for Computational Cosmology, Department of Physics, Durham University, South Road, Durham DH1 3LE, UK \\
$^{2}$ Department of Physics, Lancaster University, Lancaster, LA1 4YB, UK \\
$^{3}$Centre for Extragalactic Astronomy, Department of Physics, Durham University, South Road, Durham DH1 3LE, UK \\
$^{4}$ Department of Physics and Astronomy, Texas A\&M University, 4242 TAMU, College Station, TX 77843-4242, USA \\
$^{5}$ Argelander-Institut f\"{u}r Astronomie, University at Bonn, Auf dem H\"{u}gel 71, 53121 Bonn, Germany \\
$^{6}$ Department of Astronomy \& Astrophysics, The Pennsylvania State University, 525 Davey Lab, University Park, PA 16802, USA \\
$^{7}$ Institute for Gravitation and the Cosmos, The Pennsylvania State University, University Park, PA 16802, USA \\
$^{8}$ Department of Physics, The Pennsylvania State University, University Park, PA 16802, USA \\
$^{9}$ The University of Texas at Austin, 2515 Speedway Boulevard, Stop C1400, Austin, TX 78712, USA \\
$^{10}$ Department of Physics and Atmospheric Science, Dalhousie University, Halifax, Halifax, NS B3H 3J5, Canada \\
$^{11}$ Academia Sinica Institute of Astronomy and Astrophysics (ASIAA), No. 1, Section 4, Roosevelt Road, Taipei 10617, Taiwan \\
$^{12}$ Sorbonne Université, CNRS UMR 7095, Institut d’Astrophysique de Paris, 98bis bvd Arago, 75014 Paris, France \\
$^{13}$ International Centre for Radio Astronomy Research, University of Western Australia, 35 Stirling Hwy., Crawley, WA 6009, Australia \\
$^{14}$ Research School of Astronomy and Astrophysics, Australian National University, Canberra, ACT 2611, Australia \\
$^{15}$ ARC Centre of Excellence for All Sky Astrophysics in 3 Dimensions (ASTRO 3D), Australia \\
$^{16}$ Instituto de Astrofísica de Canarias (IAC), E-38205 La Laguna, Tenerife, Spain \\
$^{17}$ Universidad de La Laguna, Dpto. Astrofísica, E-38206 La Laguna, Tenerife, Spain \\
$^{18}$ Max-Planck-Institut f\"{u}r Astronomy,
K\"{o}nigstuhl 17, D-69117, Heidelberg, Germany \\
$^{19}$ Onsala Space Observatory, Chalmers University of Technology, SE-439 92 Onsala/85748, Sweden \\
$^{20}$ Leiden Observatory, Leiden University, P.O. Box 9513, 2300 RA Leiden, The Netherlands \\
$^{21}$ Max-Planck-Institut f\"{u}r Radioastronomie, Auf dem H\"{u}gel 69, 53121 Bonn, Germany \\
}

\date{Accepted ---. Received ---; in original form ---}

\pubyear{2023}

\begin{document}
\pagenumbering{arabic} 
\label{firstpage}
\maketitle


\begin{abstract} 
We present a new method for modelling the kinematics of galaxies from interferometric observations by performing the optimization of the kinematic model parameters directly in visibility-space instead of the conventional approach of fitting velocity fields produced with the {\sc clean} algorithm in  real-space. We demonstrate our method on ALMA observations of $^{12}$CO (2$-$1), (3$-$2) or (4$-$3) emission lines from an initial sample of 30 massive 850$\mu$m-selected dusty star-forming galaxies with far-infrared luminosities $\gtrsim$$\,10^{12} $\,L$_{\odot}$ in the redshift range $z \sim$\,1.2--4.7. Using the results from our modelling analysis for the 12 sources with the highest signal-to-noise emission lines and disk-like kinematics, we conclude the following: (i) Our sample prefers a CO-to-$H_2$ conversion factor,  of $\alpha_{\rm CO} = 0.92 \pm 0.36$; (ii) These far-infrared luminous galaxies follow a similar Tully--Fisher relation between the circularized velocity, $V_{\rm circ}$, and baryonic mass, $M_{\rm b}$, as more typical star-forming samples at high redshift, but extend this relation to much higher masses -- showing that these are some of the most massive disk-like galaxies in the Universe; (iii) Finally, we demonstrate support for an evolutionary link between massive high-redshift dusty star-forming galaxies and the formation of local early-type galaxies using the both the distributions of the  baryonic and kinematic masses of these two populations on the $M_{\rm b}$\,--\,$\sigma$ plane and their relative space densities.
\end{abstract}


\section{Introduction} \label{sec:section_1}

The most massive galaxies in the Universe form from the highest density peaks in the primordial matter distribution \citep[][]{1978MNRAS.183..341W}. Galaxy interactions and mergers are expected to be frequent in such environments and  contribute to the growth of the most massive galaxies, as well as potentially imprinting variations in the kinematic structures of galaxies as a function of their mass (and environment): with the most massive galaxies exhibiting pressure-supported spheroidal morphologies \citep[e.g.][]{2019ApJ...884L..11O}. At the present day the majority of these massive systems lack significant on-going star formation; they correspond to the red-and-dead elliptical galaxies that dominate the cores of massive clusters of galaxies \citep[][]{1973ApJ...179..731F, 1980ApJ...236..351D}.

A number of mechanisms have been suggested to explain why the supply of gas from the intergalactic medium, needed to fuel star formation, is interrupted for many of the most massive  galaxies. The accretion of cold gas may cease when the galaxy's  halo becomes massive enough that an accretion shocks develops, interrupting the inflow of cold gas streams necessary to replenish star-forming disks \citep[][]{2006MNRAS.368....2D}. In addition, AGN feedback from a growing supermassive black hole may heat or expel cool gas, further suppressing star formation \citep[][]{2006MNRAS.370..645B, 2006ApJS..163....1H}.

Tracing the physical processes driving galaxy evolution is important for understanding the diversity of the galaxy populations we see at the present day. The most massive galaxies may be particularly useful in this regard, as they represent the high-mass limit for individual stellar systems. Connecting massive galaxy populations at different epochs may thus be more straightforward than attempting to track the formation and growth (through mergers and accretion onto larger galaxies) of less-massive systems.

Despite the complex nature of the physical processes that regulate star formation and lead to galaxy morphological transformations, a number of simple scaling relations can be used to investigate the effects of these mechanisms. For example, disk galaxies exhibit a empirical correlation between their rotational velocity and their baryonic mass, otherwise known as the Tully--Fisher relation \citep[TFR;][]{1977A&A....54..661T}, while elliptical galaxies exhibit a similar relationship between their baryonic mass and the central stellar velocity dispersion, also known as the Faber-Jackson relations \citep[FJR;][]{1976ApJ...204..668F}. Studying these relations for massive galaxy populations across cosmic time can help us understand how these galaxies evolved and potentially link populations at different epochs which are observed at different stages in their evolution.

Among the high-redshift galaxy populations, dusty star-forming galaxies (DSFGs, also referred to as sub-millimetre galaxies, SMGs) originally selected as sources with flux densities $S \gs 1$ mJy at 850$\mu$m \citep[][]{1997ApJ...490L...5S, 1998Natur.394..248B, 1998Natur.394..241H, 1999ApJ...515..518E}, are among the most massive active star-forming systems that have been observed. The redshift distribution of sub-millimetre-selected DSFGs peaks around $z \sim $\,2--3 \citep[][]{2005ApJ...622..772C, 2019MNRAS.487.4648S, 2020MNRAS.494.3828D}, where the star-formation activity and black-hole accretion peak \citep[][]{2014ARA&A..52..415M} and the gas accretion onto galaxies reaches its maximum \citep[][]{2020ApJ...902..111W}. Various studies have now established that these DSFGs have high stellar masses \citep[$M_{\star} \sim $\,10$^{10}$--10$^{11}$\,M$_{\odot}$; ][]{2014ApJ...788..125S, 2015A&A...576A.127S, 2015ApJ...806..110D, 2017A&A...597A...5M, 2020MNRAS.494.3828D}, are gas \citep[$M_{\rm gas} \sim $\,10$^{10}$--10$^{11}$\,M$_{\odot}$; ][]{2013MNRAS.429.3047B, 2021MNRAS.501.3926B} and dust rich \citep[$M_{\rm dust} \sim 10^{8}$--$10^{10} \, \rm M_{\odot}$; ][]{2020MNRAS.494.3828D} and form stars at extreme rates \citep[$\rm SFR \sim $\,10$^{2}$--10$^{3}$\,M$_{\odot} \rm yr^{-1}$][]{2014MNRAS.438.1267S, 2017A&A...597A...5M}, contributing up to $\sim$ 20\% to the total star-formation rate density \citep[][]{2014MNRAS.438.1267S}.

Having such high star-formation rates, DSFGs can substantially increase their stellar mass on a very short timescale \citep[$\sim 100$ Myr;][]{2013MNRAS.429.3047B, 2021MNRAS.501.3926B}. Considering  that they already have significant mass in stars, that will result in the descendants being very massive systems. It follows that this high-redshift population will evolve to form the most massive galaxies, which are mostly early type galaxies  in our Universe today. Several studies in the literature have proposed an evolutionary link between these two galaxy populations using a variety of arguments: (i) Clustering \citep[e.g.][]{2012MNRAS.421..284H, 2016ApJ...831...91C, 2017MNRAS.464.1380W, 2018MNRAS.475.4939A, 2020ApJ...904....2G, 2021MNRAS.504..172S}, (ii) space densities \citep[e.g.][]{2014ApJ...788..125S}, (iii) stellar ages \citep[e.g., ][]{2014ApJ...788..125S, 2020MNRAS.494.3828D, 2021arXiv210813430C} among others.

One piece of information that we so far have had only limited insight into is the dynamical state of these dusty high-redshift sources. The vast majority of studies in the literature on the dynamics of high-redshift star-forming galaxies have mostly focused on more ``typical''\footnote[1]{We use the term ``typical'' to characterize star-forming galaxies that have lower dust masses and star-formation rates compared to the average population of DSFGs selected at sub-millimetre wavelengths.} sources \citep[e.g.][]{2009ApJ...706.1364F, 2015ApJ...799..209W, 2019ApJ...886..124W, 2016ApJ...819...80P, 2019MNRAS.485..934T, 2023A&A...672A.106L, 2023MNRAS.521.1045R}, although small numbers of more active, either lensed or unlensed, galaxies have been studied \citep[see][]{2013ApJ...767..151M, 2016ApJ...827...57O, 2017ApJ...846..108C, 2021MNRAS.503.5329H, 2020Natur.584..201R, 2021MNRAS.507.3952R}. All these large surveys aim to measure the kinematics of star-forming galaxies from observations of the ionized gas, targeting the redshifted H$\alpha$ emission line. The limitation of using the H$\alpha$ line as a tracer to study the kinematics is that its susceptible to dust attenuation and the influence of ionized outflows. As a result, these studies have missed the most dust rich systems, which also happen to be the most massive at high redshift \citep[][]{2020MNRAS.494.3828D}. In addition, these studies currently extend out to $z \sim 2.5$, beyond which point we can no longer observe the H$\alpha$ line from the ground and therefore miss the high redshift tail of the  star-forming galaxies redshift distribution \citep[e.g; ][]{2023arXiv230105720B}. 

In order to study the kinematics of a representative and complete sample of these more dusty and active sources we need to rely on tracers of the ISM that are less influenced by dust obscuration. One such molecule is carbon monoxide (CO), specifically its low- to mid-$J$ transitions, which is considered a good tracer of the bulk of the molecular gas in these systems. With facilities such as the Atacama Large Millimeter/submillimeter Array (ALMA) or the Northern Extended Millimeter Array (NOEMA) studies of CO in samples high-redshift dusty galaxies have become possible over the last few years \citep[e.g.][]{2005MNRAS.359.1165G, 2013MNRAS.429.3047B, 2018ApJ...853..179T, 2021MNRAS.501.3926B}. We need to note, however, that besides targeting the various CO transitions, the [C{\sc ii}] transition can also be used for dynamical studies. In fact, the [C{\sc ii}]  emission line has recently gathered significant attention, since it is the brightest emission line in the far-IR part of the spectrum. Large surveys targeting this line have now been completed for samples of normal star-forming galaxies up to redshift $z\sim $\,4--6 \citep[][]{2020A&A...643A...1L} which allow the study of their dynamical properties \citep[][]{2021MNRAS.507.3540J}. 

In our work, in order to study their kinematics of DSFGs, we focus on samples with CO detections which is historically considered the best tracer of the molecular gas in these systems. Specifically, we make use of a sample of massive DSFGs that was presented in \cite{2021MNRAS.501.3926B}. We will use this sample to model the dynamics of this population and then use these results to study some scaling relations between the dynamical properties we inferred.

The outline of this paper is as follows: In Section~\ref{sec:section_2} we introduce the sample we will use in this work and discuss its properties. In Section~\ref{sec:section_3} we describe the method we use to model the kinematics for our sources, which is specifically tailored for interferometric observations. In Section\,\ref{sec:section_4} we discuss our main scientific results and finally, in Section\,\ref{sec:section_5} we give a summary of our findings. Throughout this work, we adopt a spatially-flat $\Lambda$-CDM cosmology with H$_0=67.8 \pm 0.9$\,km s$^{-1}$ Mpc$^{-1}$ and $\Omega_{\rm M}=0.308 \pm 0.012$ \citep{2016A&A...594A..13P}.

\section{Data} \label{sec:section_2}

\begin{table*}
	\centering{
	\caption{Properties of the \textit{parent} sample: (1) ID, (2) spectroscopic redshift, (3) observed CO transition, (4) major $\times$ minor axis of the synthesized beam, (5) stellar mass, (6) molecular gas mass (derived from the CO luminosity), (7) star-formation rate, (8) infrared luminosity, (9) velocity resolution of the cube used in the modelling and producing velocity maps, (10) integrated signal-to-noise ratio of the CO emission line (using the velocity-integrated intensity and its associated error), (11) classification based on the morpho-kinematical features of the observed 3D cubes (see Section~\ref{sec:SNR_and_classification}). The properties of the sources listed here are taken from Birkin et al. (2021) \iffalse\cite{2021MNRAS.501.3926B}\fi, unless the source was not included in that work in which case the properties are taken from various studies in the literature indicated by the footnotes.}
	\begin{tabular}{lcccccccccc}
	\\
	\hline\\[-6.0mm]
    \hline
	\\[-3.0mm]
	  ALESS & $z$ & CO & Beam & $ \log\big(M_{\star}\big)$ & $ \log\big(M_{\rm gas}\big) $ & $\rm \log\big(SFR \big)$ & $ \log\big(L_{\rm IR}\big)$ & $\Delta v$ & SNR & Class  \\[1.0mm]
	  ~~~ID & & transition & (arcsec) & $\big[ \rm M_{\odot} \big]$ & $\big[ \rm M_{\odot} \big]$ & $\big[ \rm M_{\odot} \, yr^{-1} \big]$ & $\big[ \rm L_{\odot} \big]$ & km s$^{-1}$ & &  \\[1.0mm]
	\hline
	\\[-3.0mm]
	
	\textbf{003.1} & 3.375 & 4$-$3 & 1.77 $\times$ 1.32 & $11.28^{+0.15}_{-0.25}$ & $10.96 \pm 0.06$ & $2.85^{+0.08}_{-0.07}$ & $12.93^{+0.07}_{-0.06}$ & 45 & 8.7 & \Romannum{2} 
	\\[1.5mm]
	
	\textbf{006.1} & 2.337 & 3$-$2 &  0.91 $\times$ 0.69 & $10.98^{+0.56}_{-0.50}$ & $11.00 \pm 0.06$ & $2.32^{+0.07}_{-0.21}$ & $12.43^{+0.05}_{-0.03}$ & 45 & 17.6 & \Romannum{2} 
	\\[1.5mm]
	
	\textbf{007.1} & 2.692 & 3$-$2 & 1.79 $\times$ 1.27 & $11.87^{+0.01}_{-0.02}$ & $11.22 \pm 0.06$ & $2.76^{+0.06}_{-0.01}$ & $12.96^{+0.01}_{-0.01}$ & 50 & 9.9 & \Romannum{1} 
	\\[1.5mm]
	
	\textbf{009.1} & 3.694 & 4$-$3 & 1.80 $\times$ 1.63 & $11.86^{+0.18}_{-0.09}$ & $10.99 \pm 0.05$ & $2.74^{+0.17}_{-0.15}$ & $13.01^{+0.09}_{-0.08}$ & 96 & 9.6 & \Romannum{2} 
	\\[1.5mm]
	
	\textbf{017.1} & 1.538 & 2$-$1 & 1.02 $\times$ 0.80 & $11.25^{+0.00}_{-0.00}$ & $10.49 \pm 0.09$ & $2.14^{+0.00}_{-0.00}$ & $12.25^{+0.00}_{-0.00}$ & 51 & 9.1 & \Romannum{1} 
	\\[1.5mm]
	
	\textbf{022.1} & 2.263 & 3$-$2 & 1.80 $\times$ 1.63 & $11.67^{+0.09}_{-0.07}$ & $10.89 \pm 0.04$ & $2.48^{+0.13}_{-0.16}$ & $12.66^{+0.07}_{-0.06}$ & 45 & 13.5 & \Romannum{1} 
	\\[1.5mm]
	
	\textbf{034.1} & 3.071 & 3$-$2 & 0.99 $\times$ 0.81 & $10.66^{+0.07}_{-0.10}$ & $10.62 \pm 0.08$ & $2.52^{+0.14}_{-0.18}$ & $12.66^{+0.16}_{-0.18}$ & 55 & 8.4 & \Romannum{2} 
	\\[1.5mm]
	
	\textbf{041.1} & 2.547 & 3$-$2 & 1.24 $\times$ 0.95 & $11.83^{+0.16}_{-0.17}$ & $10.57 \pm $ 0.08& $2.44^{+0.15}_{-0.26}$ & $12.64^{+0.08}_{-0.08}$ & 48 & 29.1 & \Romannum{1} 
	\\[1.5mm]
	
	\textbf{049.1$^{\alpha,\beta}$} & 2.945 & 3$-$2 & 1.24 $\times$ 0.95 & $10.58^{+0.12}_{-0.22}$ & $10.89 \pm 0.02$ & $2.83^{+0.09}_{-0.05}$ & $12.83^{+0.04}_{-0.07}$ & 54 & 27.9 & \Romannum{1} 
	\\[1.5mm]
	
	\textbf{062.2} & 1.362 & 2$-$1 & 0.87 $\times$ 0.69 & $10.68^{+0.02}_{-0.00}$ & $10.75 \pm  0.07$ & $2.68^{+0.01}_{-0.01}$ & $12.52^{+0.01}_{-0.00}$ & 48 & 23.8 & \Romannum{1} 
	\\[1.5mm]
	
	\textbf{065.1} & 4.445 & 4$-$3 & 0.99 $\times$ 0.81 & $10.48^{+0.19}_{-0.13}$ & $11.05 \pm 0.06$ & $2.48^{+0.10}_{-0.15}$ & $12.64^{+0.14}_{-0.18}$ & 55 & 10.4 & \Romannum{1} 
	\\[1.5mm]
	
	\textbf{066.1} & 2.553 & 3$-$2 & 0.86 $\times$ 0.69 & $11.42^{+0.33}_{-0.44}$ & $10.83 \pm 0.04$ & $2.66^{+0.10}_{-0.13}$ & $12.78^{+0.05}_{-0.08}$ & 48 & 13.3 & \Romannum{1} 
	\\[1.5mm]
	
	\textbf{067.1} & 2.122 & 3$-$2 & 2.18 $\times$ 1.59 & $11.25^{+0.00}_{-0.01}$ & $10.78 \pm 0.04$ & $2.19^{+0.01}_{-0.01}$ & $12.55^{+0.01}_{-0.01}$ & 42 & 9.5 & \Romannum{1} 
	\\[1.5mm]
	
	\textbf{071.1} & 3.709 & 4$-$3 & 1.15 $\times$ 0.96 & $12.31^{+0.00}_{-0.03}$ & $11.20 \pm 0.04$ & $3.42^{+0.06}_{-0.00}$ & $13.48^{+0.06}_{-0.00}$ & 48 & 17.6 & \Romannum{1} 
	\\[1.5mm]
	
	\textbf{075.1$^{\alpha,\beta}$} & 2.552 & 3$-$2 & 1.24 $\times$ 0.95 & $10.48^{+0.00}_{-0.21}$ & $10.84 \pm 0.02$ & $2.65^{+0.13}_{-0.14}$ & $12.58^{+0.17}_{-0.11}$ & 48 & 38.7 & \Romannum{1} 
	\\[1.5mm]
	
	\textbf{088.1} & 1.206 & 2$-$1  & 0.90 $\times$ 0.69 & $9.892^{+0.00}_{-0.00}$ & $10.65 \pm 0.07$ & $1.83^{+0.00}_{-0.00}$ & $12.22^{+0.00}_{-0.00}$ & 45 & 15.4 & \Romannum{2} 
	\\[1.5mm]

	\textbf{098.1} & 1.374 & 2$-$1 & 0.86 $\times$ 0.69 & $11.48^{+0.06}_{-0.05}$ & $11.13 \pm 0.07$ & $2.43^{+0.04}_{-0.04}$ & $12.55^{+0.03}_{-0.01}$ & 48 & 29.7 & \Romannum{1} 
	\\[1.5mm]
	
	\textbf{101.1} & 2.353 & 3$-$2 & 0.90 $\times$ 0.69 & $11.20^{+0.18}_{-0.20}$ & $10.93 \pm 0.04$ & $2.16^{+0.13}_{-0.18}$ & $12.30^{+0.09}_{-0.09}$ & 46 & 13.9 & \Romannum{2} 
	\\[1.5mm]
	
	\textbf{112.1} & 2.316 & 3$-$2 & 0.91 $\times$ 0.69 & $11.21^{+0.11}_{-0.14}$ & $11.01 \pm 0.04$ & $2.44^{+0.07}_{-0.10}$ & $12.50^{+0.05}_{-0.05}$ & 44 & 16.3 & \Romannum{2} 
	\\[1.5mm]
	
	\textbf{122.1$^{\alpha,\gamma}$} & 2.024 & 3$-$2 & 0.45 $\times$ 0.35 & $10.89^{+0.21}_{-0.21}$ & $11.30 \pm 0.09$ & $2.84^{+0.17}_{-0.16}$ & $12.92^{+0.13}_{-0.25}$ & 123 & 13.9 & \Romannum{1} 
	\\[1.5mm]
	
	\hline

    \textbf{001.1} & 4.674 & 5$-$4 & 1.78 $\times$ 1.33 & $10.93^{+0.12}_{-0.18}$ & $11.09 \pm 0.09$ & $2.82^{+0.13}_{-0.16}$ & $12.94^{+0.17}_{-0.21}$ & 90 & 4.8 & \Romannum{3} 
	\\[1.5mm]
	
	\textbf{001.2} & 4.669 & 5$-$4 & 1.78 $\times$ 1.33 & $11.06^{+0.11}_{-0.13}$ & $11.04 \pm 0.06$ & $2.56^{+0.17}_{-0.16}$ & $12.66^{+0.16}_{-0.18}$ & 90 &  7.3 & \Romannum{3} 
	\\[1.5mm]
	
	\textbf{005.1} & 3.303 & 4$-$3 & 1.80 $\times$ 1.63 & $11.37^{+0.35}_{-0.01}$ & $10.78 \pm 0.07$ & $2.97^{+0.01}_{-0.19}$ & $12.95^{+0.01}_{-0.13}$ & 87 & 6.6 & \Romannum{3} 
	\\[1.5mm]
	
	\textbf{019.2} & 3.751 & 4$-$3 & 1.80 $\times$ 1.63 & $11.43^{+0.17}_{-0.19}$ & $10.96 \pm 0.07$ & $2.88^{+0.17}_{-0.07}$ & $13.02^{+0.17}_{-0.06}$ & 96 & 7.8 & \Romannum{3} 
	\\[1.5mm]
	
	\textbf{023.1} & 3.332 & 4$-$3 & 1.86 $\times$ 1.56 & $11.45^{+0.21}_{-0.25}$ & $10.74 \pm 0.08$ & $2.75^{+0.11}_{-0.11}$ & $12.84^{+0.10}_{-0.07}$ & 88 & 6.0 & \Romannum{3} 
	\\[1.5mm]
	
	\textbf{031.1} & 3.712 & 4$-$3 & 1.80 $\times$ 1.63 & $11.52^{+0.10}_{-0.14}$ & $10.93 \pm 0.07$ & $2.81^{+0.11}_{-0.11}$ & $12.94^{+0.09}_{-0.09}$ & 96 & 4.8 & \Romannum{3} 
	\\[1.5mm]
	
	\textbf{035.1} & 2.974 & 3$-$2 & 2.14 $\times$ 1.60 & $11.63^{+0.20}_{-0.24}$ & $10.89 \pm 0.06$ & $2.64^{+0.12}_{-0.15}$ & $12.79^{+0.05}_{-0.04}$ & 108 & 7.2 & \Romannum{3} 
	\\[1.5mm]
	
	\textbf{061.1} & 4.405 & 4$-$3 & 0.98 $\times$ 0.81 & $10.33^{+0.18}_{-0.01}$ & $11.14 \pm 0.07$ & $2.42^{+0.16}_{-0.01}$ & $12.54^{+0.20}_{-0.01}$ & 55 & 7.9 & \Romannum{3} 
	\\[1.5mm]
	
	\textbf{068.1} & 3.507 & 4$-$3 & 1.85 $\times$ 1.56 & $10.97^{+0.26}_{-0.26}$ & $10.42 \pm $ 0.08& $2.54^{+0.11}_{-0.11}$ & $12.65^{+0.09}_{-0.11}$ & 92 & 5.2 & \Romannum{3} 
	\\[1.5mm]
	
	\textbf{079.1} & 3.901 & 4$-$3 & 1.86 $\times$ 1.56 & $11.41^{+0.11}_{-0.14}$ & $10.48 \pm 0.09$ & $2.32^{+0.12}_{-0.12}$ & $12.47^{+0.08}_{-0.06}$ & 100 & 3.9 & \Romannum{3} 
	\\[1.5mm]
	
	\hline\\[-6.0mm]
	\hline
	\label{tab:properties}
	\end{tabular}
	\begin{flushleft}                                      
    $^{\alpha)}$ Stellar mass, far-infrared luminosity and SFR from \cite{2015ApJ...806..110D}; $^{\beta)}$ gas mass from \cite{2018MNRAS.479.3879W}; $^{\gamma)}$ gas mass from  \cite{2018ApJ...863...56C}; $^{\delta)}$ SED fits are poorly constrained or the SED is potentially contaminated by an AGN.
    \end{flushleft} 
	}
\end{table*}

\begin{figure*}
    \centering
    \includegraphics[width=0.95\textwidth,height=0.375\textheight]{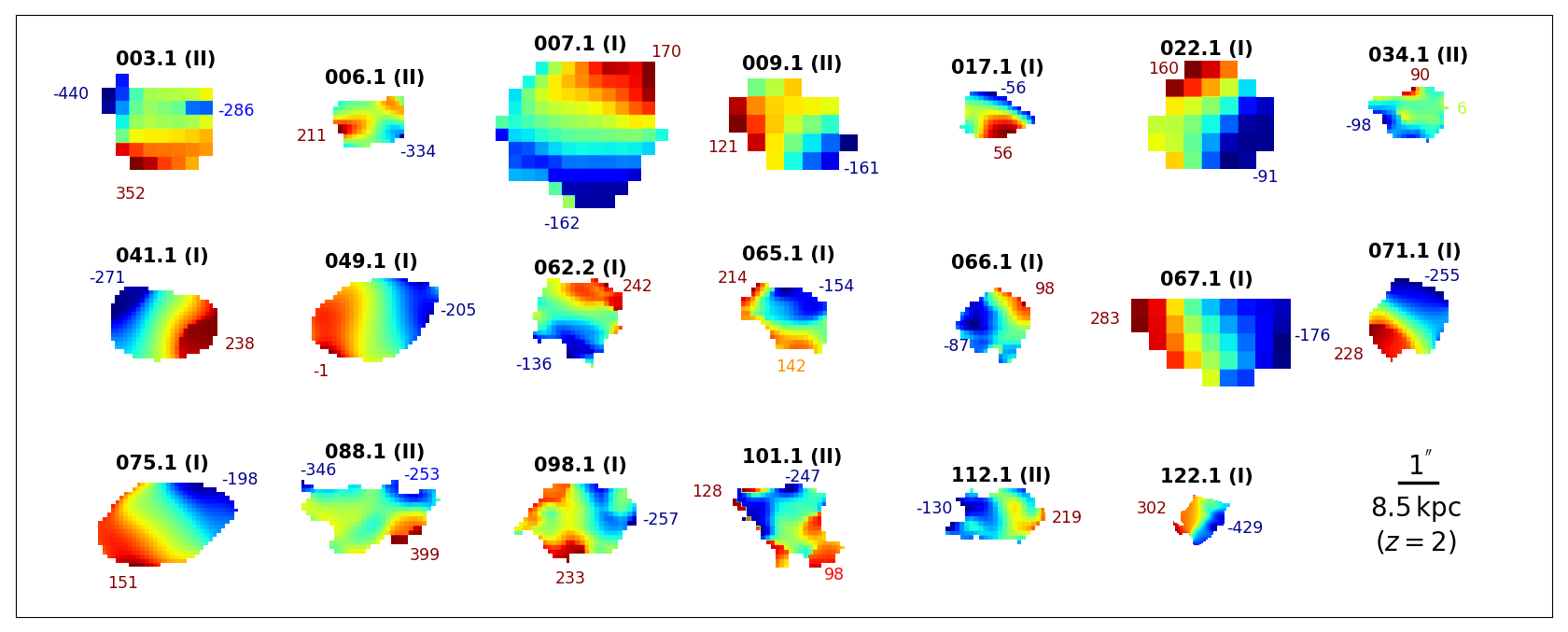}
    \caption{Velocity fields for the 20 DSFGs in our sample with CO emission lines that have integrated SNR\,$>$\,8. The velocity fields are derived from the observed CO emission lines using a pixel-by-pixel spectral fitting method. The kinematic class  assigned to each source, from our visual classification, is shown next to the source ID.  Sources that fall in Class \Romannum{1} display smooth velocity gradients across the source while Class \Romannum{2} sources have complex features in the velocity maps. All sources are shown on the same angular scale, where the black bar at the bottom right corner of the figure corresponds to 1$\arcsec$ or $\sim$\,8.5 kpc at $z = $\,2.}
    \label{fig:velmaps}
\end{figure*}

In this section we introduce the sample of sources that we use in this work. As already mentioned in the introduction, these sources primarily come from the recent study of \cite{2021MNRAS.501.3926B}, which presents a large CO survey of massive DSFGs. We focus on the analysis of ALMA observations of sources located in the ECDFS field from that study, which are part of the ALESS survey \citep{2013ApJ...768...91H}, as the observations of sources in other fields are generally at much lower spatial resolution. We complement this sample with three additional sources that were not in the \cite{2021MNRAS.501.3926B} sample, but  have data available of similar quality in the ALMA archive (ALESS 049.1, 075.1, 122.1). 

We begin by discussing our sample selection including defining a signal-to-noise (SNR) cut for our analysis. At the end of this section we discuss some of the physical properties (i.e., stellar and gas masses) of our sample. We also compare the properties of our sample to other, more ``typical'', star-forming galaxies at high redshift. We focus the comparison on samples for which large follow-up surveys, targeting the H$\alpha$ emission line, have been carried out with the aim of modelling their kinematics.

\subsection{Sample}\label{sec:data_sample}

The sources used in this work were first discovered in the LABOCA ECDFS Submillimeter Survey \citep[LESS;][]{2009ApJ...707.1201W} which is a large homogeneous 870$\mu$m survey of the ECDFS conducted with the Large Apex BOlometer Camera (LABOCA) on the APEX telescope. The LESS survey resulted in a sample of 126 sources with flux densities $S_{870 \rm \mu m} > 3.6$ mJy. These sources were subsequently followed-up with ALMA during Cycle 0 in the ALESS survey \citep{2013ApJ...768...91H}.  The MAIN sample from the ALESS survey comprises 99 sources from the maps of 69 LESS sources with high quality ALMA observations, with a considerable fraction ($\gtrsim 35\%$) of the LABOCA sources are being resolved into multiple sources in the ALMA maps \citep[][]{2013MNRAS.432....2K}.

A series of campaigns have been conducted in the optical, infrared (IR) and millimeter (mm) wavelengths to obtain spectroscopic redshifts for these ALESS sources \citep[eg.][]{2017ApJ...840...78D, 2018MNRAS.479.3879W, 2021MNRAS.501.3926B}. These redshift identifications come by targeting various CO transitions ($J_{\rm up} = $\,2--5) and the [C{\sc i}] ($\rm ^{3}P_1 - ^{3}P_0$) line in the mm wavebands \citep[see][]{2018MNRAS.479.3879W, 2021MNRAS.501.3926B} or rest-frame ultraviolet (UV) and optical emission lines \citep[][]{2017ApJ...840...78D}. From the \textit{parent} sample of 99 ALESS sources that have been detected in continuum, 30 now have robust CO line detections \citep{2018MNRAS.479.3879W, 2018ApJ...863...56C, 2021MNRAS.501.3926B}. 

The sources that we analyze in this work will be selected from this sample of 30 sources (which were observed as part of the following ALMA proposal IDs: 2016.1.00564.S, 2016.1.00754.S, 2017.1.01163.S, 2017.1.01512.S). As noted earlier we also include ALESS 049.1, 075.1, 122.1, previously discussed in \cite{2018MNRAS.479.3879W} and \cite{2018ApJ...863...56C}. In Table~\ref{tab:properties} we list these 30 sources for which we report their redshifts, observed CO transition, the size of the beam\footnote{Some sources were observed in multiple projects (e.g., ALESS 041.1, 075.1). In these cases we report the synthesized beam of the best available dataset (i.e., highest resolution and/or signal-to-noise ratio), which are the ones we use in our modelling analysis.} (major $\times$ minor axis) as well as some other properties which we discuss later in the section.

\subsection{Signal-to-Noise ratio selection \& Classification}\label{sec:SNR_and_classification}

We already mentioned that the aim of this work is to model the dynamics of the DSFGs in our sample. In order to perform such an analysis, the data we work with need be of sufficient signal-to-noise ratio (SNR). Here we refer to the integrated SNR which is defined as the ratio of the velocity integrated flux to its associated error\footnote{The error on the velocity integrated flux is computed as the quadratic sum of the errors in all channel of the cube that were used to measure the flux (i.e., the width of the spectrum). The error on the flux in each channel is computed as the standard deviation of fluxes estimated in $N = $\,100 regions, that have the same size as the aperture that was used to measure the flux, which do not contain any of the emission.}. We measured the integrated SNR for each source in our sample, which we report in Table~\ref{tab:properties}.

Before we go into the various details of our modelling analysis we want to make some useful clarifications. We attempted to model all 30 sources in our sample, however, we found that the parameters of our model were effectively unconstrained when modelling sources with an integrated SNR\,$< 8$. We therefore applied a further selection cut based on SNR, considering only sources with SNR\,$> 8$, which results in a sample of 20 sources.

We can now go a step further and divide the sources that satisfy our SNR selection criteria in two classes. We follow a classification approach that previous studies in the literature have used when dealing with 3D integral field spectroscopic data \citep[e.g.][]{2009ApJ...706.1364F, 2020A&A...643A...1L}. This classification is based on inspecting both the individual channel maps as well as the velocity maps. The velocity maps of our sources with SNR\,$> 8$ are shown in Figure~\ref{fig:velmaps}. These were produced by fitting a single Gaussian function to the spectrum in each pixel of our 3D ``dirty'' cubes and using the inferred mean of the fitted Gaussian as the value of that pixel in our 2D velocity maps. The characteristic difference between these two classes is whether the observed emission varies smoothly across the different channels of the cube, resulting in a well defined gradient in the velocity maps (Class \Romannum{1}; this is a necessary condition for a source to be considered a rotating disk) or shows a complex behavior (Class \Romannum{2}; which can be considered as an indication that the source is undergoing a minor/major merging event). Two out of these 20 sources, ALESS 003.1/009.1, are placed in class \Romannum{2} due to the lack of sufficient resolution. Finally, all sources that do not satisfy our SNR selection criteria (SNR\,$ < 8$), for which we lack sufficient SNR to unambiguously determine their kinematic nature, are put in a third class (Class \Romannum{3}). These three classes are summarized as follows: Class \Romannum{1} -- smooth transition between channel maps resulting in a well defined gradient for the velocity map; Class \Romannum{2} -- complex velocity map or galaxy-pairs or low resolution; Class \Romannum{3} -- low SNR. The specific class assigned for each source is given in the last column of Table~\ref{tab:properties}.


\subsection{Properties} \label{sec:properties}

The properties of our sample including redshifts, stellar masses ($M_{\star}$),  gas masses ($M_{\rm gas}$), far-infrared luminosities ($L_{\rm IR}$) and star-formation rates (SFR), were taken from \cite{2021MNRAS.501.3926B} or if not available, then from \cite{2015ApJ...806..110D}, \cite{2018MNRAS.479.3879W} or \cite{2018ApJ...863...56C}. We briefly summarize here how these quantities were computed, but we refer the reader to \cite{2021MNRAS.501.3926B} or the other studies for more details.

Stellar masses, far-infrared luminosities and star-formation rates were computed using the spectral energy distribution (SED) fitting code {\sc magphys} \citep[][]{2015ApJ...806..110D, 2019ApJ...882...61B}, keeping the redshifts fixed to the values derived from the observed CO lines. Gas masses were computed from the velocity-integrated intensity, $I_{\rm CO}$, of the observed transition lines. This was done as follows: The CO line luminosities were computed using the recipe from \cite{2005ARA&A..43..677S}. The measured line luminosities of each transition were converted to CO(1$-$0) luminosities using the excitation corrections that were measured for SMM\,J2135$-$0102 \citep[][]{2011MNRAS.410.1687D}. Finally, gas masses were calculated from the CO(1--0) luminosities assuming a CO--H$_2$ conversion factor, $\alpha_{\rm CO} = 0.92 \pm 0.36$, which, as we show later in Section~\ref{sec:sec_alphaCO}, is the value preferred for our sample and is consistent with previous estimates of this quantity for starburst galaxies \citep[][]{2013ARA&A..51..207B, 2018ApJ...863...56C, 2021MNRAS.501.3926B}.

We caution  that the results from the SED fitting analysis with {\sc magphys} could potentially be inaccurate for some sources in our sample. Specifically, the presence of a luminous AGN \citep[][]{2013ApJ...778..179W}, which is not accounted for in the SED model, can contaminate the UV to mid-infrared part of the SED and lead to a biased estimate of the stellar mass. However, the one source which is confirmed to host an AGN, ALESS\,17.1, shows no indication of AGN contamination of their mid-infrared SEDs.  In addition, we note that there are poor {\sc magphys} SED fits for two sources:  ALESS\,66.1 and 75.1, that make their derived properties uncertain (ALESS\,66.1 has a nearby foreground quasar \citep[][]{2020A&A...635A.119C} which could be the cause of the poor SED fit).  We have confirmed that the removal of these sources does not qualitatively change any of our conclusions and so we have retained them in our analysis.

\begin{figure}
    \centering
    \includegraphics[
        width=0.495\textwidth,
        height=0.3\textheight
    ]{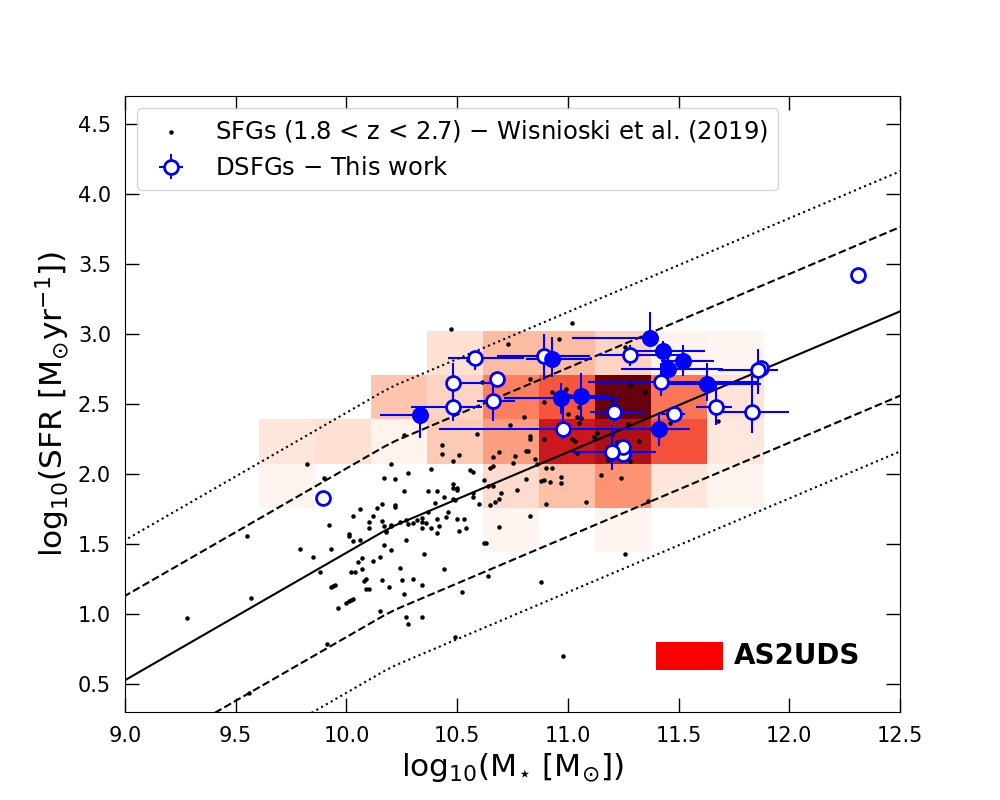}
    \caption{Star-formation rate as a function of stellar mass for galaxies in our sample, illustrating that the DSFGs in our study are more massive and more strongly star-forming than ``typical'' star-forming galaxies at these redshifts. Blue points correspond to sources in our sample, of which those with SNR $> 8$ are shown as open symbols. Black points correspond to typically less active star-forming galaxies from the KMOS$^{\rm 3D}$ \citep[][]{2019ApJ...886..124W} in the redshift range $z=$\,1.8--2.7.  The distribution for the SMGs from the AS2UDS survey \citep[][]{2020MNRAS.494.3828D} are shown as the red 2D histogram. The broken power-law model drawn as the black solid line, was optimized using 3D-HST data in all CANDELS fields selected in the redshift range $z=$\,2.0--2.5, where UV+IR SFRs were used \citep[][]{2014ApJ...795..104W}. The dashed and dot-dashed curves are $\times$4 and $\times$10 away (below/above) the mean relation.}
    \label{fig:figure_1}
\end{figure}

In order to place our sample into perspective, in Figure~\ref{fig:figure_1} we plot our sources on the SFR versus $M_{\star}$ plane along with other samples of high-redshift star-forming galaxies. We also show the parent sample of ALESS sources as presented in \cite{2021MNRAS.501.3926B} and SMGs in a similar redshift range to our sample ($z\sim $\,1--5) selected from the AS2UDS \citep{2020MNRAS.494.3828D}. In addition we also show SFGs in the redshift range $z=$\,1.8--2.7 from the KMOS$^{\rm 3D}$ survey \citep[][]{2019ApJ...886..124W}. This last sample represents more typical population of star-forming galaxies at high redshift, although with a sample selection biased against the most obscured and potentially massive galaxies. Compared to our sample these are less massive and form stars at much lower rates. We chose to use the KMOS$^{\rm 3D}$ sample for comparison as this is the largest sample with resolved kinematics from observations of the H$\alpha$ emission line. We will use this sample in later sections for comparing to the kinematical properties of our sample.

As we can see from Figure~\ref{fig:figure_1} our sources have a similar distribution on the $M_{\star}$--SFR plane to the much larger AS2UDS sample. In order to determine if our ALESS sample is representative of the DSFGs population, we performed a series of two-sample Kolmogorov-Smirnov (KS) tests for some of the properties listed in Table~\ref{tab:properties}. For these tests we matched the AS2UDS sample based on the flux at 870$\mu m$ and found probabilities consistent with them being drawn from the same distribution, indicating that our sample is representative of the DSFG population with $S_{870 \rm \mu m} \gs 1$ mJy.

\begin{figure*}
\includegraphics[width=0.95\textwidth,height=0.275\textheight]{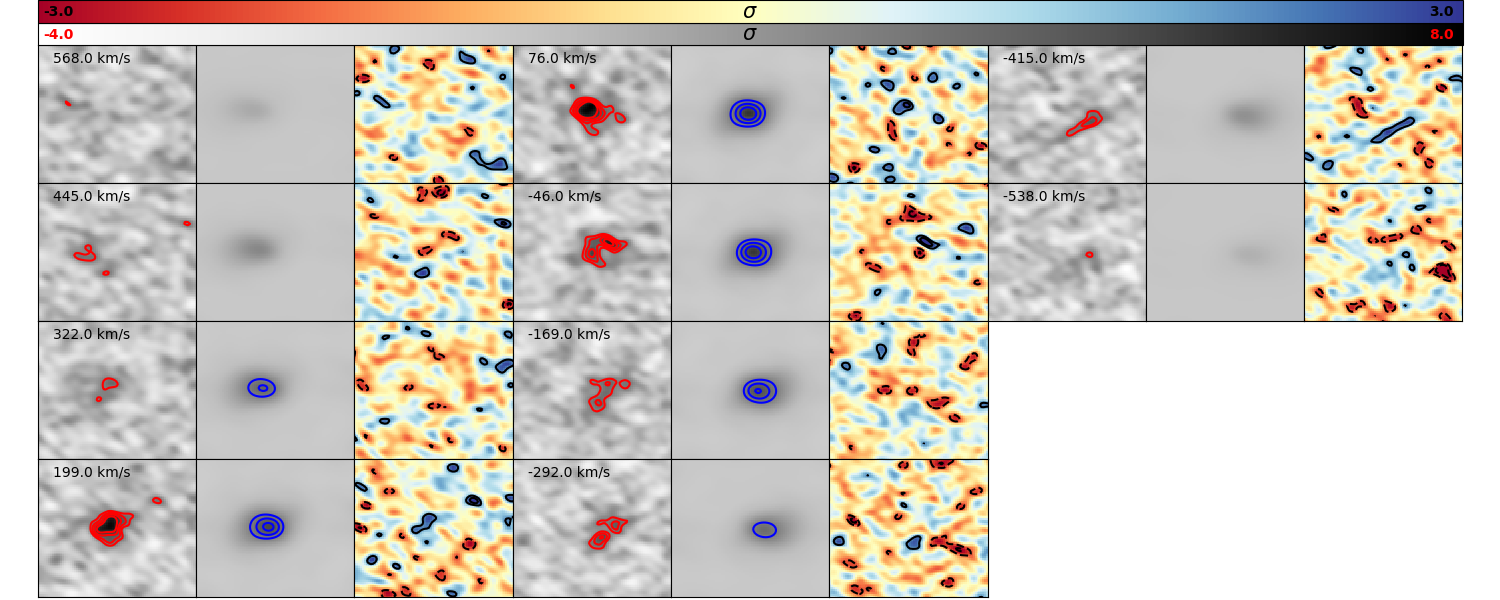}
\caption{Dirty channel maps of the normalized data, model and residuals of residuals of the CO(3--2) emission line for ALESS 122.1. The velocity of each channel, computed with respect to the redshifted CO emission line ($z = 2.024$), is indicated at the top left corner in each channel map of the data panels. The red and blue contours in the data and model panels, respectively, are drawn from $3\sigma$ to $6\sigma$ and increase in steps of $1\sigma$, where the rms noise is measured considering all channels of the cube. This source was previously analysed in ed in Calistro Rivera et al. (2018)\iffalse\cite{2018ApJ...863...56C}\fi, where the modelling was carried out in the real-plane, and found similar values for the model parameters.}
\label{fig:figure_3}
\end{figure*}

\section{Method \& Analysis} \label{sec:section_3}

In this section we describe the method we use to model the kinematics of galaxies using data that come from interferometric observations. The data we use to constrain the parameters of the models, are called visibilities. Each data point, $\mathbf{d}_{\rm ij}$, is a complex number (the $i,j$ subscript denote the visibility measured by the pair of antennas $i$ and $j$). Each visibility corresponds to a point in $uv$-space (i.e., a sample in Fourier space) which is defined by its $(u, v)$ coordinates (in units of wavelengths, $\lambda$). In the rest of this section we will describe the dynamical 3D models we use in this work to constrain the kinematics, the method to convert a 3D model cube ($x$, $y$, $\lambda$) to a set of visibility points ($u$, $v$, $\lambda$) and the framework we use to perform the parameter inference.

\subsection{Data preparation} \label{sec:section_3_1}

The raw data were calibrated using the Common Astronomy Software Applications \cite[CASA;][]{2007ASPC..376..127M}, employing the standard pipelines provided with each dataset. The un-flagged calibrated visibilities were then exported to a fits format for further analysis with our own {\sc PYTHON} routines.

{The modelling analysis, which we will describe later in this section, is performed in the $uv$-plane and therefore requires accurate weights associated with each visibility data point. The weights of the visibilities that come us a product of CASA's calibration pipeline are computed in a relative manner and so do not reflect the observed scatter of the data. In order to fit models to the data (i.e., visibilities) we require weights to be computed in an absolute manner. We therefore recompute the weights of the calibrated visibilities using {\sc casa} task {\sc statwt}, which re-calculates the weights the visibilities according to their scatter as a function of time and baseline. In practice what this task does is to select visibilities in a user-defined time interval for each pair of antennas and compute the scatter in the real and imaginary components of the visibilities. The scatter in the two components of the visibilities now corresponds to the error associated with those visibilities that were used to compute it.}

\subsection{Kinematical model}\label{sec:kin_model}

In order to extract the dynamical properties of our sources we need to fit appropriate models to the data and constrain their parameters. The velocity maps (Figure~\ref{fig:velmaps}) of most of our sources indicate the presence of order circular motions, therefore the appropriate choice of a model is that of rotating disk (deviations from ordered circular motions in the data can perhaps result in unphysical model parameters or ``poor'' fits, which we will discuss later).

We generate 3D parametric kinematical models of a  thick rotating-disk using the publicly available software {\sc Galpak3D} \footnote[1]{Available at http://galpak.irap.omp.eu/} \citep[][]{2015AJ....150...92B}. In total, our model is comprised of 10 free parameters: the $x$, $y$ and $z$ positions of the source centre in the cube, a scaling factor for the flux, effective radius, inclination, position angle (measured clockwise from East to North), turnover radius, maximum circular velocity (not projected, but asymptotic irrespective of inclination) and velocity dispersion (isotropic and constant over the disk). 

In order to generate a model we need to specify the flux, disk-thickness and velocity profiles. For the flux-profile of the disk we use an exponential profile, which is a special case of the more generic Sersic profile \citep{1963BAAA....6...41S} which is given by,
\begin{equation}
    I(r)=I_e \exp \left\{ -b_{n} \left[ \left(\frac{r}{r_e}\right)^{1/n} - 1 \right] \right\} \, ,
\end{equation}
where $r_e$ is the effective radius of the disk, $I_e$ is the surface brightness of the disk at $r_e$, and $b_n$ is a constant which given by $b_n \approx 1.9992n - 0.3271$ where $n$ is the S\'ersic index (the definition of $b_n$ is such that $r_e$ is equivalent to the half-light radius, $R_{1/2}$). For an exponential profile the S\'ersic index is set to $n=1$. For the disk thickness profile we use a Gaussian profile, 
\begin{equation}
    I(z) \propto \exp \left\{ -\frac{z^2}{2 h_z^2} \right\} \, ,
\end{equation}
which is defined perpendicular to the disc and is added to the disk component. The disk thickness is set to $h_z = 0.15 R_{1/2}$, where $R_{1/2}$ is the half-light radius of the disk\footnote{This is hard-coded in the source code and can not be used as a free parameter in our model. We note that we have not explored how this parameter affects our inference for the rest of the model parameters.}. Finally, for the velocity profile we use an isothermal model, which is given by
\begin{equation}\label{eq:velocity_profile}
V(r) = V_{\rm max} \left[1 - \arctan\left( \frac{r}{r_t} \right) \Big/ \left( \frac{r}{r_t} \right) \right] \, ,
\end{equation}
where $r_t$ is the turnover radius and $V_{\rm max}$ is the maximum circular velocity.  In what follows, we define the model cube in real-space, given a set of parameters, $\theta$, as $\mathbf{s}(\theta)$.

\subsection{The Non-Uniform Fast Fourier Transform}

In order to compare our model (i.e., the surface brightness in each channel of the cube) with our data we need to convert our model cubes defined in the real-space to visibilities defined in the Fourier space. An image of the surface brightness of a source, $I(x, y)$, in the $xy$-space and the visibilities, $V(u, v)$, in the $uv$-space form a 2D Fourier transform pair,
\begin{equation}
    V(u, v) = \int \int I(x, y) \exp^{ -2 \pi i (ux + vy) } dx dy \,
\end{equation}
However, since the samples in the $uv$-space are not uniformly distributed this operation reduces to a direct discrete Fourier transform (DFT). The DFT is a memory heavy and time-consuming operation making the analysis of large interferometric data very prohibitive. This problem can be alleviated by the use of a non-uniform fast Fourier transform (NUFFT). The use of NUFFTs for analysing astronomical interferometric data has been discussed in recent works \citep[e.g., ][]{2020Natur.584..201R, 2021MNRAS.501..515P} and a plethora of literature exist on the theoretical side \citep[e.g.][]{1435541}. In our work we use the publicly available software {\sc pyNUFFT} \footnote[2]{Available at https://github.com/jyhmiinlin/pynufft} \citep[][]{jimaging4030051} to perform this computation. In order to set up our NUFFT operator we need to pass it a real-space grid and the $(u, v)$ coordinates of our Fourier samples.

We determine the number of image-plane pixels and the size of each pixel based on the $uv$-coverage and the Nyquist criterion. We require that the size of each pixel is at least half the resolution of our observation which we approximate as the inverse of the maximum $uv$ distance (in arcsec). The number of image-plane pixels and the size of each pixel are determined by choosing the size of the field-of-view (FoV) and rounding up the number of image-plane pixels to the closest power of 2 so that the pixel scale is at least half the resolution.

Once we have defined the image-plane grid, following the procedure that was described above, we can then initialize our NUFFT operator. The NUFFT operator, $\mathbf{D}$, will have dimensions ($2 N_{\rm vis} \, \times \, N_{\rm p}$), where $N_{\rm p}$ is the number of image-plane pixel and $N_{\rm vis}$ are the number of visibilities that we use in our analysis. To initialize the NUFFT operator, besides the image-plane grid, we also use the $uv$-coordinates of our visibilities. Here we need to note that the $(u, v)$ position of a visibility recorded at time, $\textbf{t}_0$, shifts radially outwards from the phase-center as the frequency increases. We therefore define a separate NUFFT operator, $\mathbf{D}_n$ (where $n = 1, 2, 3, ..., N_c$), for each of the $N_c$ channels of our cube (each of the $\mathbf{D}_n$ operators on the respective channel image of the cube $\mathbf{s}_{n}(\theta)$). Therefore, applying the NUFFT operator to our model cube, $\mathbf{D} \mathbf{s}(\theta)$, has an effect equivalent to convolving the model cube with the dirty beam (which can be thought of as the Point Spread Function) in the $xy$-plane and the Line Spread Function (LSF) along the $z$-axis (i.e.\ the frequency axis).

\subsection{Constraining the model parameters}

In order to fit our model to the data we use the publicly available software {\sc PyAutoFit}\footnote[3]{Available at https://github.com/rhayes777/PyAutoFit} (Nightingale et al.\ in prep;). {\sc PyAutoFit} is probabilistic programming language (PPL) that is designed to provide the user with a flexible interface to fit a model to set of data points by defining the likelihood function of this model given the data.

The optimization of the model parameters, $\theta$, is performed in a Bayesian framework. The posterior probability distribution, $P(\theta | \textbf{d})$, is given by
\begin{equation}
    P(\theta | \textbf{d}) \propto P(\textbf{d} | \theta) P(\theta) \, ,
\end{equation}
where $P(\textbf{d} | \theta)$ is the likelihood and $P(\theta)$ the prior distribution of our model parameters. It is well known that the noise dominates over the signal for an individual visibility point and its nature is random (thermal noise), which results in the distributions of both the real and imaginary components of our visibilities being well described by a Gaussian distribution \citep[][]{2017isra.book.....T}. We can therefore assume that the likelihood function is also Gaussian and write it as,
\begin{equation}\label{eq:likelihood}
\begin{split}
    P(\textbf{d} | \theta) & = \frac{1}{\sqrt{\det(2 \pi C)}} \times \\ 
    & \exp\Big\{ -\frac{1}{2} \mathop{\sum}_{n = 0}^{N_c} \left( \mathbf{D}_n \mathbf{s}_n(\theta) - \mathbf{d}_n \right)^T \mathbf{C}_n^{-1} \left( \mathbf{D}_n \mathbf{s}_n(\theta) - \mathbf{d}_n \right) \Big\} \, ,
\end{split}
\end{equation}
where $\textbf{d}_n$ are the observed visibilities, $\textbf{C}_n^{-1}$ are the errors of the observed visibilities (see Section \ref{sec:section_3_1}), $\textbf{s}_n(\theta)$ are the images for each channel of the model cube given a set of model parameters $\theta$ and $\textbf{D}_n$ is the NUFFT operator. The index $n$ corresponds to a channel of the cube. The matrix $C$ is a block diagonal matrix where each block correspond to the individual $\textbf{C}_n$ matrices. 

We use a Gaussian prior for the geometric centre, which is centered on the peak value of the intensity ($0^{\rm th}$ moment) map and has a standard deviation of 1 arcsec. For the rest of our model parameters we use uniform priors, 0 $\rm km \, s^{-1}$ < $V_{\rm max}$ < 600 $\rm km \, s^{-1}$; 0 $\rm km \, s^{-1}$ < $\sigma$ < 200 $\rm km \, s^{-1}$; 0 arcsec < $r_e$ < 2 arcsec; 0 arcsec < $r_t$ < 1 arcsec; 0\textdegree < $\theta$ < 90\textdegree; 0\textdegree < $i$ < 90\textdegree.

We note that for the optimization we actually minimize the log of the likelihood. Having defined our log-likelihood function, {\sc PyAutoFit} allows the user to choose between a collection of different non-linear samplers to be used to constrain the parameters of the model by sampling the posterior distribution of these parameters. In our analysis we use a nested sampling Monte Carlo technique of Skilling (2006) implemented in the {\sc PyMultinest} algorithm, which itself is a {\sc python} wrapper of the {\sc Multinest} \citep{2014A&A...564A.125B} algorithm \citep{2008MNRAS.384..449F, 2009MNRAS.398.2049F}. Finally, in order to speed up the likelihood evaluation we create manual masks in the velocity axis, ignoring channels that do not contain any emission. This essentially reduces the number of NUFFTs that we have to perform each time we evaluate a likelihood for a given combination of model parameters.


\begin{figure*}
\begin{tabular}{cc}

\includegraphics[width=0.475\textwidth,height=0.125\textheight]{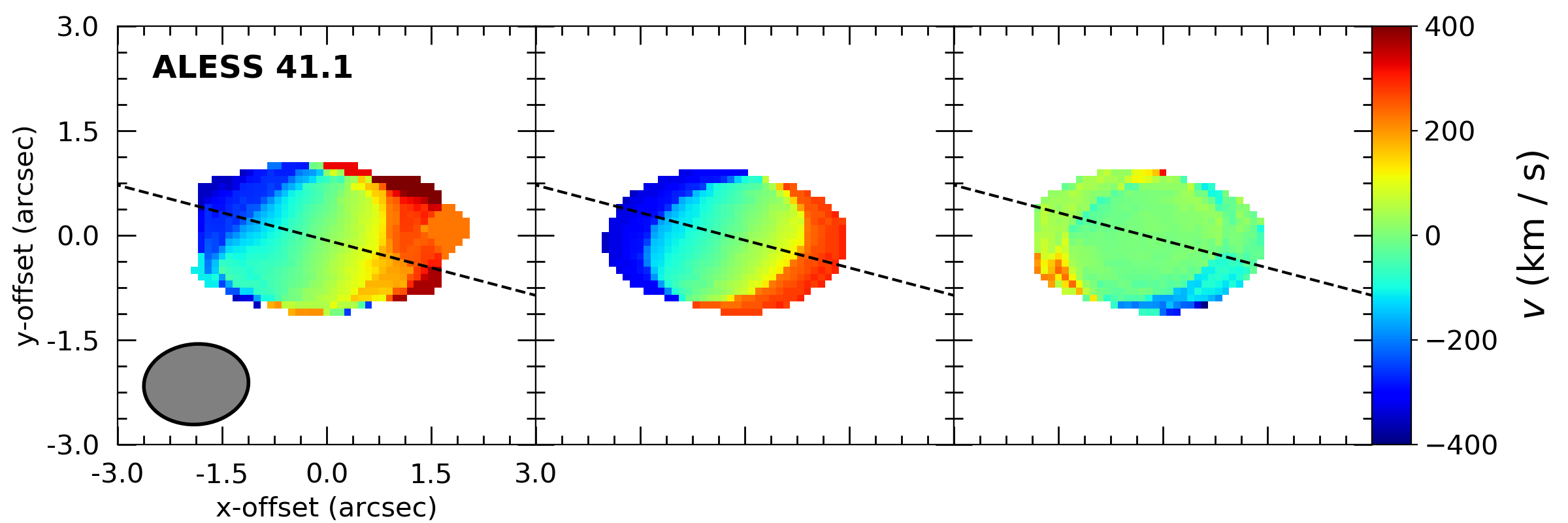} & \includegraphics[width=0.475\textwidth,height=0.125\textheight]{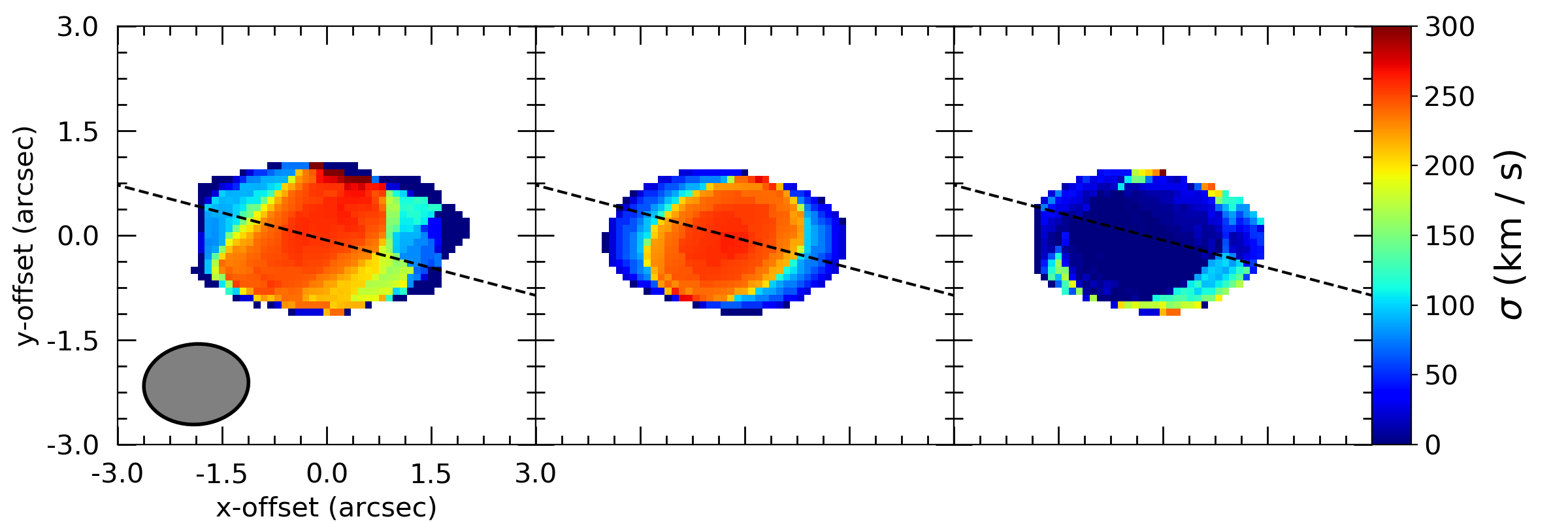} \\[1.0mm]

\includegraphics[width=0.475\textwidth,height=0.125\textheight]{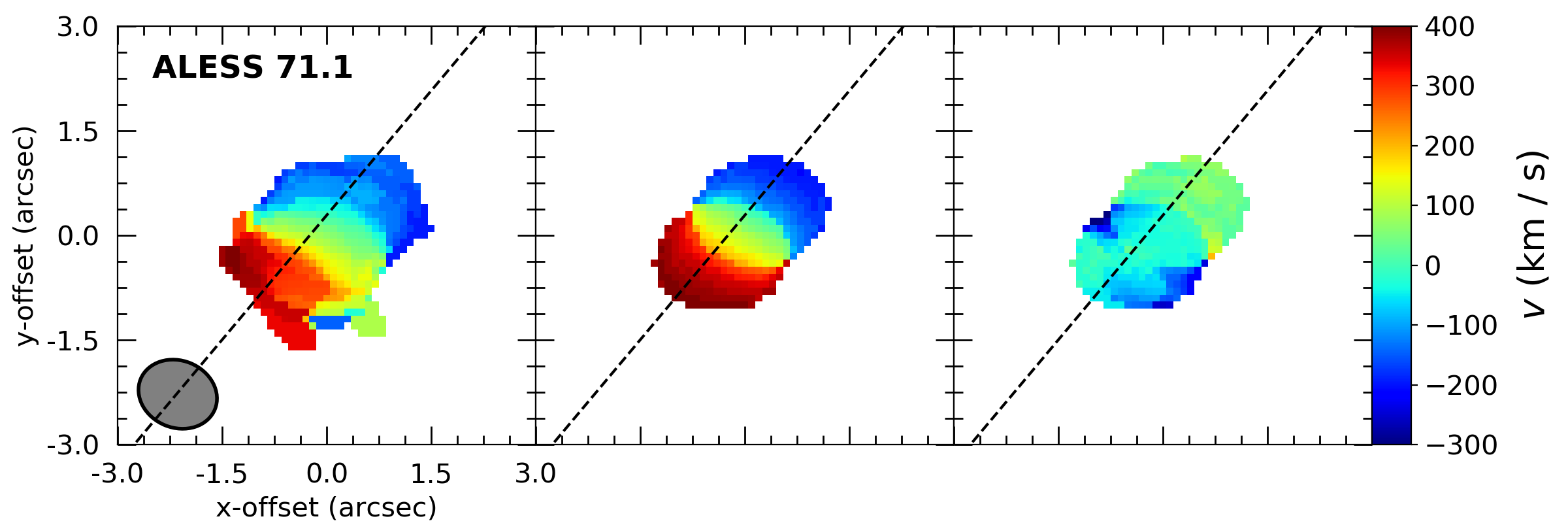} & \includegraphics[width=0.475\textwidth,height=0.125\textheight]{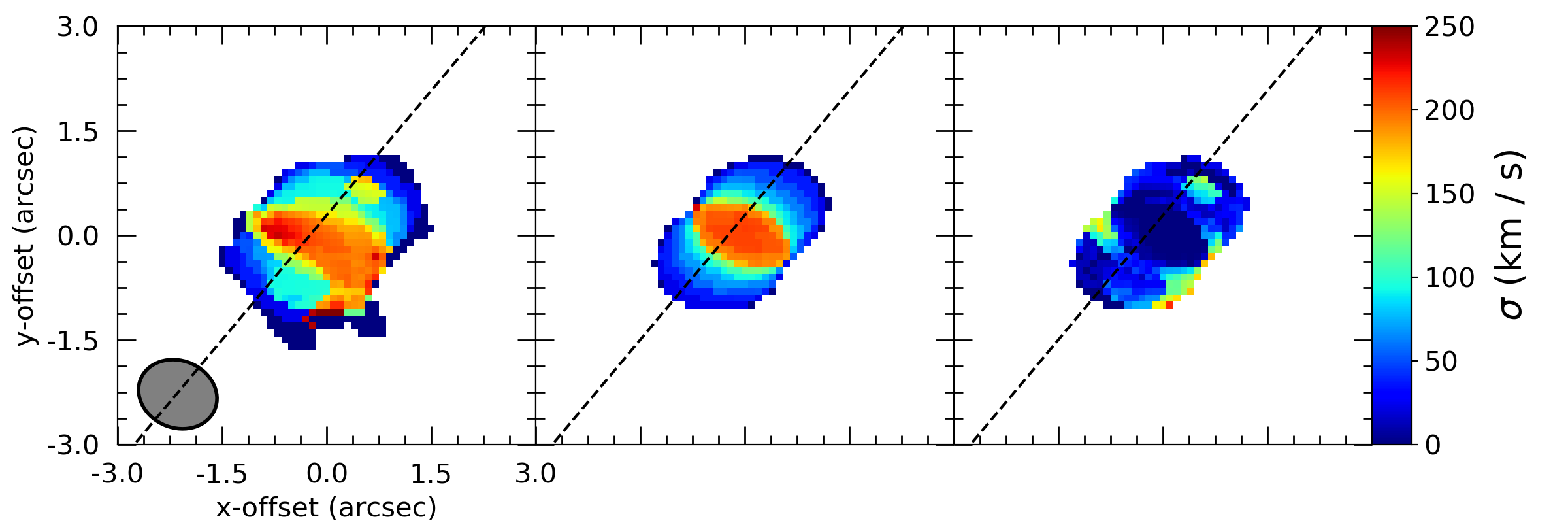} \\[1.0mm]

\includegraphics[width=0.475\textwidth,height=0.125\textheight]{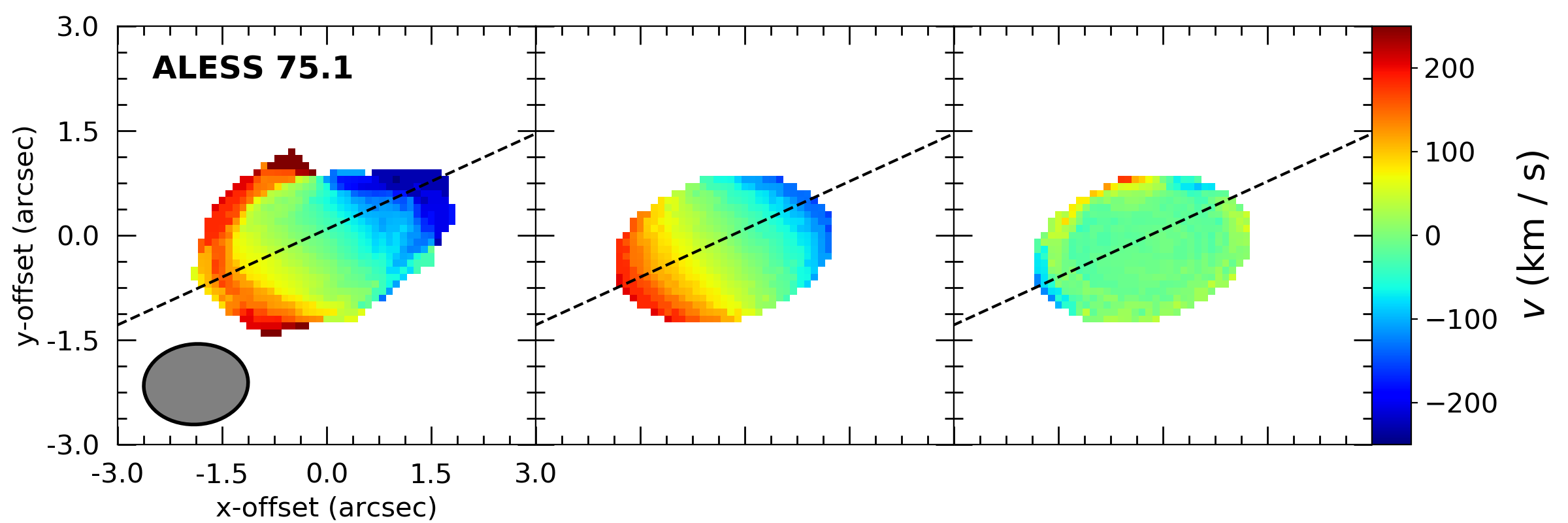} & \includegraphics[width=0.475\textwidth,height=0.125\textheight]{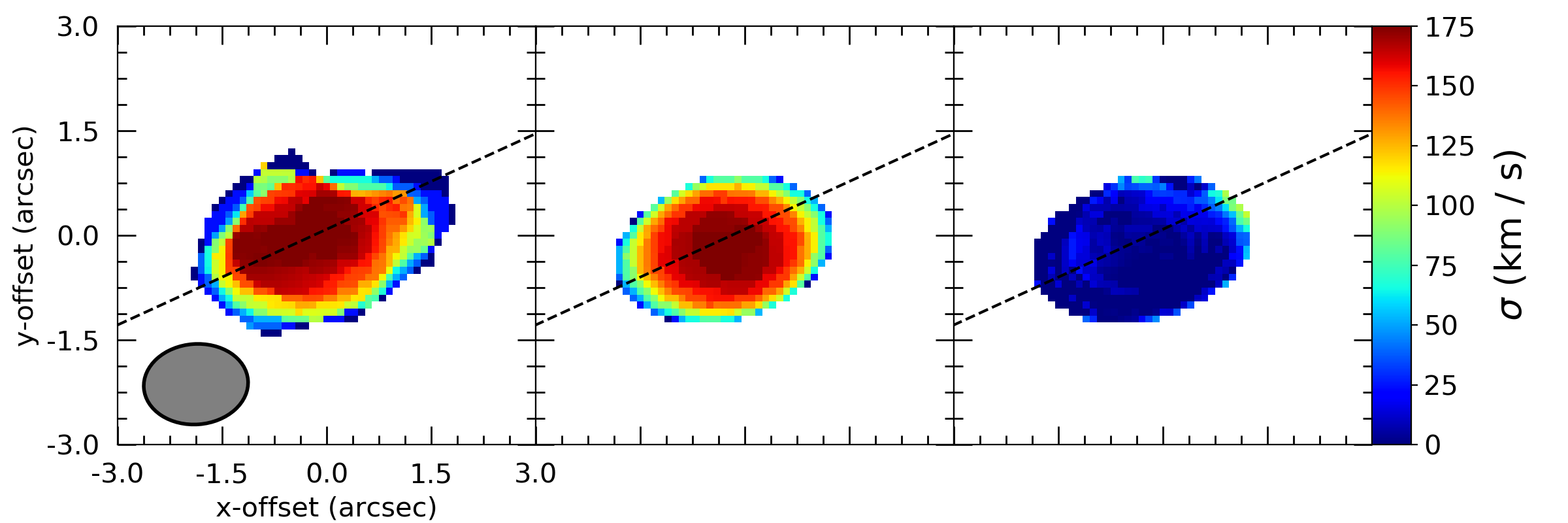} \\[1.0mm]

\end{tabular}
\caption{
  Moment 1 (velocity; left column) and 2 (dispersion; right column) maps for three example sources in our sample (the name of each source is indicated at the top left corner). From left to right in each row we show the observed data, the maximum likelihood model and residual dirty moment maps (the moment maps were computed after masking any emission below $3\sigma$). In these example sources (as well as for the other sources regarded as well described by our rotating disk model; see Section~\ref{sec:model_observations}) the model does a good job at reproducing the observed data. The black dotted line in each panel corresponds to the best-fit position angle, $\theta$, of the major axis. The black ellipse in the bottom left corner represents the beam. }
\label{fig:figure_2}
\end{figure*}

\subsection{Best-fit models} \label{sec:model_observations}

As we already discussed in section~\ref{sec:SNR_and_classification}, we attempted to model all 30 sources listed in Table~\ref{tab:properties}, but here we only discuss the results for the 20 sources with SNR\,$> 8$. For the remaining ten sources with SNR\,$< 8$ the best-fit parameters were largely unconstrained due to the low SNR of our data, effectively not allowing us to characterise their kinematics (rotation/dispersion supported or merger/disrupted). These ten sources, which are listed at the bottom of Table~\ref{tab:properties} with Class \Romannum{3}, will not be considered as part of our sample for the rest of this work. 

In Figure~\ref{fig:figure_1} we show the results from our dynamical modelling analysis for one of the sources in our sample, ALESS 122.1. The different rows correspond to the dirty channels maps of the data, model (ML) and residuals (data-model). This source was previously studied in \cite{2018ApJ...863...56C} using the same dynamical modelling software but instead the analysis was carried out in image-plane rather than the $uv$-plane. Our inferred values are in good agreement with those reported in \cite{2018ApJ...863...56C}, but their claimed uncertainties are significantly lower (e.g., $V_{\rm max} = 564 \pm 8$ km\,s$^{-1}$ and $\sigma = 129 \pm 1$ km\,s$^{-1}$; Calistro Rivera, private communication). This demonstrates one of the advantages of carrying out the analysis in the $uv$-plane where the observational errors (i.e., error on the real and imaginary components of the visibilities) are better defined compared to the uncertainties defined in the image-plane (i.e., the rms noise of cleaned images). Complementary to this figure are the dirty $1^{\rm st}$-moment (intensity-weighted velocity; left column) and $2^{\rm nd}$-moment (intensity weighted dispersion; right column) maps, which we show in Figure~\ref{fig:figure_2} for three of the sources in our sample. We reiterate that while our analysis is carried out in the $uv$-plane, it is useful to visually inspect both the residual (data versus model) dirty cubes and moment maps to determine if a fit is successful.

Among the 20 sources with SNR\,$> 8$, our visual classification of their velocity maps places seven of them in Class \Romannum{2} (ALESS 003.1, 006.1, 009.1, 034.1, 088.1, 101.1, 112.1; see Section~\ref{sec:SNR_and_classification}) because they either display complex velocity fields or the resolution is too poor to characterise them. For five of these seven sources -- those with suffiecient resolution -- our model converged to a solution that was rather unphysical, specifically the velocity dispersion converged to values $>200$\,km\,s$^{-1}$. Visually inspecting the residual 3D cubes, using the best-fit model, it is not obvious that the fit was unsuccessful. However if we instead inspect the 1D spectrum we find that the model is not able to reproduce the data: the wings of the model spectrum are wider compared to the data, effectively trying to model these sources as if they are dispersion dominated.
Somewhat similar behavior of the {\sc Galpak3D} model (i.e., converging to non-physical values, in addition to the velocity dispersion, for the maximum rotational velocity and/or the effective radius as well) was also reported in \cite{2021MNRAS.503.5329H} who studied the kinematics of a sample of (ultra) luminous infrared galaxies (U/LIRGs) at $z\sim$\,2--2.5. For these five sources, the inability of our model to fit the data is not surprising. The model we use does not have the flexibility to account for complex features that deviate from the simple assumption of a regularly rotating disk and can therefore converge to solutions that are not physical, but happen to fit the data better than a model with more physical parameters.   To circumvent this we also tried different functional forms for both the flux profile (e.g., Gaussian) and the velocity profile (e.g., Freeman model), but the same behavior was observed.    As the kinematics of these systems are not well-described by a simple rotating disk, we suggest that the complex kinematics in these sources may indicate that they are currently in a merger phase or recently had a merging event which led to their gas dynamics being significantly disturbed from ordered circular motions.   Additional  support for this hypothesis may be provided by archival Hubble Space Telescope (HST) observations of some of these sources (see Figure~\ref{fig:figure_1_for_Appendix_A} in Appendix~\ref{sec:Appendix_HST}), that show   clumpy structures\footnote{We caution that dusty galaxies showing clumpy structure in the restframe UV/optical does not necessarily mean that they are in the process of merging or recently had a merging event,  but this morphology may instead reflect structured dust extinction within these systems.}.

In addition to these five sources with complex kinematics, our attempt to model a further source, ALESS 67.1, also resulted in large residuals in the velocity map. The kinematics of ALESS 67.1 were previously analysed by \cite{2017ApJ...846..108C} where the authors suggest that this source is consistent with a merger scenario. In addition, the HST image of this source reveals  tidal features that considerably strengthen this hypothesis. Therefore, it is not surprising that our thick rotating disk model did not fit the data well. In this work we also classify ALESS 67.1 as a likely merger and remove it from the sample, leaving 12 galaxies whose kinematics appear to be well described by a rotating disk model. We caution that in a recent study by \cite{2022A&A...667A...5R} the authors showed that mergers can  be missclassified as rotationally supported disks when the resolution of the data is low, which could be the case for some of our source in this work. Nevertheless, as this is the best we can do with the current data at hand for the remainder of this paper we assume that we have correctly classified these 12 sources as rotationally supported disks. We further examine the reliability of our recovered parameters by applying our modelling procedure on realistic simulated observations that mimic our observed data in both resolution and SNR. The results from our simulations are shown in Appendix~\ref{sec:Appendix_Simulations} where we find that we can accurately recover the true parameters without any biases.

\begin{table*}
	\centering{
	\caption{The best-fit maximum likelihood (ML) and maximum posterior (MP) parameters, $\boldsymbol{\theta}$, of our model. The upper and lower limits (1$\sigma$) of the best-fit MP parameters are computed from the 1$\sigma$ errors (see main text). An inclination of $i = 0$ (deg) means that the source is face-on, while $i = 90$ (deg) edge-on and the position angle, $\theta$, is defined counterclock-wise from North. In the last two columns we list the circularized velocity computed at twice the effective radius (Eq.~\ref{eq:v_circ}) and the dynamical mass computed at a radius, $r=$ 10 kpc (Eq.~\ref{eq:m_dyn}).}
	\begin{tabular}{lcccccccccccc}
	\\
	\hline\\[-6.0mm]
    \hline
	\\[-3.0mm]
	  ID & $r_{\rm e}$ & $\theta$ & $i$ & $V_{\rm max}$ & $\sigma$ & $V_{\rm circ}(r=2\,r_{\rm e})$ & $M_{\rm dyn}(r=10 \, {\rm kpc})$ \\[1.0mm]
	  & (arcsec) & (deg) & (deg) & (km / s) & (km / s) &  (km / s) & ($\rm 10^{11} \, M_{\odot}$) \\[1.0mm]
	\hline
	\\[-3.0mm]

	
	
	\textbf{007.1} & $0.61_{\, -0.07}^{\, +0.06}$ & $153_{\, -8}^{\, +8}$ & $34_{\, -7}^{\, +6}$ & $481_{\, -72}^{\, +86}$ & $42_{\, -18}^{\, +17}$ & 399 $\pm$ 68 & 3.9 $\pm$ 1.3
	\\[1.5mm]
	
	
	\textbf{017.1} & $0.41_{\, -0.07}^{\, +0.06}$ & $32_{\, -15}^{\, +13}$ & $36_{\, -15}^{\, +14}$ & $417_{\, -99}^{\, +150}$ & $73_{\, -16}^{\, +16}$ & 255 $\pm$ 81 & 2.2 $\pm$ 1.3
	\\[1.5mm]
	
	
	\textbf{022.1} & $0.44_{\, -0.06}^{\, +0.06}$ & $57_{\, -7}^{\, +7}$ & $34_{\, -10}^{\, +8}$ & $347_{\, -87}^{\, +85}$ & $15_{\, -12}^{\, +10}$ & 334 $\pm$ 84 & 2.8 $\pm$ 1.4
	\\[1.5mm]
	
	
	\textbf{041.1} & $0.29_{\, -0.03}^{\, +0.01}$ & $75_{\, -2}^{\, +3}$ & $69_{\, -4}^{\, +4}$ & $385_{\, -8}^{\, +10}$ & $50_{\, -7}^{\, +7}$ & 400 $\pm$ 15 & 4.0 $\pm$ 0.3
	\\[1.5mm]
	
	\textbf{049.1} & $0.19_{\, -0.07}^{\, +0.05}$ & $120_{\, 5}^{\, -5}$ & $90_{\, -13}^{\, +0}$ & $347_{\, -55}^{\, +25}$ & $91_{\, -30}^{\, +23}$ & 350 $\pm$ 55 & 5.1 $\pm$ 1.7
	\\[1.5mm]
	
	\textbf{062.2} & $0.28_{\, -0.15}^{\, +0.06}$ & $166_{\, -6}^{\, +7}$ & $73_{\, -9}^{\, +17}$ & $254_{\, -167}^{\, +63}$ & $99_{\, -66}^{\, +23}$ & 334 $\pm$ 135 & 4.5 $\pm$ 3.2
	\\[1.5mm]
	
	\textbf{065.1} & $0.21_{\, -0.5}^{\, +0.04}$ & $158_{\, -10}^{\, +14}$ & $58_{\, -17}^{\, +13}$ & $431_{\, -81}^{\, +79}$ & $41_{\, -28}^{\, +24}$ & 371 $\pm$ 69 & 4.8 $\pm$ 1.7
	\\[1.5mm]
	
	\textbf{066.1} & $0.43_{\, -0.09}^{\, +0.06}$ & $109_{\, -46}^{\, +77}$ & $52_{\, -14}^{\, +15}$ & $420_{\, -123}^{\, +138}$ & $92_{\, -18}^{\, +18}$ & 344 $\pm$ 85 & 3.8 $\pm$ 1.8
	\\[1.5mm]
	
	
	\textbf{071.1} & $0.48_{\, -0.05}^{\, +0.06}$ & $138_{\, -4}^{\, +5}$ & $71_{\, -8}^{\, +7}$ & $368_{\, -19}^{\, +22}$ & $67_{\, -24}^{\, +24}$ & 381 $\pm$ 25 & 3.6 $\pm$ 0.5
	\\[1.5mm]
	
	\textbf{075.1} & $0.32_{\, -0.02}^{\, +0.02}$ & $115_{\, -2}^{\, +3}$ & $55_{\, -6}^{\, +4}$ & $386_{\, -37}^{\, +27}$ & $108_{\, -10}^{\, +9}$ & 389 $\pm$ 29 & 4.8 $\pm$ 0.6
	\\[1.5mm]
	

	\textbf{098.1} & $0.22_{\, -0.01}^{\, +0.02}$ & $100_{\, -3}^{\, +3}$ & $41_{\, -13}^{\, +7}$ & $555_{\, -24}^{\, +48}$ & $26_{\, -17}^{\, +17}$ & 519 $\pm$ 44 & 7.1 $\pm$ 1.2
	\\[1.5mm]
	
	
	
	\textbf{122.1} & $0.62_{\, -0.08}^{\, +0.07}$ & $86_{\, -6}^{\, +7}$ & $55_{\, -6}^{\, +8}$ & $564_{\, -17}^{\, +44}$ & $157_{\, -18}^{\, +19}$ & 533 $\pm$ 37 & 6.5 $\pm$ 0.8
	\\[1.5mm]
	
	\hline\\[-6.0mm]
	\hline
	\label{tab:table1}
	\end{tabular}
	}
      \end{table*}

      
\section{Results and Discussion} \label{sec:section_4}

To summarise the modelling in the previous section: from our parent sample of 30 DSFGs, we find that we are unable to constrain the kinematics of the ten galaxies with integrated CO emission line signal-to-noise ratios of SNR $<$ 8 (Class \Romannum{3}).  From the remaining 20 DSFGs with CO SNR $>$ 8, we have 12 where our  rotating disk model provides a good description of the kinematics, indicating that the molecular gas reservoirs in these systems are likely to be in rotationally supported disks. 

\subsection{Dynamical parameter relations}

In this section we discuss the dynamical properties of our sample using various scaling relations involving their observable properties. We  also compare them with less extreme star-forming samples from the literature to see how they relate to the wider population in terms of their dynamics.

The best-fit maximum posterior (MP) model parameters from the non-linear searches for the 12 galaxies, for which our thick rotating disk model provides a good fit, are presented in Table~\ref{tab:table1}. We used uniform priors for all the model parameters except for the total intensity, $I$, for which we used a log-uniform prior. The MP solution is computed as the median of the 1-dimensional (1D) marginalised posterior distributions of each parameter. The upper and lower limits on the MP solution reflect the 1$\sigma$ error of these parameters and are computed from the 68\% percentile of the posterior distributions. 

Along with the best-fit model parameters in Table~\ref{tab:table1} we list some additional quantities, namely the circularized velocity computed at a twice the effective radius and the dynamical mass computed within a radius of $r=$\,10 kpc. The circularized velocity, $V_{\rm circ}$, is defined as the rotational velocity corrected for asymmetric drift \citep[][]{2010ApJ...725.2324B}. Several studies have suggested that star-forming disk galaxies at high redshift are typically more turbulent \citep[e.g.][]{2015ApJ...799..209W, 2016MNRAS.460.1059J} compared to local analogues. This turbulent motion needs to be accounted for as it contributes to the dynamical support of a system. Assuming a constant velocity dispersion profile and considering that the vertical profile of our thick disk model does not depend on the radius, one has 
\begin{equation}
    V_{\rm circ} (r) = \sqrt{ \, V_{\rm rot}^2(r) - \sigma^2 \frac{dln \Sigma}{dlnr}} \, ,
\end{equation}
which for an exponential profile, $\Sigma \propto \exp{\left(-r / r_e\right)}$, reduces to
\begin{equation}\label{eq:v_circ}
    V_{\rm circ} (r) = \sqrt{ \, V_{\rm rot}^2(r) + 1.68 \sigma^2 \left( \frac{r}{r_e} \right)} \, ,
\end{equation}
where $V_{\rm rot}$ is the rotational speed of the gas (its functional form is defined in Section~\ref{sec:kin_model}), $\sigma$ is the velocity dispersion and $r_e$ is the effective radius of the disk. The dynamical mass, $M_{\rm dyn}$, which is a measure of the total mass of the system, is computed as 
\begin{equation}\label{eq:m_dyn}
    M_{\rm dyn}(r) = \frac{r V^2_{\rm circ}(r)}{G} \,
\end{equation}
where again we use the circularized rotational velocity to account for the effect on turbulent motions.

\begin{figure}
    \centering
    \includegraphics[width=0.95\columnwidth, height=0.28\textheight]{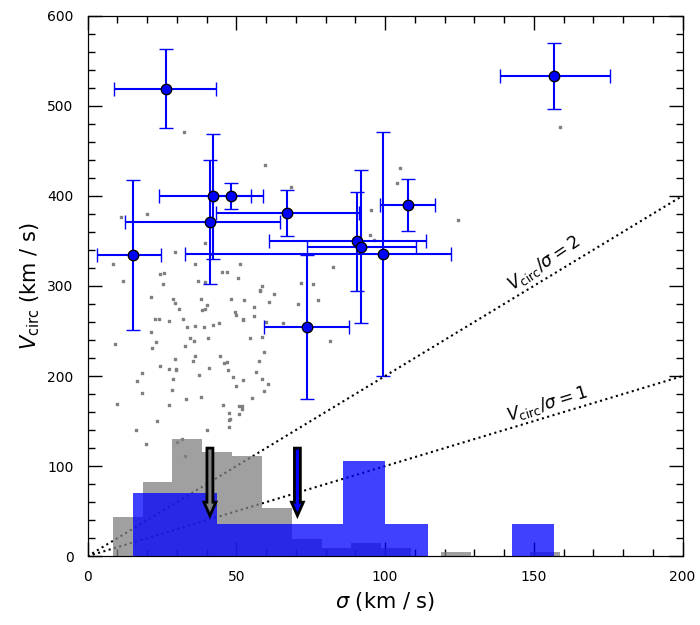}
    \caption{Circular rotational velocity, $V_{\rm circ}$, as a function of the velocity dispersion. The blue points correspond to our sample of massive dusty star-forming galaxies while the gray points correspond to sources from the KMOS$^{\rm 3D}$ survey which are more representative of the ``typical'' star-forming galaxies at $z \sim 2$ ($V_{\rm circ}$ is computed at twice the effective radius, $r = 2 \times R_{\rm eff}$, for both samples). The histograms at the bottom of the figure correspond to the distribution of velocity dispersions for the two populations (arrows indicate the medians). The dotted lines are the $1:1$ and $2:1$ relation between the two properties, highlighting that all source shown in this figure are classified as rotationally supported.}
    \label{fig:Vcirc_vs_sigma}
\end{figure}

\begin{figure*}
    \centering
    \includegraphics[width=0.6\textwidth,height=0.245\textheight]{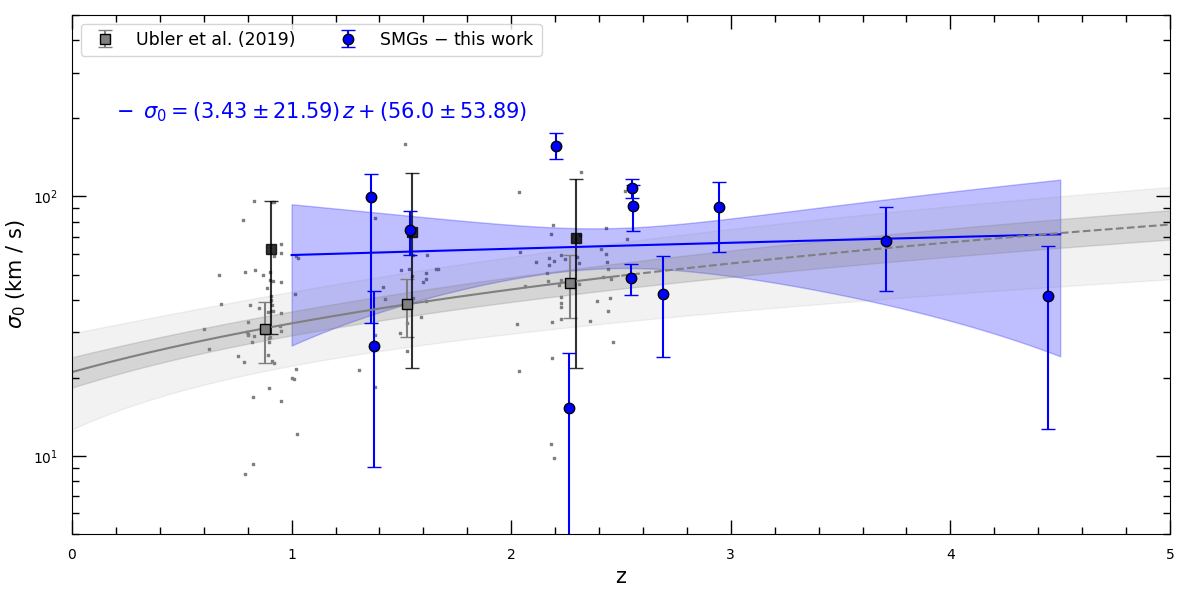}
    \caption{Velocity dispersion, $\sigma_0$, as a function of redshift, $z$, for the 12 sources with SNR $>$ 8 in our sample that have kinematics that are well described by a rotating disk model. These show a modest increase in $\sigma$ with redshift, but this is not statistically significant. We also show points from the KMOS$^{\rm 3D}$ survey (grey; \"{U}bler et al. (2019))\iffalse\citep[grey;][]{2019ApJ...880...48U}\fi, where the grey line with its corresponding uncertainty (grey shaded regions; 1 and 2 $\sigma$) is the best-fit linear relation to these points as derived by \"{U}bler et al. (2019)\iffalse\cite{2019ApJ...880...48U}\fi. The blue line and its corresponding uncertainty (blue shaded region; 1$\sigma$) is the best-fit linear relation to our sample of sources. We also re-calculated averages from the KMOS$^{\rm 3D}$ points, in the same three redshift bins, using only those sources with $V_{\rm circ} > $\,325 km s$^{-1}$ to match  our sample. These are shown as the square black points and agree with the relation derived from our sample (i.e. consistent with no redshift evolution).}
    \label{fig:sigma_vs_z}
  \end{figure*}

\begin{figure}
    \centering
    \includegraphics[
        width=0.45\textwidth,
        height=0.5\textheight
    ]{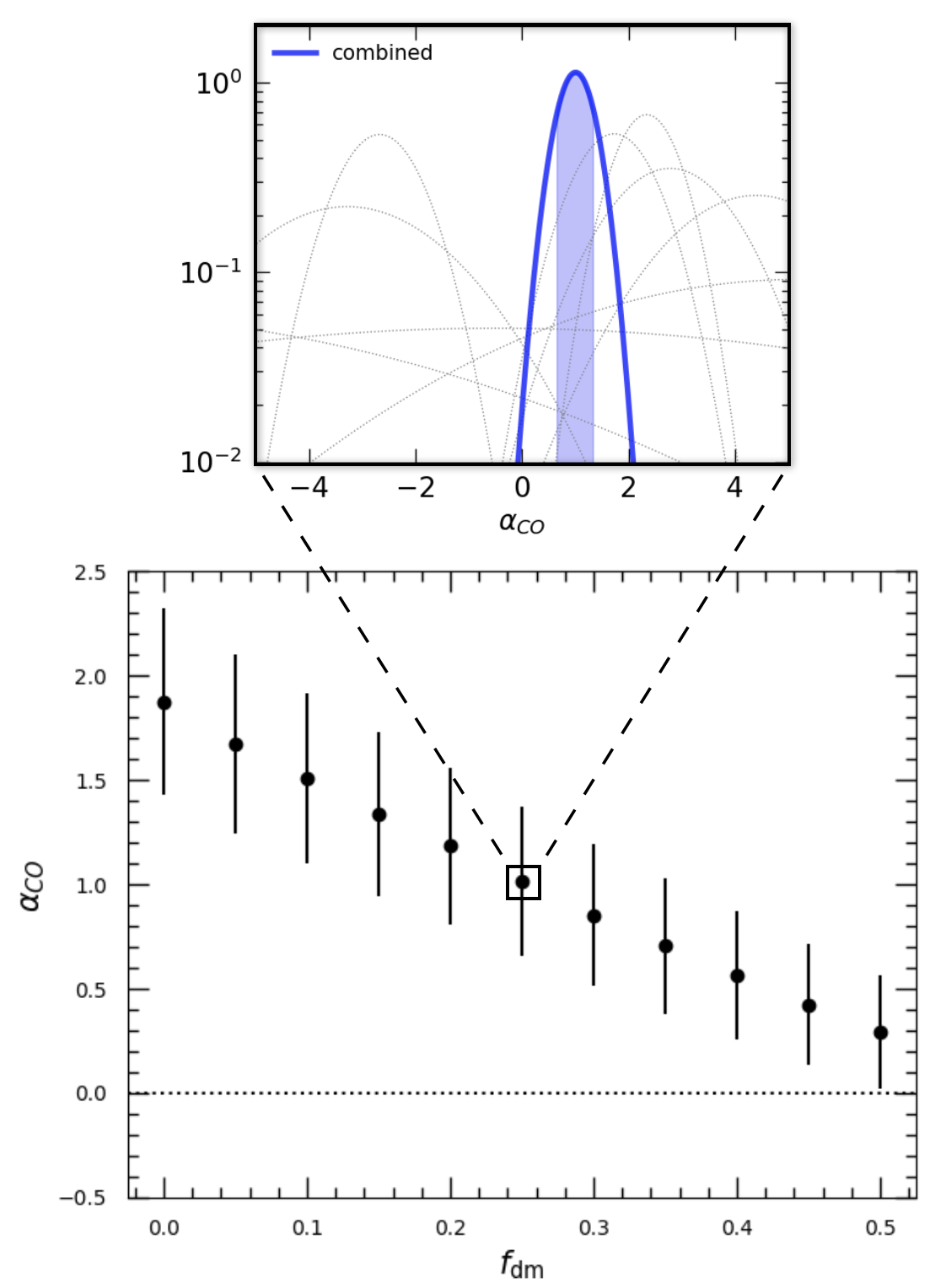}
    \caption{The CO--H$_2$ conversion factor, $\alpha_{\rm CO}$, as a function of the dark-matter fraction, $f_{\rm DM}$ (within a radius of $ r < 10$\,kpc). The zoom in plot at the top of the figure shows the individual posteriors distributions (grey), for a dark matter fraction of  , of the $\alpha_{\rm CO}$ value for each source in our sample as well as the combined distribution (blue) which is computed by multiplying together all the individual distributions.}
    \label{fig:alpha_CO}
\end{figure}

\subsubsection{$V_{\rm circ} - \sigma$} 

We first look at the relation in our sample between the circularized velocities, $V_{\rm circ}$, and the velocity dispersion, $\sigma$, which is shown in Figure~\ref{fig:Vcirc_vs_sigma}  including a comparison sample of star-forming galaxies from KMOS$^{\rm 3D}$. 
The figure demonstrates that our sample is not comparable to``typical'' star-forming galaxies, but instead represents the most massive sources as judged by their dynamics. The median velocity dispersion of our sample ($\sim$ 75 km s$^{-1}$; blue arrow) is significantly higher than seen in the KMOS$^{\rm 3D}$ sample ($\sim $ 40 km s$^{-1}$; grey arrow). We argue that this to be a consequence of our selection; the most massive star-forming disk galaxies at any redshift also have the higher velocity dispersions. Indeed, if we apply a cut in circularized velocity, $ V_{\rm circ}$, to the KMOS$^{\rm 3D}$ sample above 325 km s$^{-1}$ we find that the median velocity dispersion increases and becomes consistent with our sample (this is important to keep in mind when later we look at the evolution of the velocity dispersion with redshift).

We also looked at the evolution with redshift of the ratio of the circularized velocity with the velocity dispersion, $V / \sigma$. Previous studies have suggested that this ratio decreases with redshift and this evolution is a consequence of the increased velocity dispersion of galaxies at higher redshifts \citep[][]{2016ApJ...819...80P, 2017MNRAS.471.1280T, 2019ApJ...886..124W, 2021MNRAS.503.5329H}. However, we do not see a clear trend in our sample (a very mild negative trend is observed when we averaged points in bins of redshift, but this was not statistically significant).

\subsubsection{$\sigma - z$} 

We next take a look at the evolution of the intrinsic velocity dispersion, $\sigma$, with redshift, $z$, for our sample.   Again we use sources from the KMOS$^{\rm 3D}$ survey as the reference sample of typical star-forming galaxies at high redshift (grey points).   Previous studies have claimed that the velocity dispersion of star-forming, rotationally supported galaxies increases with redshift \citep[e.g.][]{2019ApJ...880...48U} and proposed that the increased turbulence is mainly the result of gravitational instabilities \citep[e.g.][]{2019ApJ...880...48U}. 
In order to quantify the redshift dependence of the velocity dispersion for our sample we fit a linear relation, $\sigma = a z + b$ and plot this in Figure~\ref{fig:sigma_vs_z}, where the  shaded region corresponds to the 1$\sigma$ error. 

Our fitting suggests no evolution for the velocity dispersion with redshift for our sample (the slope is positive but not statistically significant). Furthermore, we note that our relation is above the one reported with the KMOS$^{\rm 3D}$ sample, although still consistent within the 1$\sigma$ errors. If one ignores selection effects, it will seem surprising that our CO-based relation is above the H$\alpha$-derived one from \cite{2019ApJ...880...48U}. This is because most previous studies claim that velocity dispersions derived from observations of the molecular gas are typically lower than those derived from ionized gas \citep{2018ApJ...860...92L, 2019A&A...631A..91G, 2021ApJ...909...12G}. Here, however, we stress that we are not comparing the same galaxy populations in terms of their instrinsic dynamical properties. As we showed in Figure~\ref{fig:Vcirc_vs_sigma} our sources are more massive than the typical star-forming galaxy population, and so it is not surprising that their velocity dispersions are also elevated. In order to make a more fair comparison we apply a cut in circularized velocity, $V_{\rm circ} > 325 $ km s$^{-1}$ to the KMOS$^{\rm 3D}$ sample and re-calculate the average points in the same bins of redshift as \cite{2019ApJ...880...48U}. There are shown as the black points in  Figure~\ref{fig:sigma_vs_z} and agree with the relation derived from our sample (also suggesting no redshift evolution).   

We note that in recent years there have been many high-resolution studies on the dynamics of both SFGs and DSFGs based on observations of various CO emission lines \citep[][e.g.]{2018ApJ...854L..24U, 2022A&A...664A..63X, 2023A&A...672A.106L, 2023arXiv230316227R} as well as [C{\sc ii}] \citep[][e.g.]{2020Natur.584..201R, 2021MNRAS.507.3952R, 2021Sci...371..713L, 2023MNRAS.521.1045R}. We opted not to include these sources in Figure~\ref{fig:sigma_vs_z} for two reasons: First, the majority of these studies focus on individual sources whose selection is different to our ALESS sample. Secondly, the main point of that figure is to exercise caution when comparing velocity dispersion estimates for different populations. In our case the sample we are studying perhaps represents a biased subset of the star-forming galaxy populations, specifically the most massive ones.

\subsection{Dynamical constraints on $\alpha_{\rm CO}$}\label{sec:sec_alphaCO}

In addition to the dynamical scaling relations discussed above, our kinematic estimates of the masses of these galaxies also allow us to constrain a critical parameter used to estimate the gas masses of these systems: $\alpha_{\rm CO}$. 

The bulk of the molecular gas in the interstellar medium of galaxies is in the form of molecular hydrogen, H$_2$. Unfortunately, due to H$_2$ being symmetric molecule, the conditions needed to observe its transitions in emission are extreme ($T > 500 $\,K or a strong UV radiation field), compared to  typical ISM conditions ($T \sim 20 $\,K). In order to overcome this limitation most studies typically use CO, which is the second-most abundant molecule,  as an indirect tracer of the total molecular gas in the ISM. The problem with using the CO molecule to estimate molecular gas masses is that one has to assume a conversion factor, otherwise known as the CO-to-$H_2$ conversion factor, $\alpha_{\rm CO}$ \citep[][]{2013ARA&A..51..207B}. 
        
There have been several attempts in previous studies to estimate the value of $\alpha_{\rm CO}$. The general picture that emerged from these studies is that there is a dichotomy in the measured value of $\alpha_{\rm CO}$ between ``normal'' star-forming galaxies \citep[e.g.][]{2010ApJ...713..686D} and ``starburst'' galaxies \citep[e.g.][]{1998ApJ...507..615D, 2012ApJ...760...11H, 2013MNRAS.429.3047B}. In addition, the value of $\alpha_{\rm CO}$ has been shown to strongly depend on the gas phase metallicity \citep[e.g.][]{2013ARA&A..51..207B, 2018A&ARv..26....5C}, among other factors \citep[][]{2020ARA&A..58..157T}. 
       
Here we attempt to estimate  $\alpha_{\rm CO}$ following the same approach as previous studies in the literature \citep[e.g.][]{2018ApJ...863...56C}. Specifically, we use the dynamical mass within a fixed radius, $M_{\rm dyn}$, and equate it to the sum of the masses of the different galactic components (dark matter and baryons), 
\begin{equation}
    M_{\rm dyn} = M_{\rm gas} + M_{\star} + M_{\rm DM},
\end{equation}
Expressing the molecular gas mass as, $M_{\rm gas} = \alpha_{\rm CO} \, L_{CO(1-0)}$, and the dark-matter mass as, $M_{\rm DM} = f_{\rm dm} * M_{\rm dyn}$, the mass equation above, becomes,
\begin{equation} \label{eq:a_CO}
    \alpha_{\rm CO} = (M_{\rm dyn} - M_{\star} - f_{\rm dm} * M_{\rm dyn}) / L_{\rm CO(1-0)} \, ,
\end{equation}
where $L_{\rm CO(1-0)}$ is the CO (1$-$0) line luminosity and $f_{\rm dm}$ is the dark-matter fraction.
        
Before we calculate the value of $\alpha_{\rm CO}$ we need to make some assumptions about the various terms on the right-hand side of Eq.~\ref{eq:a_CO}. First, we chose a fixed radius, $r = $ 10 kpc, within which we calculate the dynamical mass.  
This value is close to twice the median effective radius, $r_e$, for the sample and thus should contain the bulk of the baryonic material in these systems.
Secondly, we assume a fixed dark matter fraction, $f_{\rm dm} = 0.25$, which is a reasonable value at the radius we choose to estimate the dynamical mass \citep[e.g.][]{2017Natur.543..397G, 2021A&A...653A..20S}. In theory, instead of adopting a fixed value for the dark matter fraction, one could directly constrain this value by incorporating the dark-matter contribution into the dynamical modelling analysis, as is done in some recent studies \citep[e.g.][]{2020ApJ...902...98G, 2022A&A...658A..76B}. However, our observations lack the necessary resolution and SNR to be able to  independently constrain the contribution from both the baryonic and dark matter components. Finally, we use the stellar masses obtained via SED-fitting \citep[][]{2015ApJ...806..110D} which are listed in Table~\ref{tab:properties}. We note, however, that stellar-mass estimates can be uncertain, especially in starburst systems for which star formation histories (among other ingredients of the SED models) are poorly constrained. One approach to account for our lack of knowledge in the assumptions we make during the SED-fitting is to substitute the stellar mass in our mass equation with the term $L_{\rm H} \left(\rm M_{\star} / L_{\rm H}\right)$, where $L_{\rm H}$ is the rest frame $H$-band luminosity corrected for dust obscuration. If we were to make this substitution, we could allow the ratio, $M_{\star} / L_{\rm H}$, to be a free parameter and constrain it simultaneously with the $\alpha_{\rm CO}$ as was done in a recent study by \cite{2018ApJ...863...56C}. Exploring this possibility is outside the scope of this paper, however, we note that this ratio is degenerate with the $\alpha_{\rm CO}$ and will result in larger uncertainties for each individual source.
        
We compute the value of $\alpha_{\rm CO}$ and its associated error (via bootstrapping) for each of the sources in our sample. We show these as posterior distributions in the upper panel of Figure~\ref{fig:alpha_CO}. We then combined the individual distribution to compute the  joint constraint on $\alpha_{\rm CO}$ for our sample, shown in the same figure. Our approach yields an estimate of $\alpha_{\rm CO} = 0.92 \pm 0.36$, for a dark-matter fraction of, $f_{\rm dm} = 0.25$, which is consistent with estimates from previous studies \citep[e.g.][]{2008ApJ...680..246T, 2012ApJ...760...11H, 2013MNRAS.429.3047B, 2013ARA&A..51..207B, 2018ApJ...863...56C, 2021MNRAS.501.3926B}. We note, however, that in a recent study by \cite{2022MNRAS.517..962D} the authors report a value of $\alpha_{\rm CO} = 4.0 \pm 0.1$, which is in tension with the limits we place here. Given the differences in sample selection between our work and theirs, where the later included both lensed and unlensed sources, understanding the origin of this difference is not straightforward and outside the scope of this work.

We also explored how the value of $\alpha_{\rm CO}$ changes when varying the adopted dark-matter fraction, which we show in the bottom panel of Figure~\ref{fig:alpha_CO}.   This indicates that we can place a firm upper limit on $\alpha_{\rm CO}$ of $\alpha_{\rm CO} \ls 2$ (assuming no dark matter within 10 kpc radius).  Similarly adopting a dark matter fraction, $f_{\rm dm} > 0.5$, results in the value of $\alpha_{\rm CO}$ becoming  negative.\footnote{We note that there is a significant scatter among the constraints from the individual sources with some of them centered on negative values. Negative $\alpha_{\rm CO}$ values occur because the stellar masses, which are estimated via SED-fitting, are in some cases higher than the estimated dynamical masses, which  can occur due to uncertainties in these measurements \citep[e.g.][]{2016ApJ...831..149W}.}


\subsection{Tully--Fisher relation at high redshift}

Next we use the results we obtained above from our dynamical modelling analysis of the 12 DSFGs with disk-like kinematics, specifically the rotational velocity corrected for inclination, the velocity dispersion and the global constraint on $\alpha_{\rm CO}$ and hence the gas masses, to  study the stellar and baryonic mass Tully--Fisher relations  \citep[sTFR/bTFR;][]{2000ApJ...533L..99M} for this population.

\begin{figure*}
\begin{tabular}{cc}
\includegraphics[width=0.495\textwidth,height=0.395\textheight]{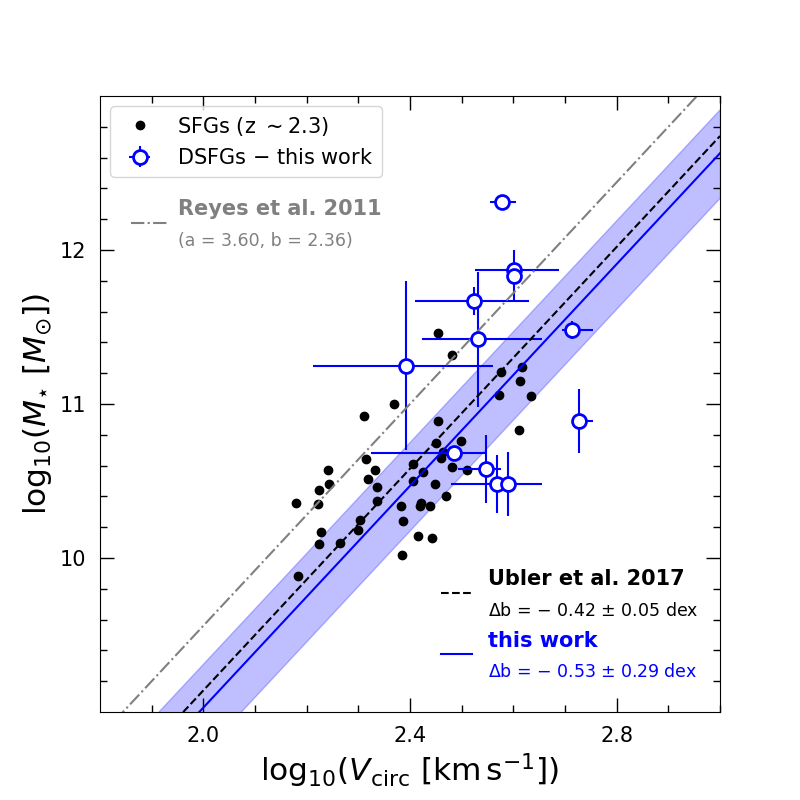} & \includegraphics[width=0.495\textwidth,height=0.395\textheight]{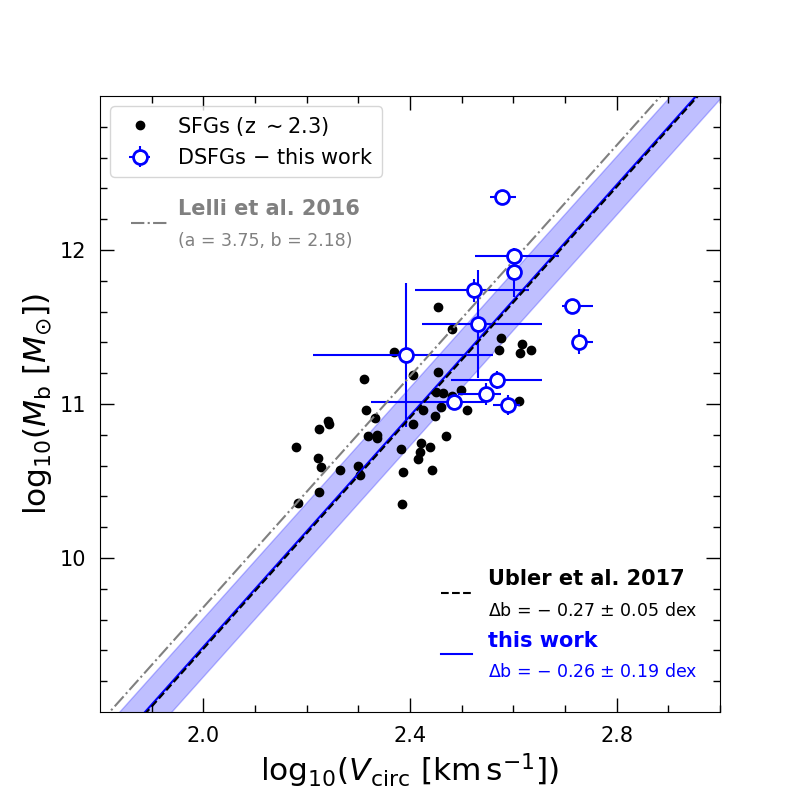}
\end{tabular}
\caption{The stellar and baryonic Tully-Fisher Relation, sTFR (left) and bTFR (right), respectively  for the 12 disk-like sources with SNR $>$ 8 in our sample (blue points) as well as for $z \sim 2.3$ sources from the KMOS$^{\rm 3D}$ (black points).  The best-fit power-law models to our sample are shown as the blue solid lines. For the sTFR and bTFR, compared to the local relations from Reyes et al. (2011)\iffalse\cite{2011MNRAS.417.2347R}\fi and Lelli et al. (2016a)\iffalse\cite{2016AJ....152..157L}\fi, we find offsets in the normalisation of $\Delta = -0.53 \pm 0.29$ and $\Delta = -0.26 \pm 0.19$, respectively, indicating modest evolution of the masses at fixed circular velocity.  These offsets are consistent with those seen by \"{U}bler et al. (2017)\iffalse\cite{2017ApJ...842..121U}\fi for less active galaxies.}
\label{fig:TFR}
\end{figure*}

The Tully-Fisher relation  holds important information about the interplay between the build up of galaxies and the dark matter halos in which they grow. It connects an observable of the baryonic mass, where in this case we consider both the stellar mass alone and the stellar + gas mass, with a proxy for the total potential of the halo, the circular velocity. Following \cite{2017ApJ...842..121U} we use the circularized velocity as it is directly connected to the total potential of the halo for sources with elevated velocity dispersions, as are the sources in this study. In the left and right panels of Figure~\ref{fig:TFR} we show the sTFR and bTFR, respectively, and compare these to the less active galaxies from the KMOS$^{\rm 3D}$ survey.

In order to quantify the relation between the observed mass (stellar or baryonic) and the circularized velocity, we fit our data points using a linear regression model of the form:
\begin{equation}\label{eq:bTFR}
    M = V_{\rm circ}^a 10^b \, ,
\end{equation}
where $a$ and $b$ are constant parameters, which we would like to constrain, and $M$ corresponds to either the stellar or the baryonic mass, where the latter is the sum of the stellar and gas mass. For the fitting, we use an orthogonal distance regression method taking into account the error in both the $x$ and $y$ coordinates (we use the {\sc scipy.odr} package). During the fitting process we actually fix the value of the slope, $a$, to values found in local studies and only optimize the offset, $b$. This is a common practice in high-redshift studies of the TFR \citep[e.g.][]{2009ApJ...697..115C, 2016MNRAS.460..103T, 2016ApJ...819...80P, 2017ApJ...842..121U} simply because the dynamic range in mass is typically very limited (which is particularly evident for our sample of sources, mainly probing the high-mass end of the TFRs) and is therefore not easy to robustly constrain. The results are then quoted in terms of a zero-point offset compared to the $z\sim$\,0 value, $\Delta b$.  For the sTFR and bTFR we adopt the local slopes of $\alpha = 3.60$ \citep[][]{2011MNRAS.417.2347R} and $\alpha = 3.75$ \citep[][]{2016ApJ...816L..14L}, respectively. These are the same values that were used in \cite{2017ApJ...842..121U} with which we want to eventually compare our results. 

The best-fit offsets, $b$, we obtain from our fitting are $b = 1.83 \pm 0.29$ and $b = 1.92 \pm 0.19$, translating to zero-point offsets relative to the $z\sim$\,0 values  of $\Delta b = -0.53 \pm 0.29$ and $\Delta b = -0.26 \pm 0.19$ for the sTFR and bTFR, respectively. The power-law models are shown in each of the panels in Figure~\ref{fig:TFR}, with the  shaded area corresponding to the 1$\sigma$ errors. The values we obtain are in good agreement with those found for less-active galaxies in \cite{2017ApJ...842..121U} for both the sTFR ($\Delta b = -0.42 \pm 0.05$) and bTFR ($\Delta b = -0.27 \pm 0.05$).
\footnote{We note that the scatter in the sTFR appears larger than the individual errors, suggesting there may be a second parameter at work.  However, we have searched for correlations of the offsets from the relation  with other  properties derived from the SED fitting (e.g., age, $A_v$) but found no significant correlations.}

We caution here that our sample of sources spans a wide range in redshift compared to the sample in \cite{2017ApJ...842..121U}. This could potentially introduce a bias in the zero-point offsets that we infer and will definitely contribute to the scatter of the relation. When splitting our sample in redshift, below and above $z \sim $\,2.5 (roughly the median for our sample), and repeating the power-law fit we find no significant differences in the inferred zero-point offsets; however, the errors are significant due to the small number of data points in each redshift bin.

The interpretation of our results is rather straightforward. Our population of massive dusty star-forming galaxies with disk-like kinematics appears to represent the high-mass end of the Tully-Fisher relation traced by the more typical star-forming galaxies at high redshift. While the DSFGs have higher star-formation rates, higher gas and stellar masses, than the more typical galaxies surveyed by KMOS$^{\rm 3D}$, they also reside in more massive dark matter haloes, which ultimately translates to baryon fractions (in the case of the bTFR for example) that are similar to these more typical star-forming galaxies.  Hence, the key conclusion from this comparison is the massive nature of these disk-like DSFGs, that appear to just extend the scaling relations found in less massive galaxies.  

\subsection{The descendants of massive DSFGs}\label{sec:SMG_descendants}

In this final subsection we use our kinematic modelling of the DSFG galaxies to  test if there is an evolutionary link between early-type galaxies in the local Universe and massive dusty star-forming galaxies at high redshift by comparing the dynamical properties of these two populations.

\begin{figure*}
    \centering
    \includegraphics[width=0.95\textwidth,height=0.55\textheight]{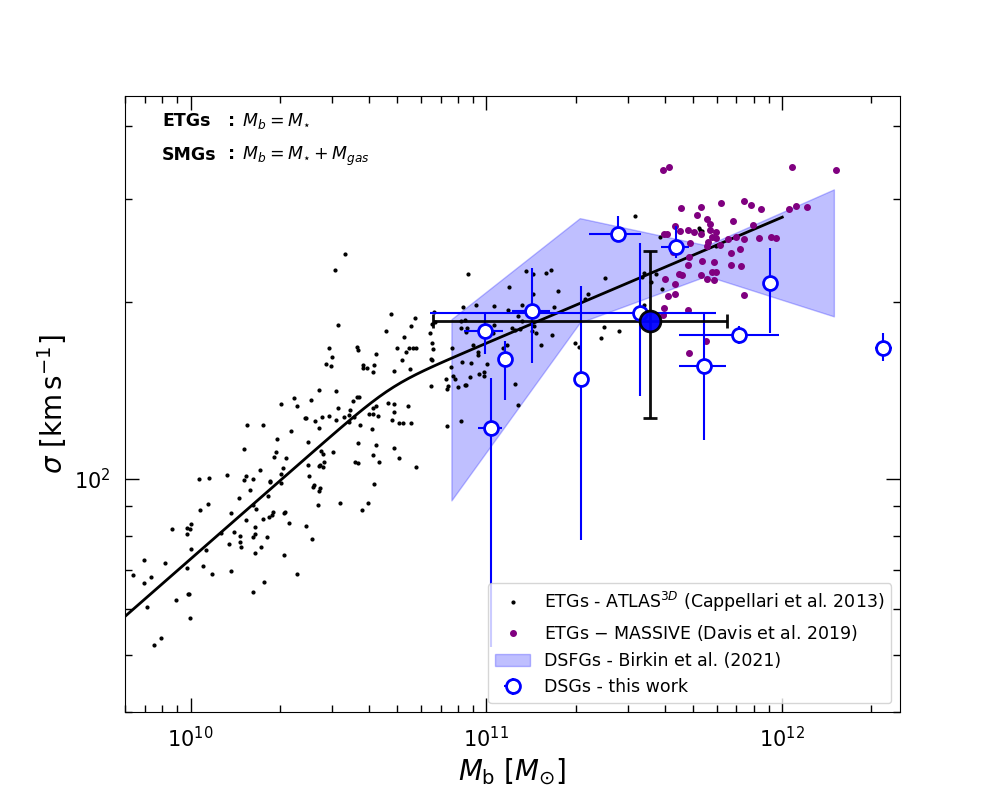}
    \caption{The $M$--$\sigma$ relation for DSFGs (blue unfilled points) and early-type galaxies from two surveys: ATLAS$^{\rm 3D}$ (black points) and MASSIVE (purple points). For early-type galaxies, $\sigma$ correspond to the central velocity dispersion and $M_b$ is the stellar mass (the gas mass in these systems is negligible compared to their stellar mass, where the latter is computed from their $K$-band luminosity). For DSFGs on the other hand, the quantity $\sigma$ is computed from Eq.~\ref{eq:sigma_SMGs} and $M_b$ is the sum of the gas and stellar masses of the galaxies (this assumes that by $z \sim $\,0 these galaxies will have converted all of their gas to stars). For comparison, the  shaded region corresponds to median trend found by Birkin et al. (2021)\iffalse\cite{2021MNRAS.501.3926B}\fi using the parent CO survey sample from which the data analysed in this work were selected (see Section~\ref{sec:section_2}). The larger blue filled point represent the median of all the blue unfilled points where the errors were calculated via boostrap. The solid line shows the trend followed by the local early-type galaxies and is computed using Eq.\ 5 from Cappellari et al. (2013)\iffalse\cite{2013MNRAS.432.1862C}\fi. The parameters of that equation were optimized from fitting directly to the early-type galaxies plotted and as can be seen from the figure it adequately describes points at the high-mass end of the relation.}
    \label{fig:figure_final}
\end{figure*}

In order to investigate the evolutionary link between between these populations we compare their distribution in terms of their baryonic mass, $M_{\rm b}$, and central velocity dispersions (i.e., velocity dispersion, $\sigma_e$, within the half-light radius, $R_e$),  frequently referred to as the $M_{\rm b}$--$\sigma$ relation. The connection between these two galaxy populations, using  their distributions on this plane, was previously studied in a statistical manner by \cite{2021MNRAS.501.3926B}. Here instead we will use the results from the dynamical modelling analysis which allow us to infer both the rotational speed of the gas as well as the velocity dispersion of individual galaxies.

For  comparison we use samples of local massive early-type galaxies from two surveys, ATLAS$^{\rm 3D}$ \citep{2011MNRAS.413..813C} and MASSIVE \citep[][]{2014ApJ...795..158M}. These two samples together span a wide range in stellar mass $M_{\star} \sim 10^{10}$--$10^{12} $\, M$_{\odot}$. The ATLAS$^{\rm 3D}$ survey \citep{2011MNRAS.413..813C} is a volume-limited ($D < $\,42\,Mpc) study of 260 early-type galaxies selected to have absolute $K$-band magnitude of $M_K < -$21.5. Stellar central velocity dispersion measurements for the ATLAS$^{\rm 3D}$ sources are taken from \cite{2013MNRAS.432.1862C}. The MASSIVE survey (Ma et al.\ 2014) is a volume-limited ($D < $\,108\,Mpc) survey of  116 early-type galaxies where galaxies were selected to have absolute $K$-band magnitude of $M_K < -$25.3 mag. Stellar central velocity dispersion for these are taken from \cite{2018MNRAS.473.5446V}. For the samples of early-type galaxies, stellar masses are computed from their $K$-band magnitude which provides a fairly robust approximation of their stellar mass (the $K$ band is less sensitive to dust absorption compared to optical wavelengths and has relatively small mass-to-light ratio variations with star-formation history). 

In order to place our sample of massive DSFGs at high redshift on the $M-\sigma$ plane, along with the samples of early-type galaxies, we first need to make three assumptions. The first assumption that we make is that by $z \sim $\,0 all of the available gas that our sources have will be depleted in order to form stars, without any loss of mass. Therefore, their total baryonic mass at $z \sim 0$ will be the sum of two components, their stellar and gas masses at the redshift of detection. We note, however, that some of their gas might be expelled due to their strong star-formation activity or if they undergo a QSO phase, although it may also subsequently be re-accreted or they may accrete stars and gas though future  mergers.

The second assumption is that the energy of a system will be conserved when transitioning from the active star-forming phase where the system is dominated by rotation, to a dispersion-dominated phase at  $z = 0$. Under this assumption, we can use the virial theorem \citep{2009A&A...504..789E} and equate their virial masses at these two redshift epochs. This allow us to estimate the velocity dispersion of the system in this dispersion-dominated phase from,
\begin{equation}\label{eq:sigma_SMGs}
    \beta \frac{\sigma^2_{\rm e, \, z_0} R_{\rm e, \, z_0}}{G} = \frac{V^2_{\rm circ, max, \, z} R_{\rm e, \, z}}{G}
\end{equation}
where $V_{\rm circ, max, \, z}$ is the maximum circular velocity of the system at the observed redshift (these are the values we inferred from our dynamical modelling analysis; see Table~\ref{tab:table1}), $\beta$ is a constant which for a spherical system takes a value of $\beta \sim 5$ \citep[][]{2013MNRAS.432.1862C}, $R_{e, \, z}$ and $R_{\rm e, \, z_0}$ are the radii before and after the system transitioned to a dispersion-dominated phase.

The final assumption is that the radius of the system after it transitions to a dispersion-dominated phase will be roughly equal the radius it had when it was still in the rotation-dominated phase. We make this assumption to further simplify Eq.~\ref{eq:sigma_SMGs} but we note that these systems potentially do evolve in size (e.g., through minor mergers, which do not significantly increase their stellar mass). However, since the velocity dispersion in early-type galaxies is measured only in the central region, our assumption can be justified.    

Using the assumptions described above, we can place our sample of DSFGs on the same $M$--$\sigma$ plot along with the samples of local early-type galaxies in Figure~\ref{fig:figure_final}.
The estimated baryonic masses of the DSFGs place them at the highest masses seen for early-type (or any) galaxies at $z\sim$\,0, $M_{\rm b} \gs$\,10$^{11}$\,M$_\odot$ \citep[e.g.,][]{2014ApJ...795..158M}, and the dynamical masses we infer for the DSFG support these high masses.  Indeed,  based on the gas dynamics, the DSFGs scatter around the trend line from \cite{2013MNRAS.432.1862C} that describes the relationship between the total stellar mass and the velocity dispersion of stars in local early-type galaxies (see Figure~\ref{fig:figure_final}). The median values of our sample, $M_{\rm b}=($3.6$\pm$2.9$)\times 10^{11}$\,M$_\odot$ and $\sigma=$\,186$\pm$59\,km\,s$^{-1}$ are within $\sim$\,1$\sigma$ of the $z\sim$\,0 trend.  We therefore conclude that there is good agreement on the $M$--$\sigma$ plane between the predicted properties of the DSFGs (subject to the assumptions listed above) and those found for the most massive early-type galaxies found in the Universe at low redshifts.   This provides further support for the existence of an evolutionary link between the formation of the most massive early-type galaxies in the local Universe and massive gas-rich and disk-like DSFGs at high redshift.

Finally, we note that a crude estimate of the space density of the sources in our sample yields a rough volume density of $\sim $\,1\,$\times$\,$10^{-5}$\,Mpc$^{-3}$ (based on a subset of $\sim$\,110 sources from the parent ALESS survey over 0.35 degrees$^2$, which matches the median $S_{\rm 870\mu m}=$\,4.5\,mJy for our sample, and adopting a FWHM of the redshift
distribution for our sources spanning $z=$\,2.0--3.9, giving a survey volume of 8.3\,$\times$\,10$^6$\,Mpc$^3$).  Taking the median lifetime of $\sim$\,400\,Myrs (twice the gas depletion timescale) for SMGs from \cite[][]{2021MNRAS.501.3926B}, we need to apply a duty cycle correction of $\sim$\,4\,$\times$ to account for the duration of the SMG phase over the 1.7\,Gyrs corresponding to  $z=$\,2.0--3.9, indicating a descendant volume density of $\sim$\,4--5\,$\times$\,10$^{-5}$\,Mpc$^{-3}$.   This falls between  the volume density of galaxies in the MASSIVE survey of $\sim $\,2\,$\times$\,10$^{-5}$\,Mpc$^{-3}$, which are typically more massive than our sources, and $\sim $\,10\,$\times$\,10$^{-5}$\,Mpc$^{-3}$ for the subset of the ATLAS$^{\rm 3D}$ galaxies more massive than  $M_\ast\geq$\,10$^{11}$\,M$_\odot$.   This suggests that a substantial fraction of local early-type galaxies more massive than $M_\ast\sim$\,10$^{11}$\,M$_\odot$ could be formed through the massive DSFG population in our sample \citep[see][]{2020MNRAS.494.3828D, 2019MNRAS.488.2440M}.

\section{Summary \& Conclusions} \label{sec:section_5}

We have presented a method to model the dynamics of sources in 3D, from interferometric observations of emission lines, that builds upon existing methods \citep[i.e., {\sc Galpak3D};][]{2015AJ....150...92B} with the added extension that  the analysis is performed directly in the $uv$-plane.  Performing the modelling in the $uv$-plane results in more realistic estimates of errors on the inferred best-fit parameters of the model we are fitting to the data (e.g.\ ALESS 122.1; see Section~\ref{sec:model_observations}), because the errors on the visibilities are well defined (i.e., Gaussian errors). Our method does not require any image preprocessing, such as the {\sc clean} task from CASA \citep[][]{2007ASPC..376..127M}, which can depend on user assumptions (e.g., masking, threshold, etc.), a process that is not reversible. Our main science results are summarized below:

We find that 12 of the 20 sources that satisfy our SNR selection criteria, SNR $>$ 8 in their integrated CO emission, can be reasonably well fit by a simple thick rotating disk model. We classify all of these sources as rotationally supported disks, suggesting that $\sim$\,60\% of massive DSFGs fall in this category.  This fraction is consistent with the behaviour seen at high masses in previous studies of more typical star-forming galaxies \citep[e.g.][]{2015ApJ...799..209W}. We note, however, that at this stage, this fraction should only be considered an upper limit because the modest resolution of our data can lead, in some cases, to a misclassification of mergers as disks \citep[see,][]{2022A&A...667A...5R}. Among the remaining eight sources in our sample, six are classified as potential mergers, with the majority of them displaying complex features in their velocity maps, and for two sources we are not able to assign a classification due to resolution limitations.

We used the best-fit parameters of our kinematic model to investigate the dynamical state of our sources and compare them with samples in the literature. First, we looked at the relation between the circularised velocity, $V_{\rm circ}$, and the velocity dispersion, $\sigma$. Comparing our sample with more typical star-forming samples from the literature (e.g., KMOS$^{\rm 3D}$) we find that the DSFGs occupy the high velocity part of the distribution on this plane, with velocity dispersion that partly overlap but extend to higher dispersions than typical star-forming galaxies. Although, we find that our sample has an elevated median velocity dispersions compared to the ``typical'' star-forming galaxy population, their similarly higher circularized velocities means that the ratio of these two quantities, $V/\sigma$, has values consistent with the less actively star-forming population. We also looked at the variation in the typical velocity dispersion, $\sigma$, with redshift and found little evidence for  evolution. We highlighted that, although our inferred $\sigma$--$z$ relation lies above the one derived from  H$\alpha$ gas kinematics, this is a consequence of the  higher masses of our sample. As a result, we cannot draw any conclusions about the difference in velocity dispersion between molecular and ionized gas kinematics.

Combining the dynamical and physical properties of our sample we make the following conclusions:
\begin{itemize}
    
    \item We were able to constrain the median $\alpha_{\rm CO}$ conversion factor, a quantity that is used to convert CO(1$-$0) line luminosities to gas masses, to have a value of $\alpha_{\rm CO} = 1.0 \pm 0.4$. This measurement is consistent with previous estimates based on samples of similarly far-infrared luminous galaxies \citep[e.g.][]{2008ApJ...680..246T, 2012ApJ...760...11H, 2013MNRAS.429.3047B, 2013ARA&A..51..207B, 2018ApJ...863...56C, 2021MNRAS.501.3926B}.  
    
    \item We studied the stellar and baryonic Tully-Fisher relations, using our sample of massive DSFGs at high redshift, and compared it a sample of more ``typical'' star-forming galaxies at $z \sim $\,2.3 from the KMOS$^{\rm 3D}$ survey.  Our DSFGs   occupy the high-mass end of these relations, but have normalisations that are consistent with those measured for the KMOS$^{\rm 3D}$ sources. This shows that these DSFGs represent some of the most massive disk-like galaxies that have existed in the Universe.
    
    \item Finally, we have shown that under three reasonable assumptions (see Section~\ref{sec:SMG_descendants}) the most massive DSFGs at high redshift ($z \sim $\,1.2--4.7) have remarkably similar distributions on the baryonic mass versus velocity dispersion plane to the most massive ($M_{\rm b} \gtrsim 10^{11} $\,M$_{\odot}$) early-type galaxies in the local Universe. The apparent agreement between the distributions of these two populations, and their similar space densities,  adds further evidence to support the hypothesis that there is an evolutionary link between these two galaxy populations \citep[e.g.,][]{2020MNRAS.494.3828D,2021MNRAS.501.3926B}.
\end{itemize}

\section*{Acknowledgements}

AA is supported by ERC Advanced Investigator grant, DMIDAS [GA 786910], to C.S.\ Frenk.  AA, JEB, IRS, AMS, JN acknowledge support from STFC (ST/T000244/1). 
This paper makes use of the DiRAC Data Centric system at Durham University,
operated by the Institute for Computational Cosmology on behalf of the STFC DiRAC HPC Facility (www.dirac.ac.uk). This equipment was funded by BIS National E-infrastructure capital grant ST/K00042X/1, STFC capital grants ST/H008519/1 and ST/K00087X/1, STFC DiRAC Operations grant ST/K003267/1 and Durham University. DiRAC is part of the
National E-Infrastructure. This paper makes use of the following ALMA data: ADS/JAO.ALMA \# 2016.1.00564.S, 2016.1.00754.S, 2017.1.01163.S, 2017.1.01471.S and 2017.1.01512.S. ALMA is a partnership of ESO (representing its member states), NSF (USA) and NINS (Japan), together with NRC (Canada), MOST and ASIAA (Taiwan), and KASI (Republic of Korea), in cooperation with the Republic of Chile. The Joint ALMA Observatory is operated by ESO, AUI/NRAO and NAOJ. This work was performed using the Cambridge Service for Data Driven Discovery (CSD3), part of which is operated by the University of Cambridge Research Computing on behalf of the STFC DiRAC HPC Facility (www.dirac.ac.uk). The DiRAC component of CSD3 was funded by BEIS capital funding via STFC capital grants ST/P002307/1 and ST/R002452/1 and STFC operations grant ST/R00689X/1. DiRAC is part of the National e-Infrastructure.
This work used the DiRAC@Durham facility managed by the Institute for Computational Cosmology on behalf of the STFC DiRAC HPC Facility (www.dirac.ac.uk). The equipment was funded by BEIS capital funding via STFC capital grants ST/P002293/1 and ST/R002371/1, Durham University and STFC operations grant ST/R000832/1. DiRAC is part of the National e-Infrastructure.

\emph{Software:} numpy \citep{2011CSE....13b..22V}, scipy \citep{jones_scipy:_2001} and Astropy \citep{2013A&A...558A..33A}.

\section*{DATA AVAILABILITY}
The ALMA data that were used in this study are available from the ALMA Science Archive at https://almascience.eso.org/asax/.


\bibliographystyle{mnras}
\bibliography{biblio}

\appendix

\section{HST Morphologies} \label{sec:Appendix_HST}

\begin{figure*}
    \centering
    \includegraphics[width=0.95\textwidth,height=0.225\textheight]{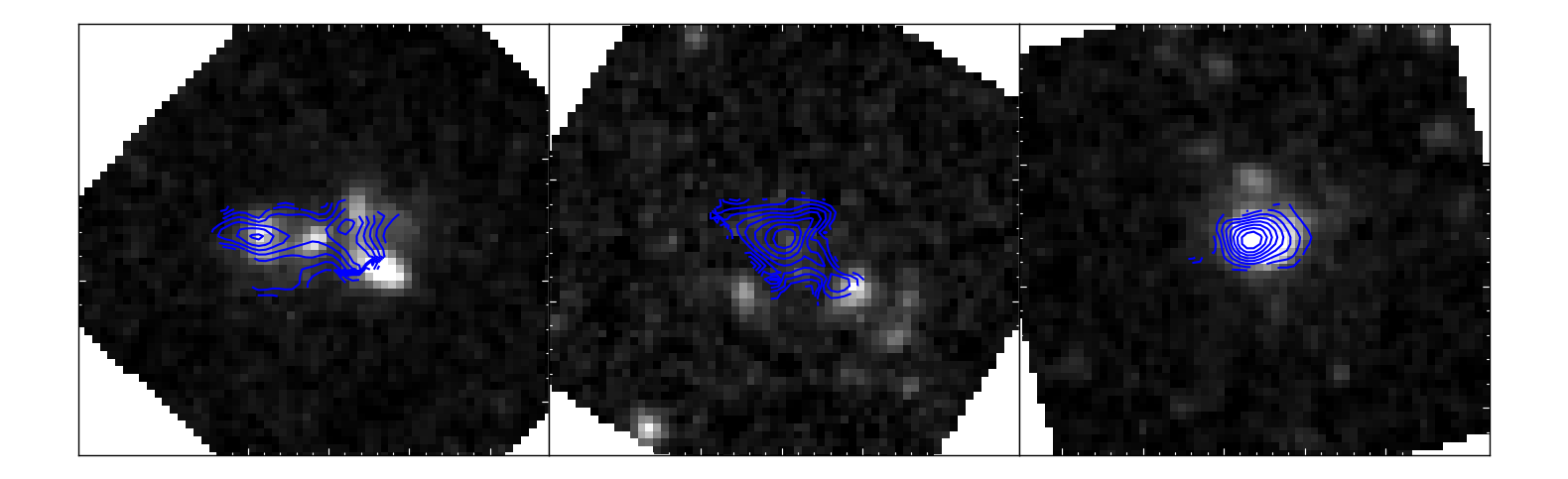}
    \caption{HST images (greyscale) of three of the Class \Romannum{2} sources in our sample (ALESS 088.1, 101.1, 112.1). The blue contours show the distribution of the CO emission, while the HST imaging suggests potentially complex morphologies for these sources.}
    \label{fig:figure_1_for_Appendix_A}
\end{figure*}

In Section~\ref{sec:model_observations} we discussed the results from our dynamical modelling analysis for the 20 sources with SNR$> 8$. We argued there that when our non-linear search convergences to unphysical values, $\sigma > 200$ km/s, this indicates that these sources can not be described by our simple rotating disk model. For three of the sources we have archival HST observations in the $H_{160}$ band \citep[ID 12866;][]{2015ApJ...799..194C, 2016ApJ...833..103H, 2019ApJ...876..130H}, which we show in Figure~\ref{fig:figure_1_for_Appendix_A}. These sources appear to be breaking into multiple components in the HST images \citep[5-6 individual clumps from fitting the images with GALFIT;][]{2015ApJ...799..194C}, a feature that can be used to strengthen our interpretation that these sources are mergers. 

We note however that previous studies have shown that in some star-forming galaxies at $z\sim 1 - 3$ even though their stellar morphologies appear clumpy, their kinematics are consistent with ordered circular motions \citep[e.g.][]{2011ApJ...733..101G}. We therefore stress that we are not claiming that clumpy looking sources in the rest-frame UV are all necessarily mergers, but when both their kinematics and their stellar morphologies appear irregular this is strong evidence in favour of a merger/interactions scenario.


\section{Exploring the effects of resolution and SNR} \label{sec:Appendix_Simulations}

In this section we explore the effects of resolution and SNR on the inferred best-fit parameters from our modelling analysis using both simulations and observations.

\subsection{Simulations}

\begin{figure*}
\begin{tabular}{c}
\includegraphics[width=0.95\textwidth,height=0.35\textheight]{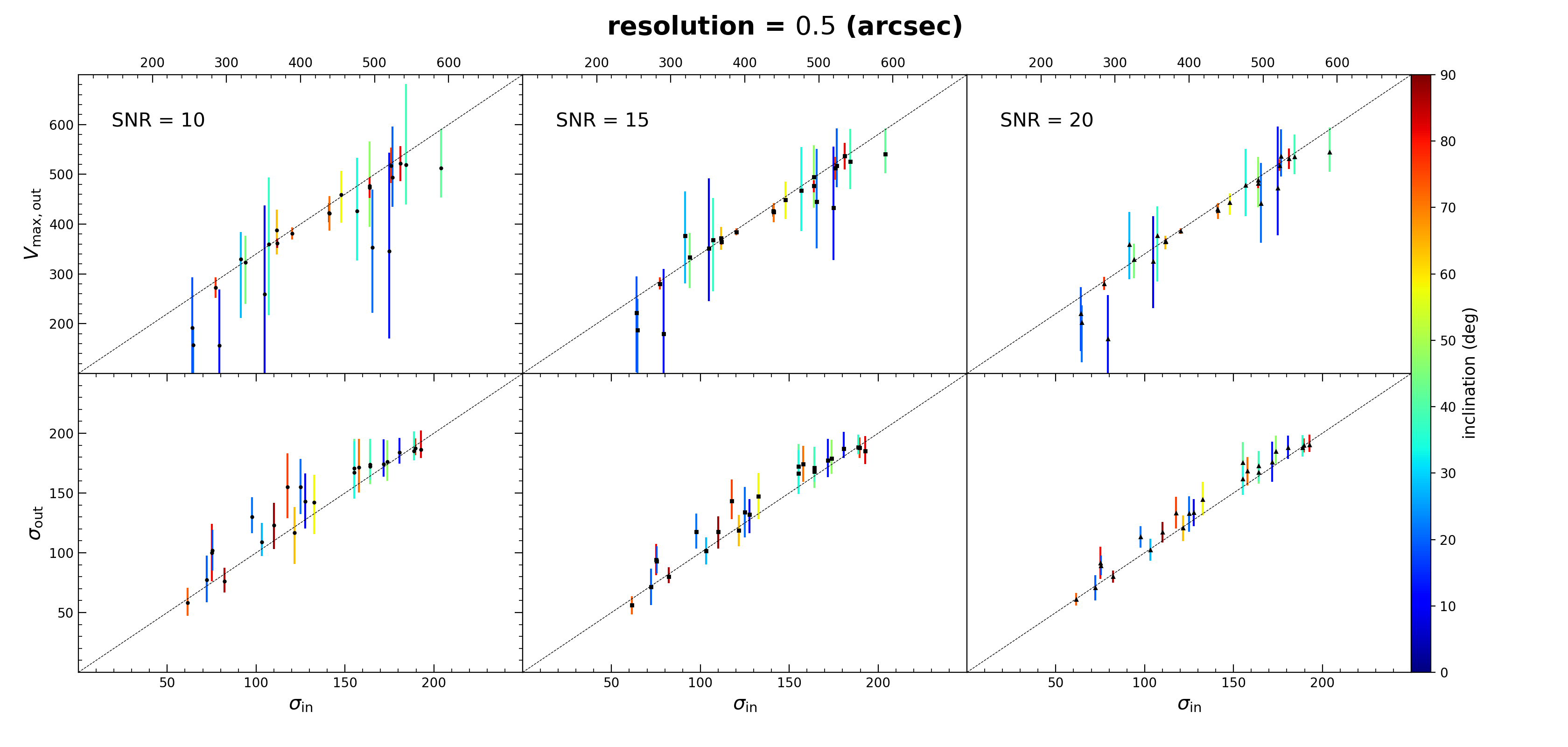} \\
\includegraphics[width=0.95\textwidth,height=0.35\textheight]{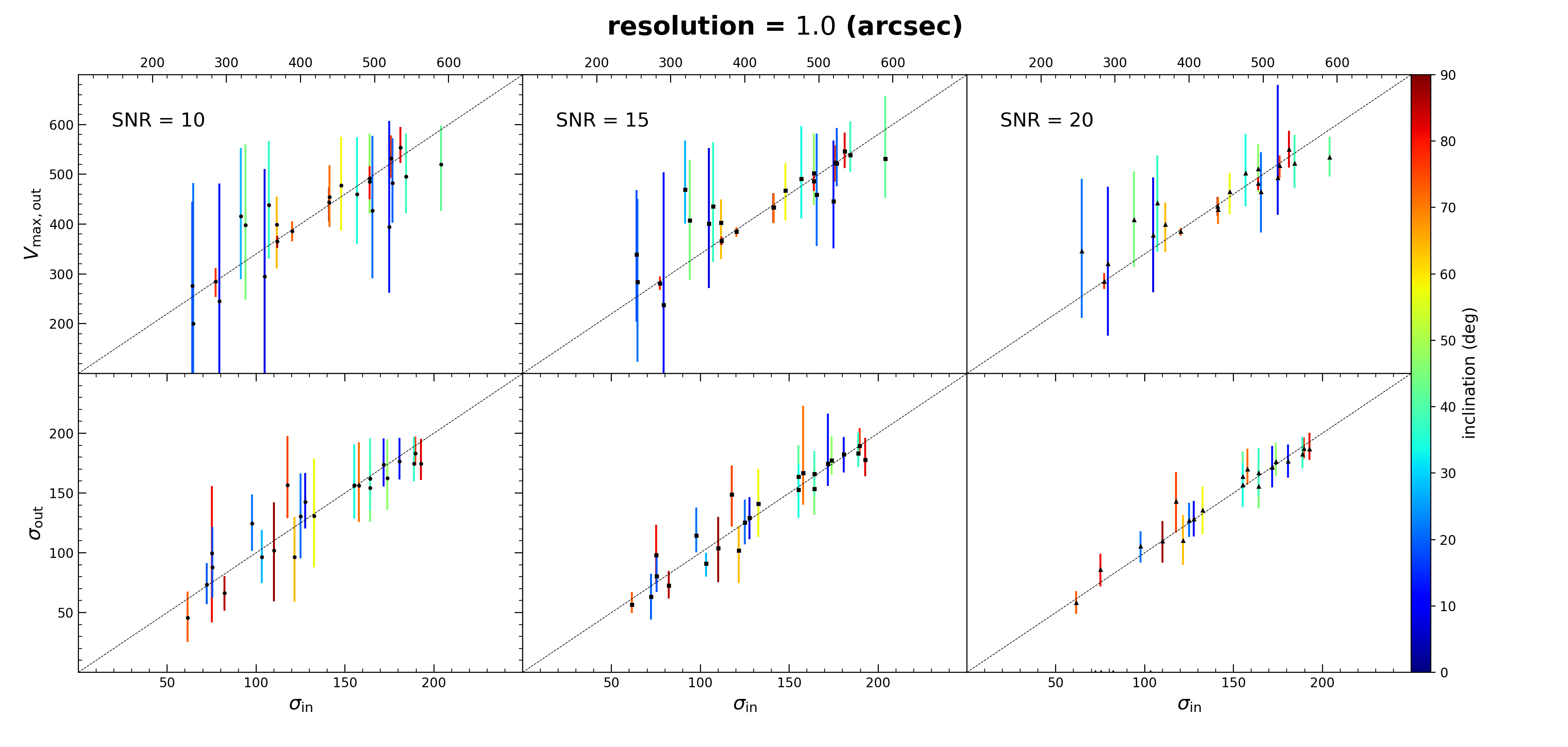} \\
\end{tabular}
\caption{Output versus input values for the two main parameters of interest, the maximum rotation velocity ($V_{\rm max}$) and the velocity dispersion ($\sigma$), color-coded by the true inclination. The different columns corresponds to datasets with different SNR, indicated at the top left corner of each panel. The top and bottom figures correspond to datasets with different resolution, 0.5 arcsec and 1.0 arcsec, respectively.}
\label{fig:sims}
\end{figure*}

We start by outlining the steps for creating simulated observations which are used to evaluate the reliability/accuracy of the best-fit model parameters. Specifically, we want to examine how well we can recover the true parameters of our model for data with different SNR and resolution. In order to create simulated visibilities for this exercise we use the observed uv-coverage of ALESS 122.1, which achieves a resolution of $\sim 0.5$ arcsec, the highest out of all datasets used in this study. This gives us the flexibility to taper the simulated data to produce lower resolution simulations. 

To create our simulations we randomly draw values for each model parameter from a wide enough range to include most of the best-fit values we infer for the sources in our sample (e.g. 200 < $V_{\rm max}$ < 600, 50 < $\sigma$ < 200). After creating the model cube and Fourier transforming it to compute the model visibilities, we need to add noise. To ensure the noise properties of our simulated data are as realistic as possible we use the observed visiblities for ALESS 122.1 in a spectral window where no emission line is observed. We treat these as the noise, which is then added to our simulated visibilities. 

Next, we need to ensure that we create simulations of a given SNR. For each combination of model parameters we first by create a simulation with arbitrarily high intensity, and therefore SNR. We then estimate the SNR of this dataset following the same approach as we used on the real observations (see Section~\ref{sec:SNR_and_classification}). Having determined the SNR of this reference dataset we can then scale the intensity of the model sources to produce datasets with the desired SNR. 

In Figure~\ref{fig:sims} we show fitting results to our simulations with different resolutions (0.5 and 1.0 arcsec) and SNR (10, 15, 20). We focus on two parameters, the maximum velocity and the velocity dispersion, which are the ones that are mostly used in the main body of this work. In general we find that we always recover the true input parameters within 3$\sigma$. The errors on the output parameters are larger for dataset with poorer resolution and lower SNR. In addition, we color-coded the points in these graphs according to their true inclinations to highlight that sources with lower inclination typically have larger errors.


\subsection{Observations: ALESS 073.1}

In this section, we explore how resolution affects the inferred parameters of our model using observed data instead of simulated data. For this exercise, we utilize data for ALESS 073.1, originally presented in \cite{2021Sci...371..713L}, achieving a resolution of approximately $0.1$ arcsec (2017.1.01471.S).

\begin{figure*}
    \centering \includegraphics[width=0.95\textwidth,height=0.5\textheight]{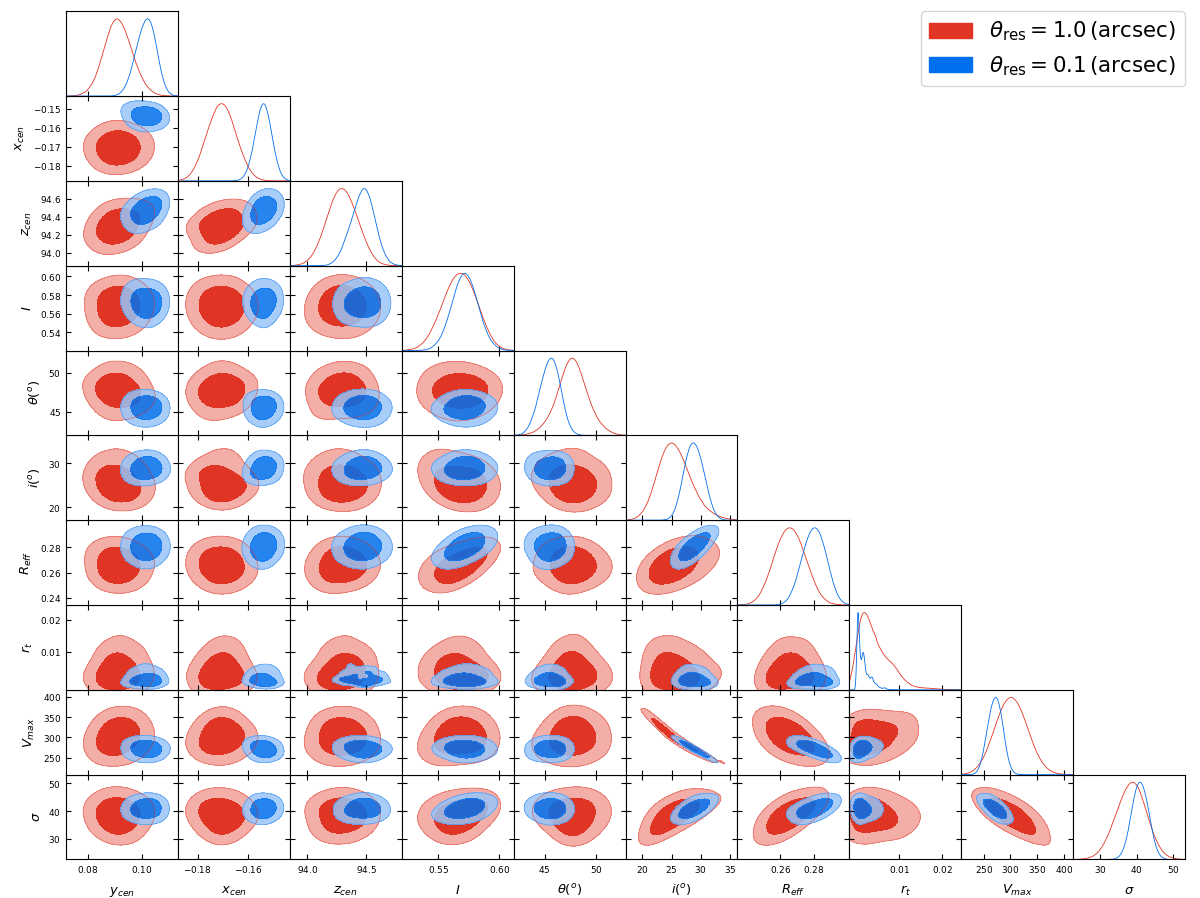}
    \caption{Corner plot for ALESS 073.1 using the data presented in Lelli et al. (2021)\iffalse\cite{2021Sci...371..713L}\fi. The blue and red contours, drawn at 2$\sigma$ and 3$\sigma$, show the posterior distribution using the data at the native resolution (0.1 arcsec) and after tapering to a resolution of $\sim 1$ arcsec, respectively. The best-fit values for all the inferred parameters from these two datasets are consistent and also in agreement with the values reported in Lelli et al. (2021)\iffalse\cite{2021Sci...371..713L}\fi.}
    \label{fig:ALESS073.1_cornerplots}
\end{figure*}

The dynamical modeling of this source strongly suggests it is a rotating disk galaxy \citep{2021Sci...371..713L}. Initially, we modeled this source following the procedure outlined in Section~\ref{sec:section_3} and used the observed data (i.e., visibilities) at the native resolution. Subsequently, we repeated the fitting process, this time tapering the data to a resolution of about $1$ arcsec, similar to the average resolution of data in our main ALESS sample. In Figure~\ref{fig:ALESS073.1_cornerplots}, the corner plot displays our fitting results for both the native (blue) and tapered (red) resolution data. As evident from this figure, the best-fit parameters inferred from these two datasets are consistent at 3$\sigma$. Notably, constraints tighten when using higher resolution data.

Lastly, the best-fit values we infer for our model parameters align with those reported in \cite{2021Sci...371..713L}. Due to differences in the modeling software used in that work, an exact direct comparison is not possible. Nevertheless, we make a rough comparison. For instance, the inclination reported in \cite{2021Sci...371..713L} (25\textdegree $\pm$ 3\textdegree) closely matches our finding (29\textdegree $\pm$ 2\textdegree). The velocity dispersion of the tilted rings in \cite{2021Sci...371..713L} varies from $\sim 50-60$ km/s in the inner part to $\sim 10$ km/s in the outer part of the disk, while we find a value of $41 \pm 3$ km/s using a uniform velocity dispersion across the disk.

\section{Fitting Diagnostics} \label{sec:Appendix_fitting_diagnostics}

In this section we show various fitting diagnostic figures for all class I sources which are considered to be well described by a disk kinematic model. These include channels maps, $0^{\rm th}$ moment maps, pv diagrams along the major/minor axis and rotation curves extracted from the velocity maps.

The rotation curves are extracted from our observed velocity maps shown in Figure~\ref{fig:velmaps}. We only show these for the purpose of visualization as the analysis is carried out directly in the uv-plane. We note that these can be subject to artifacts, at this resolution, that can arise when performing the imaging with CASA (we tested this by generating model visibility datasets from a smooth rotating disk model and repeated the rotation curve extraction in the same manner as the data). Nevertheless, one thing that we take away from this figure is that not all rotation curves flatten at large radii which can bias some of our estimates of the maximum rotational velocities.

\begin{figure*}
\begin{tabular}{c@{\hspace{0.0cm}}c@{\hspace{0.0cm}}c}

\includegraphics[width=0.325\textwidth,height=0.08\textheight]{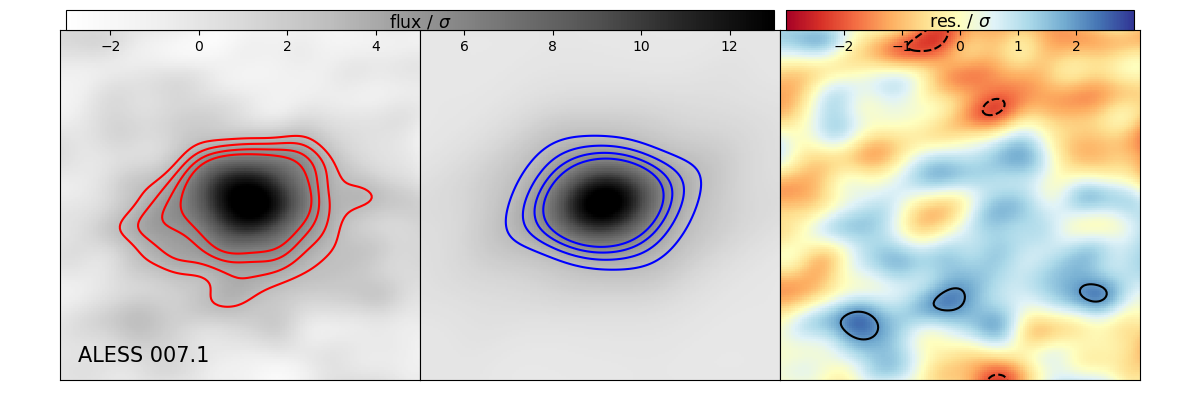}
&
\includegraphics[width=0.325\textwidth,height=0.08\textheight]{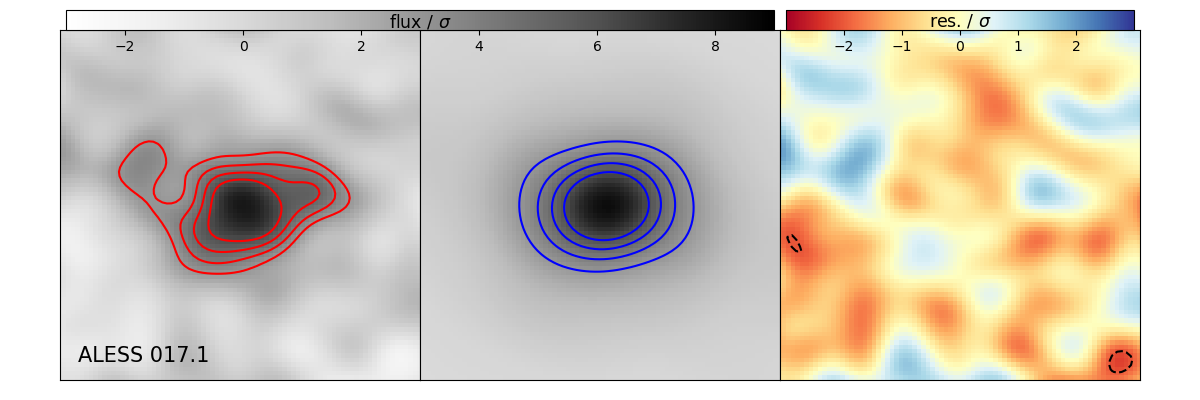} 
& 
\includegraphics[width=0.325\textwidth,height=0.08\textheight]{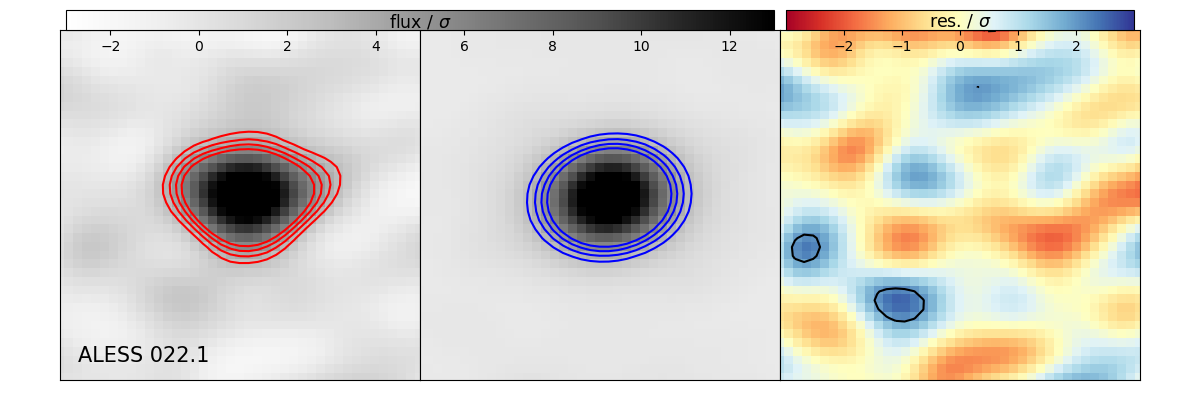}\\[1.0mm]

\includegraphics[width=0.325\textwidth,height=0.08\textheight]{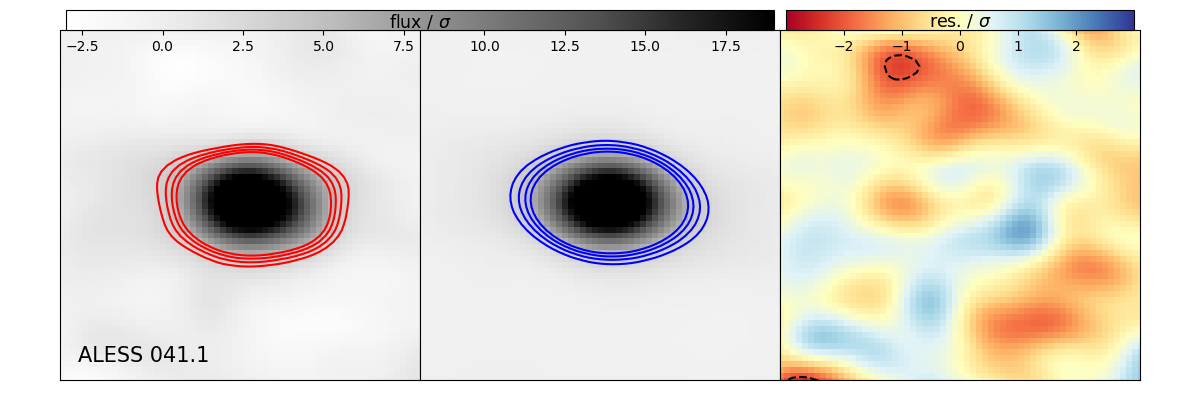} 
& 
\includegraphics[width=0.325\textwidth,height=0.08\textheight]{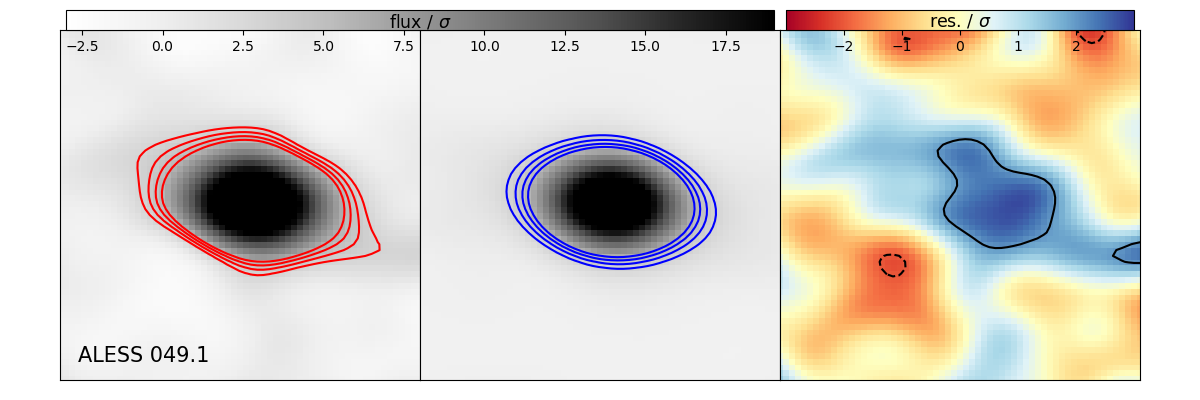}
&
\includegraphics[width=0.325\textwidth,height=0.08\textheight]{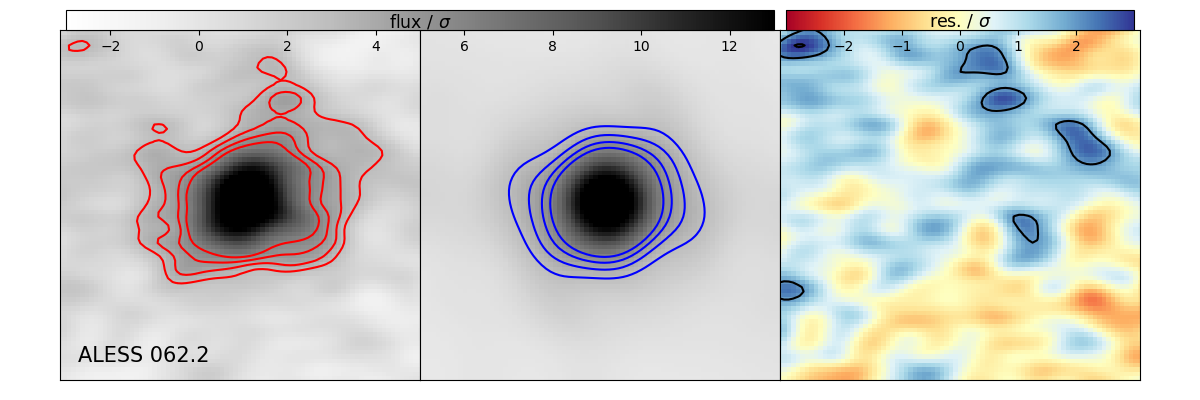}\\[1.0mm]

\includegraphics[width=0.325\textwidth,height=0.08\textheight]{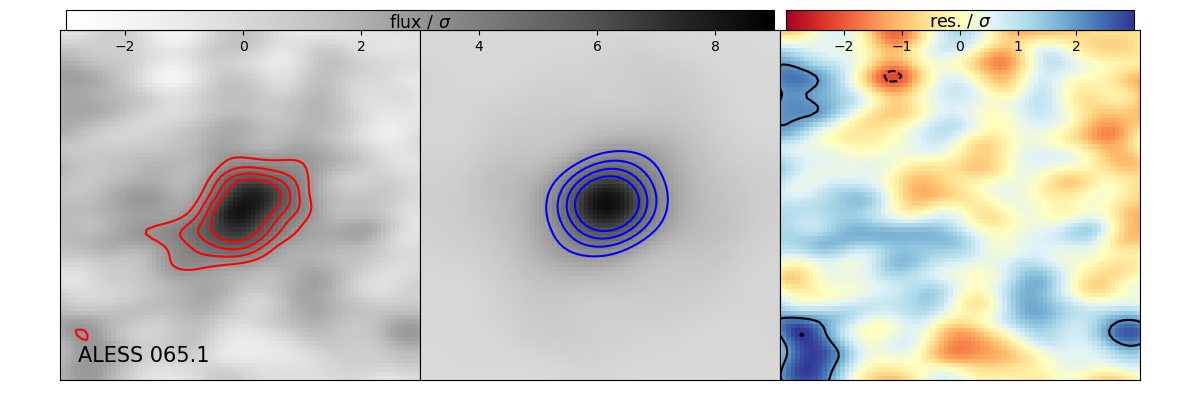} 
& 
\includegraphics[width=0.325\textwidth,height=0.08\textheight]{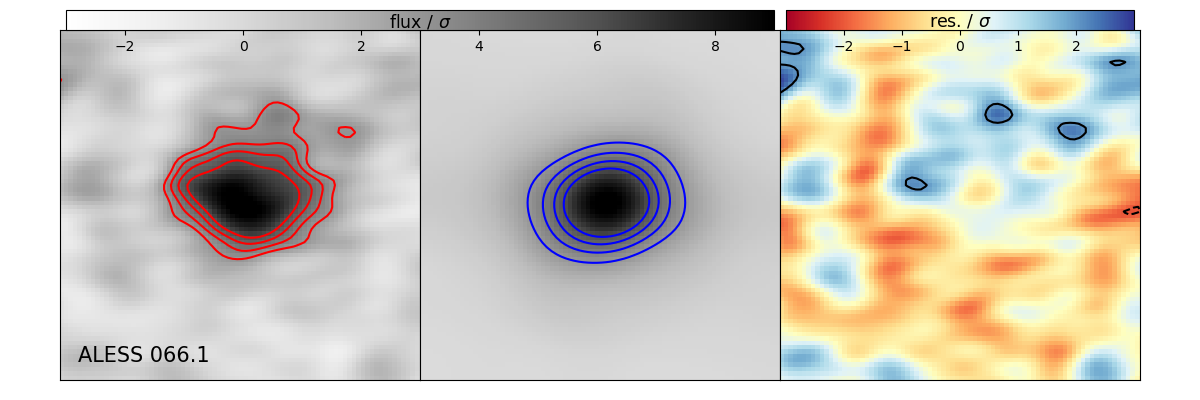}
&
\includegraphics[width=0.325\textwidth,height=0.08\textheight]{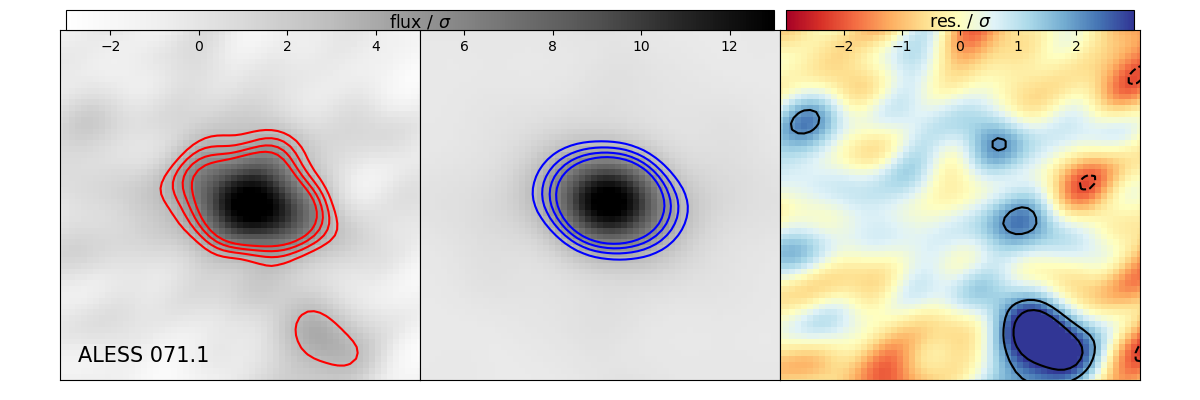}\\[1.0mm] 

\includegraphics[width=0.325\textwidth,height=0.08\textheight]{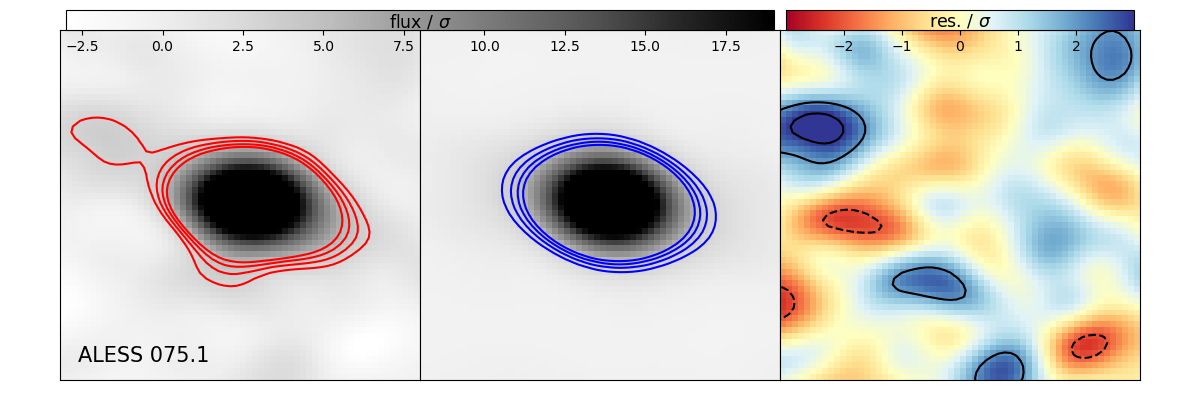}
&
\includegraphics[width=0.325\textwidth,height=0.08\textheight]{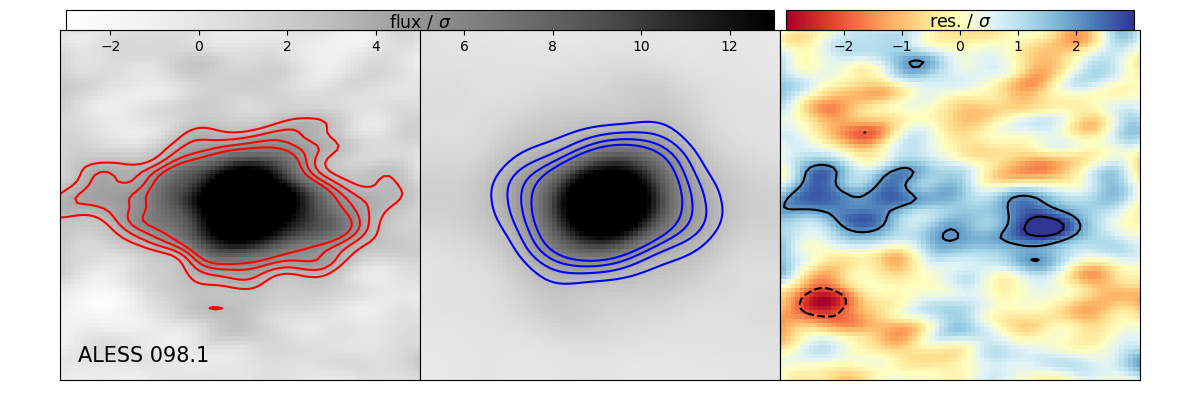}
&
\includegraphics[width=0.325\textwidth,height=0.08\textheight]{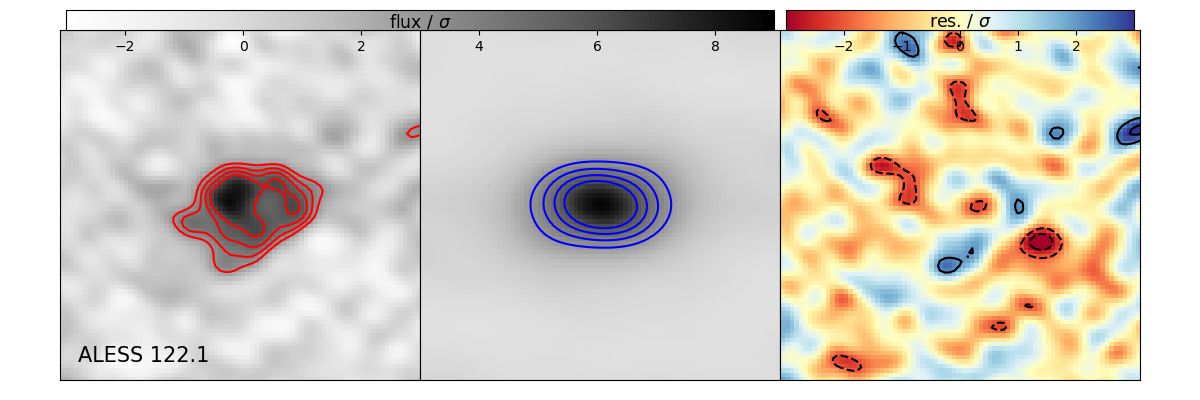}\\[1.0mm]

\end{tabular}
\caption{Normalized dirty images of the 0$^{\rm th}$ moment (intensity) of the observed, model and residual maps for each source in our sample. The red and blue contours in the first two panels show contours from 3$\sigma$ to 6$\sigma$ levels in increments of 1$\sigma$. The normalized residuals contours levels are drawn at $\pm 2 \sigma$, $\pm 3 \sigma$.}
\label{fig:figure_Appendix_C}
\end{figure*}


\begin{figure*}
\includegraphics[width=0.975\textwidth,height=0.225\textheight]{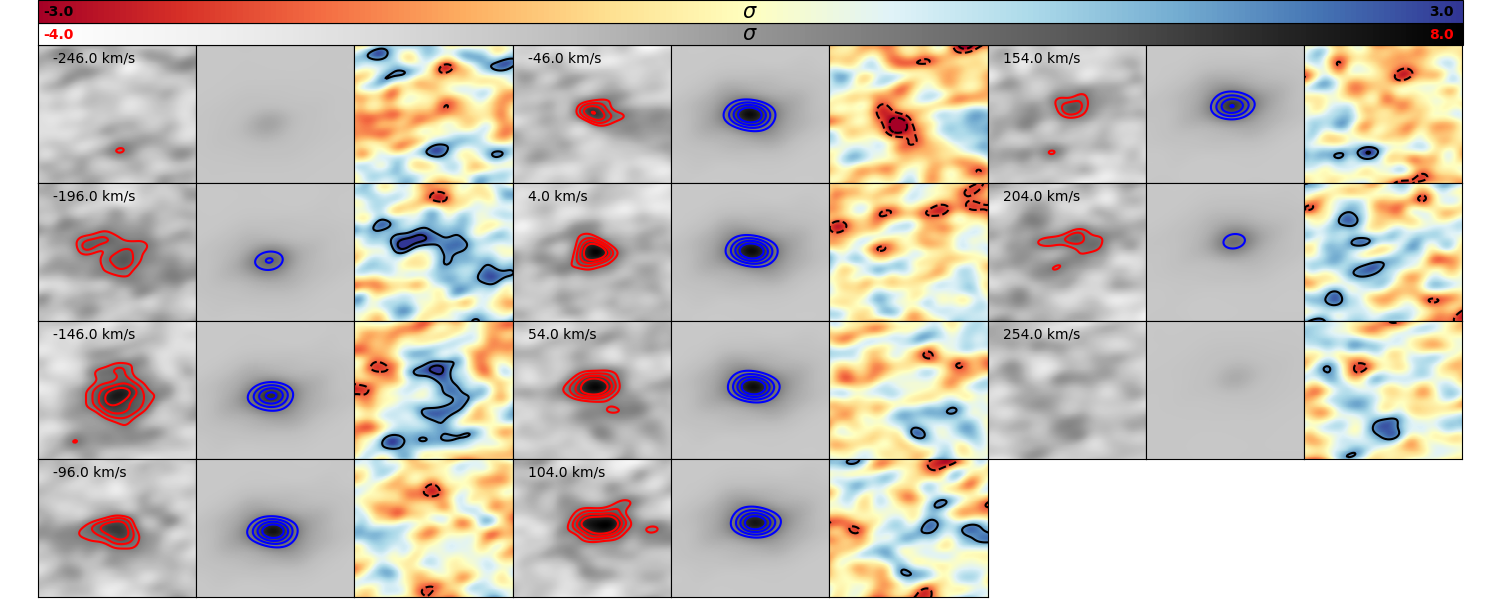} \\[1.0mm]
\end{figure*}

\begin{figure*}
\includegraphics[width=0.975\textwidth,height=0.225\textheight]{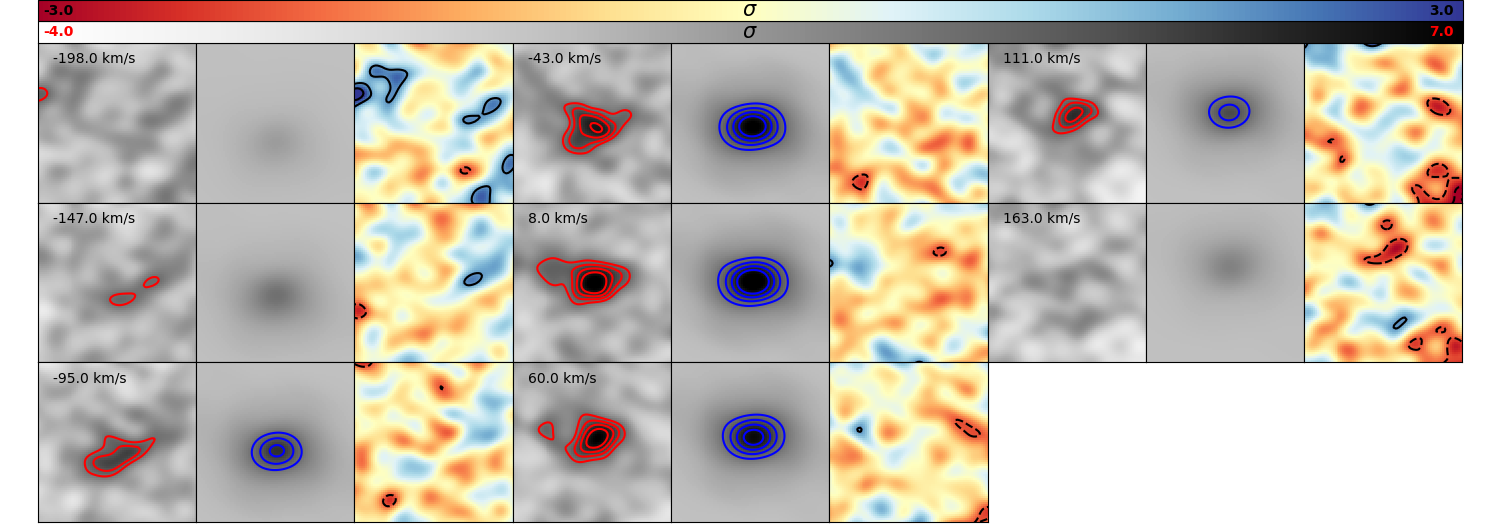} \\[1.0mm]
\end{figure*}

\begin{figure*}
\includegraphics[width=0.975\textwidth,height=0.225\textheight]{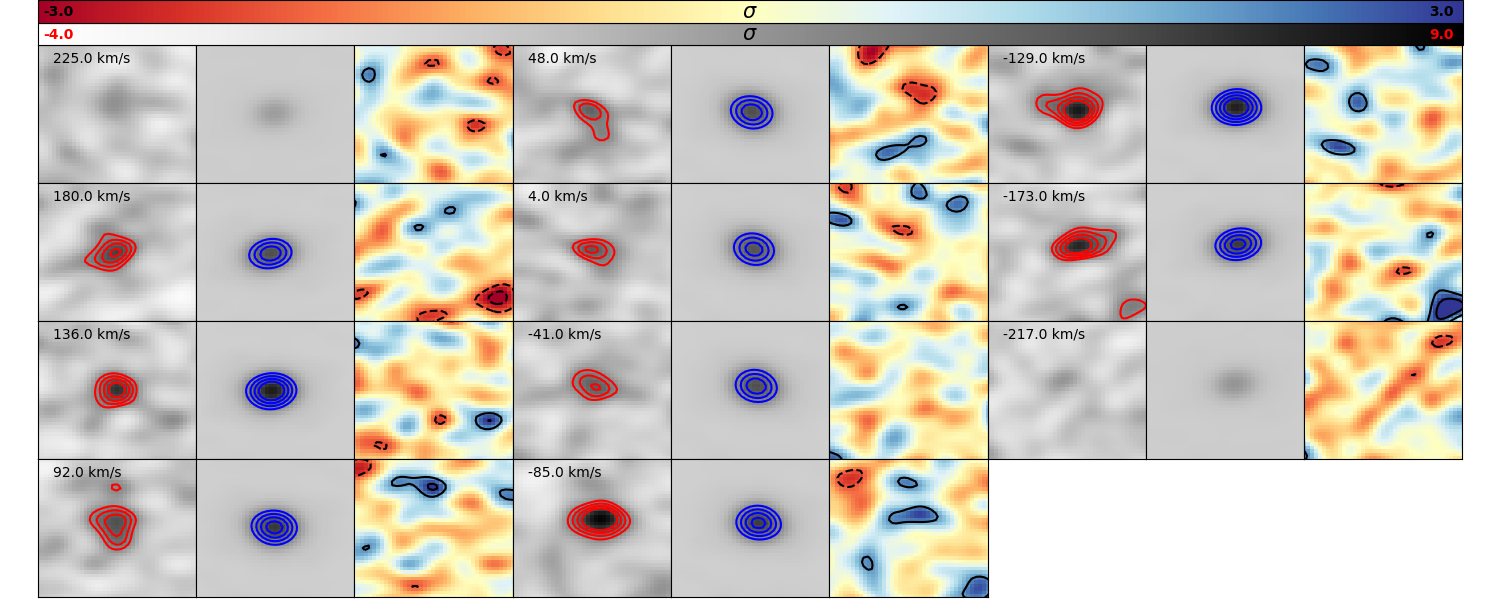} \\[1.0mm]
\end{figure*}

\begin{figure*}
\includegraphics[width=0.975\textwidth,height=0.45\textheight]{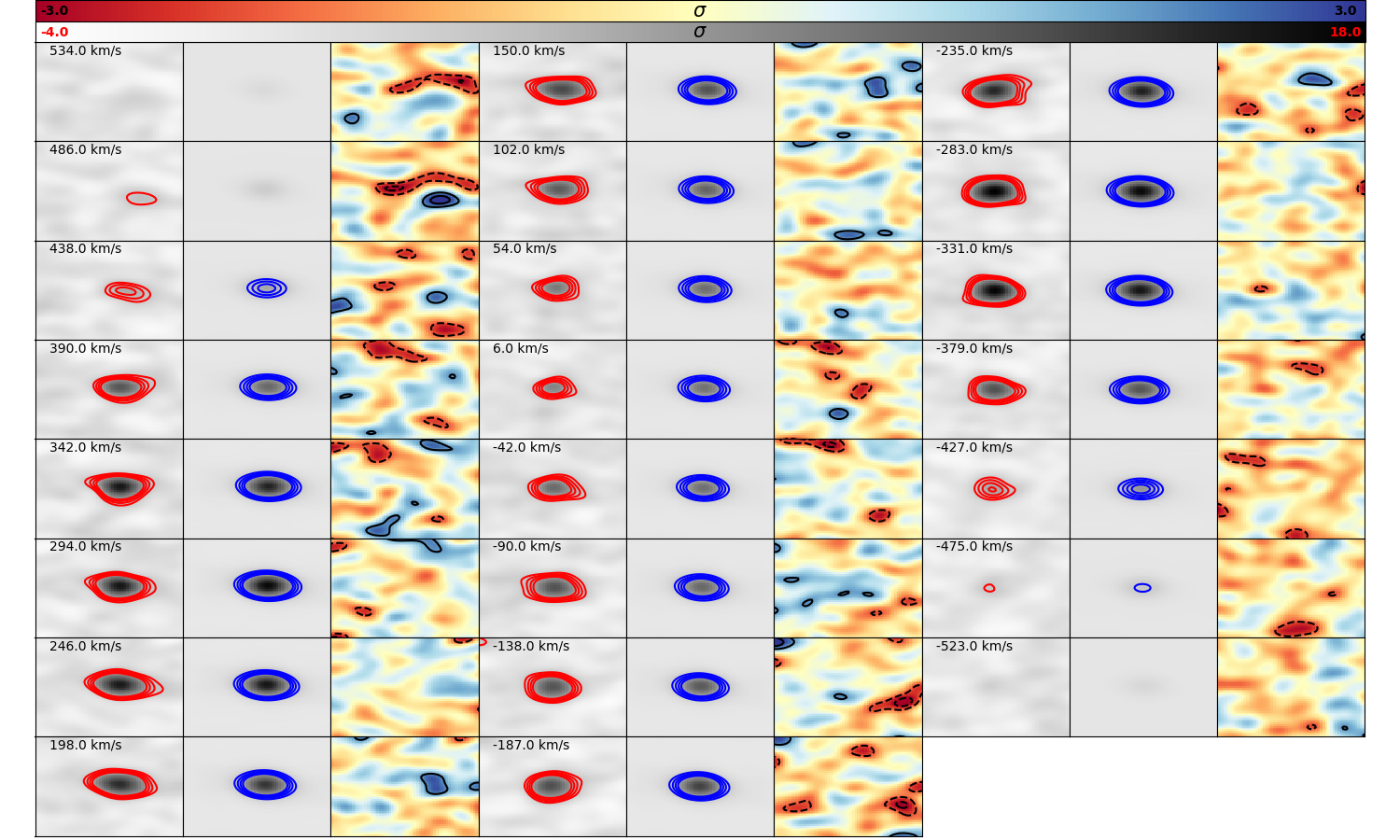} \\[1.0mm]
\end{figure*}

\begin{figure*}
\includegraphics[width=0.975\textwidth,height=0.3375\textheight]{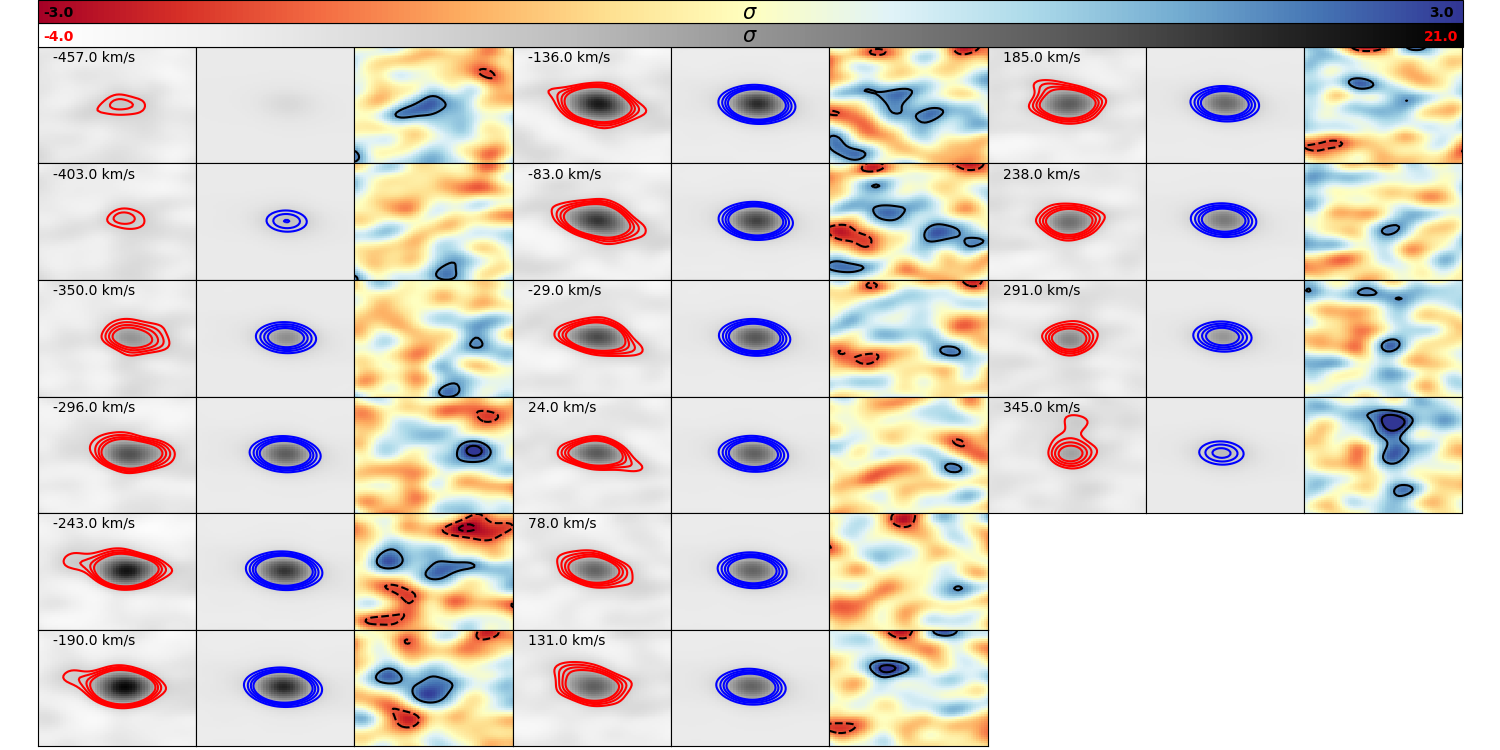} \\[1.0mm]
\end{figure*}

\begin{figure*}
\includegraphics[width=0.975\textwidth,height=0.28125\textheight]{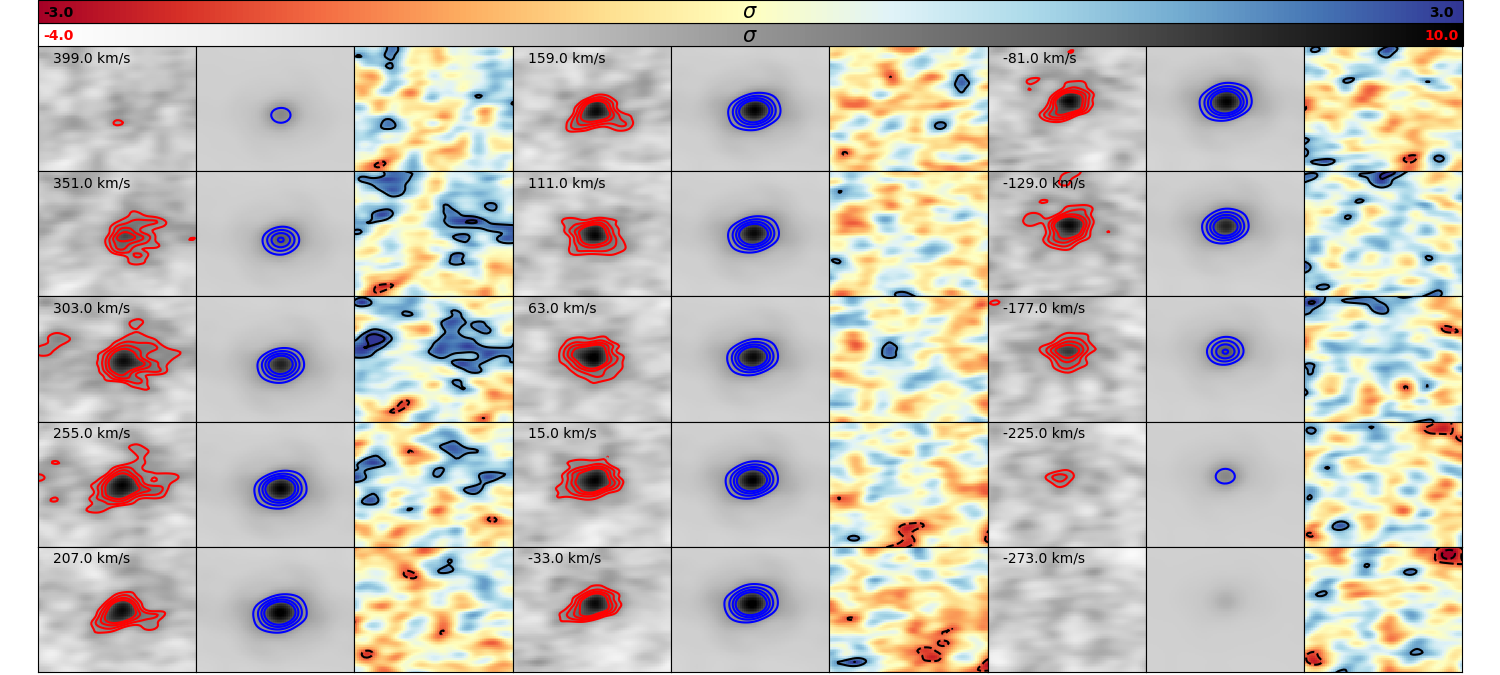} \\[1.0mm]
\end{figure*}

\begin{figure*}
\includegraphics[width=0.975\textwidth,height=0.225\textheight]{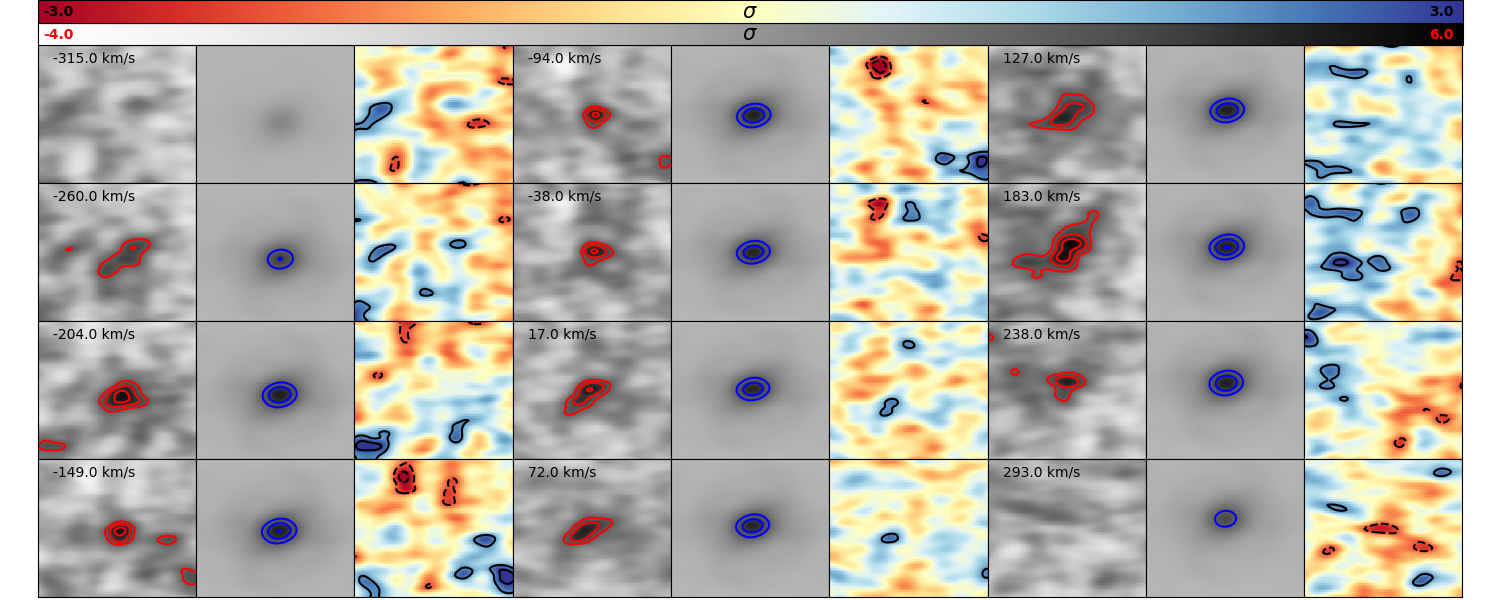} \\[1.0mm]
\end{figure*}

\begin{figure*}
\includegraphics[width=0.975\textwidth,height=0.225\textheight]{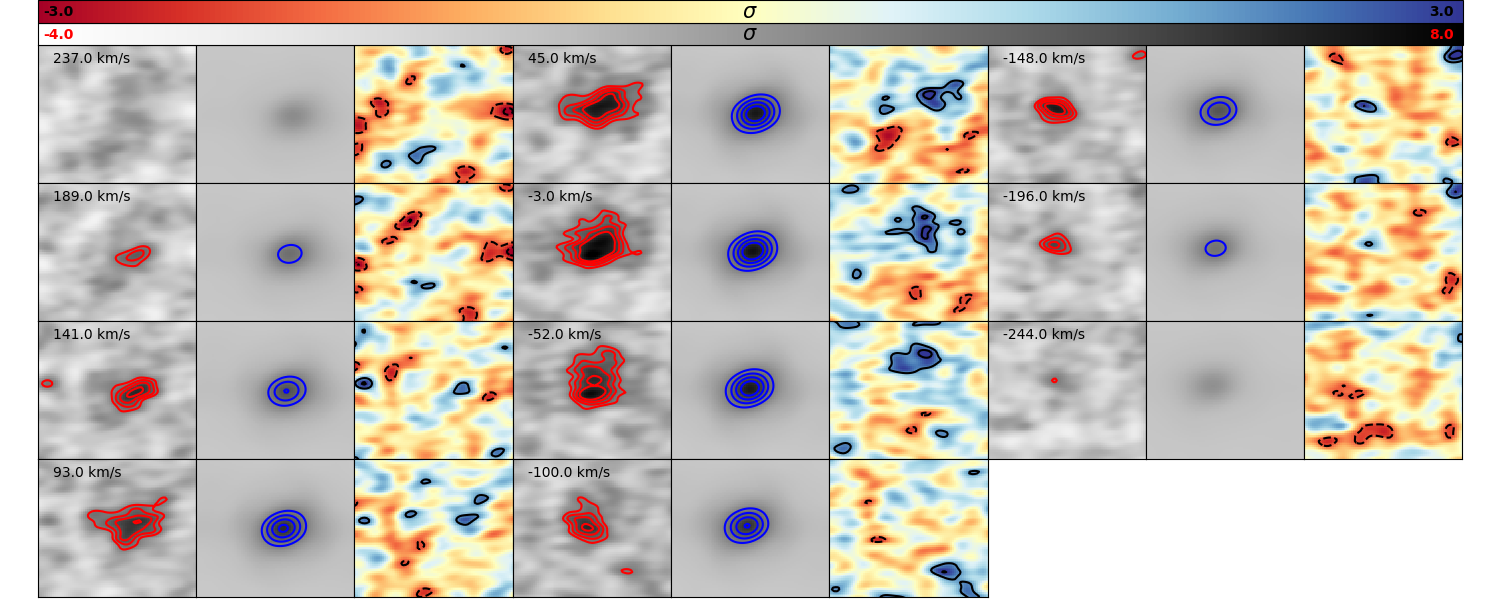} \\[1.0mm]
\end{figure*}

\begin{figure*}
\includegraphics[width=0.975\textwidth,height=0.3375\textheight]{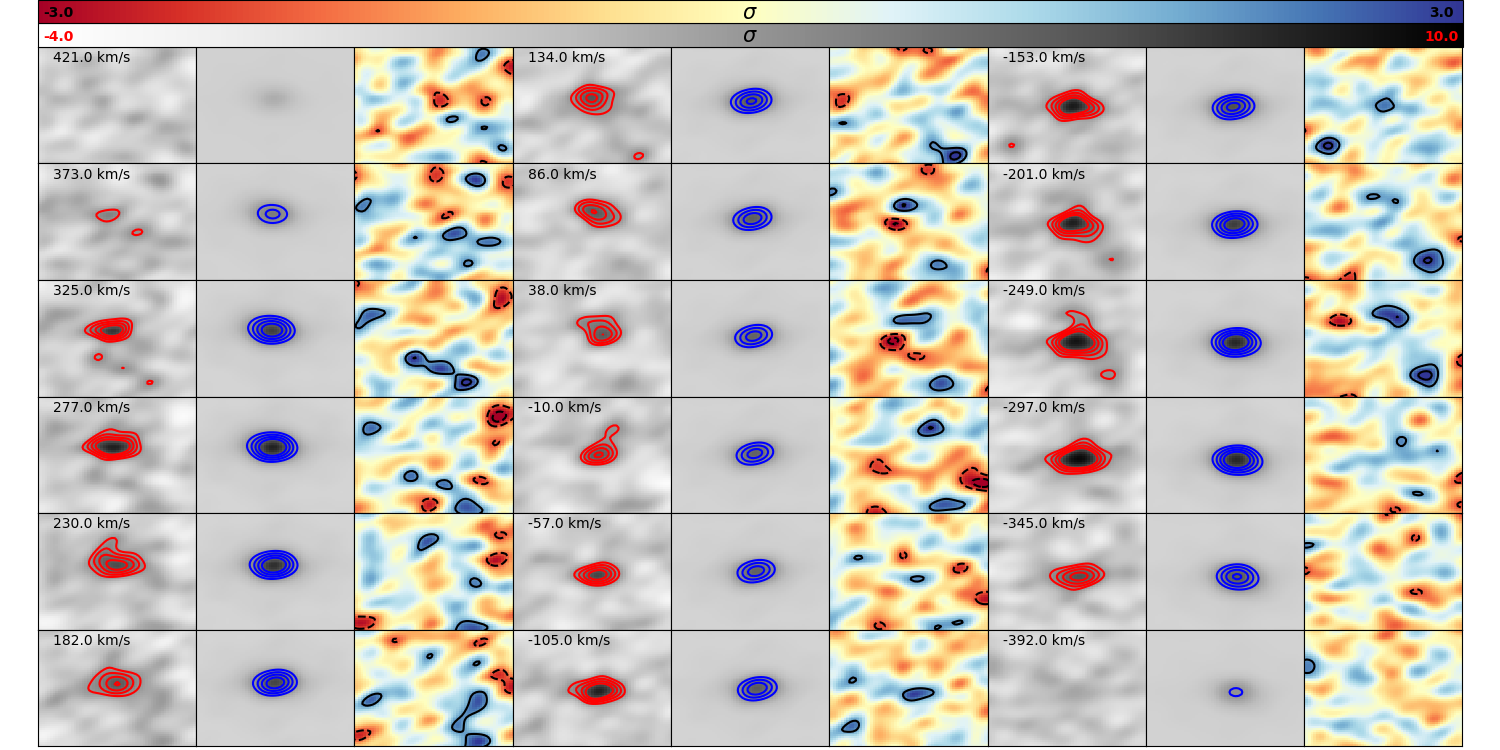} \\[1.0mm]
\end{figure*}

\begin{figure*}
\includegraphics[width=0.975\textwidth,height=0.39375\textheight]{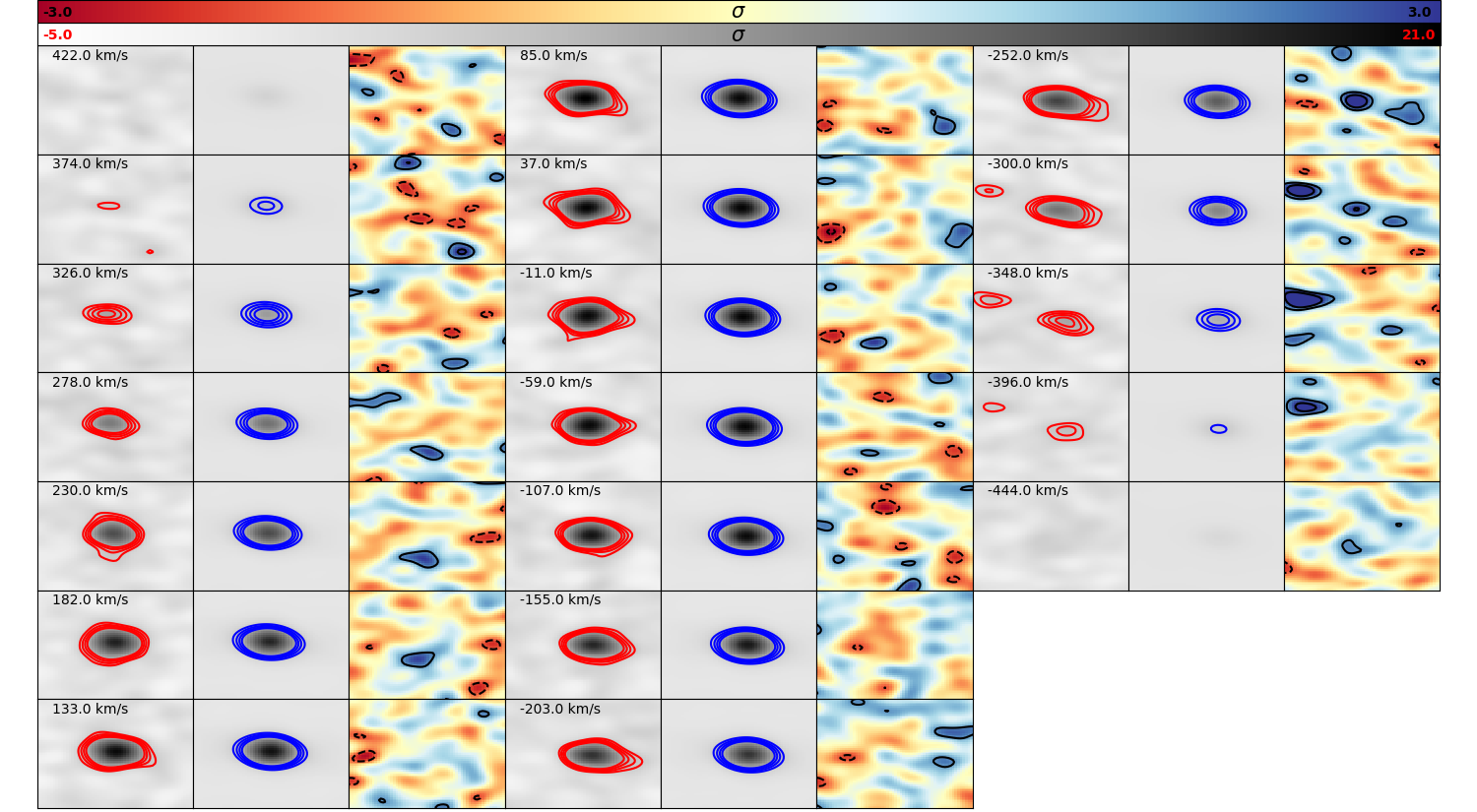} \\[1.0mm]
\end{figure*}

\begin{figure*}
\includegraphics[width=0.975\textwidth,height=0.39375\textheight]{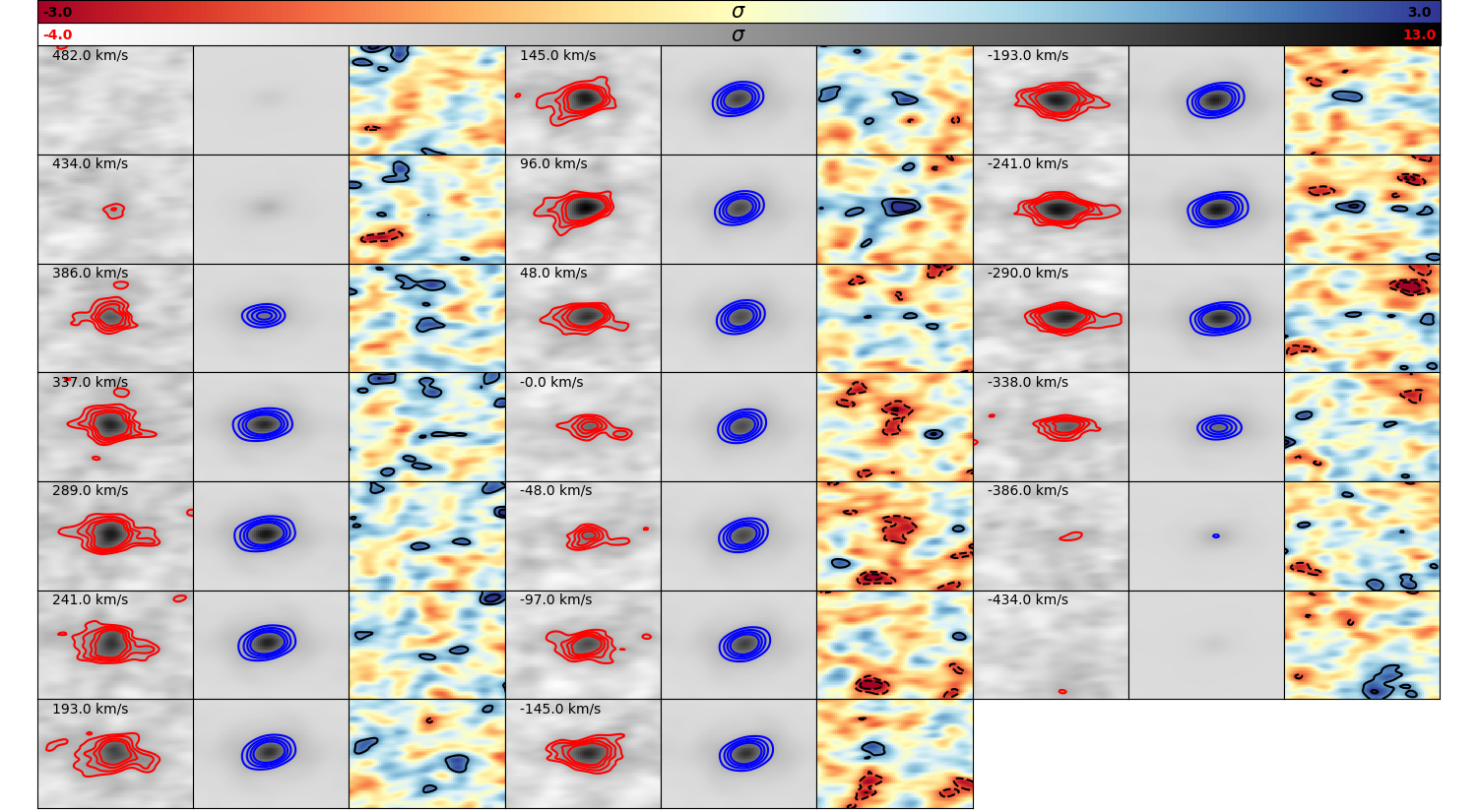} \\[1.0mm]
\caption{Dirty channel maps for the data, model and residuals, normalized by the rms noise of the cube. Contours for the data and model are drawn from $3\sigma$ to $6\sigma$ in incremenents of $1\sigma$, while contours in the residual panels are drawn at $\pm 2\sigma$ and $\pm 3\sigma$.}
\end{figure*}


\begin{figure*}
\begin{tabular}{c@{\hspace{0.0cm}}c@{\hspace{0.0cm}}c}

& \textbf{major-axis} & \\[1.0mm]

\includegraphics[width=0.325\textwidth,height=0.08\textheight]{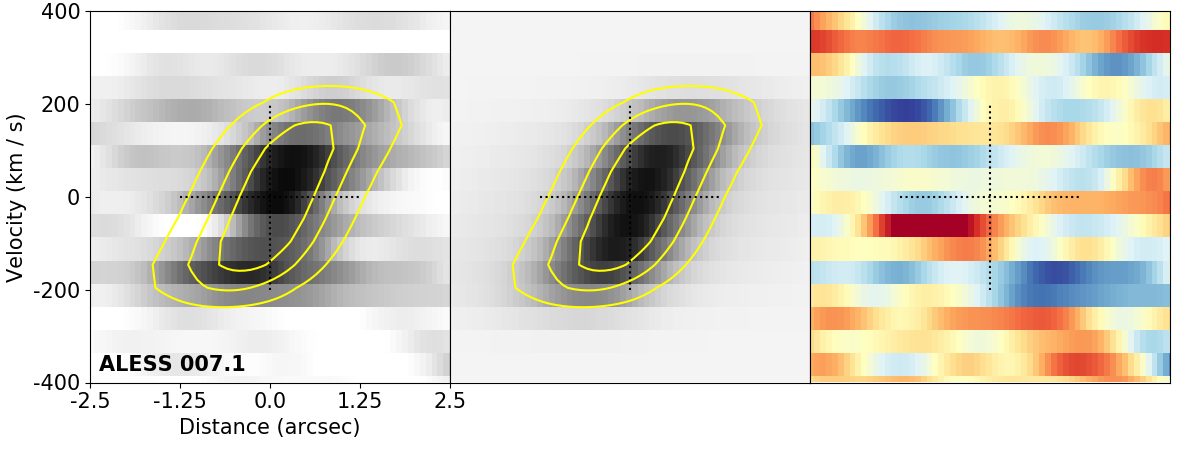}
&
\includegraphics[width=0.325\textwidth,height=0.08\textheight]{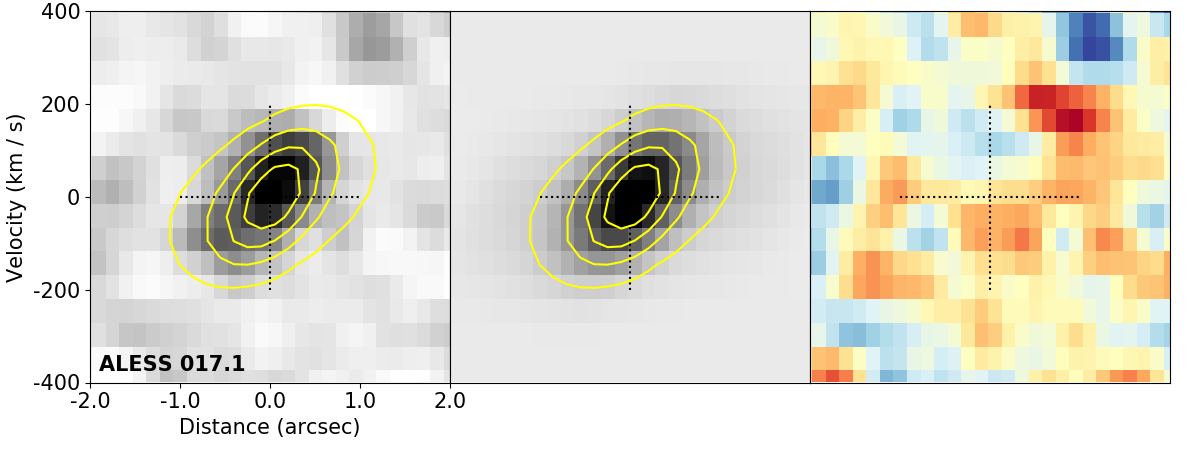}
&
\includegraphics[width=0.325\textwidth,height=0.08\textheight]{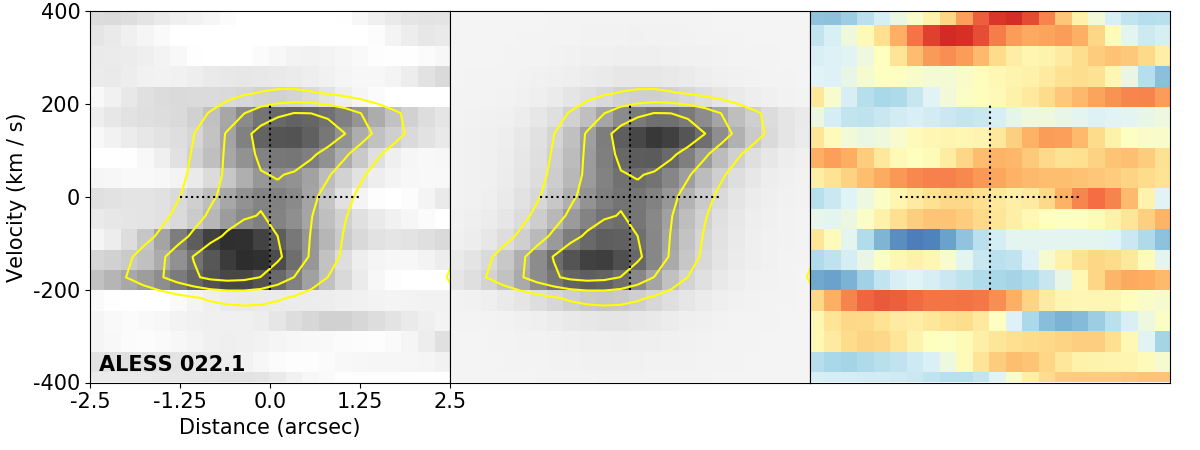}\\[1.0mm]

\includegraphics[width=0.325\textwidth,height=0.08\textheight]{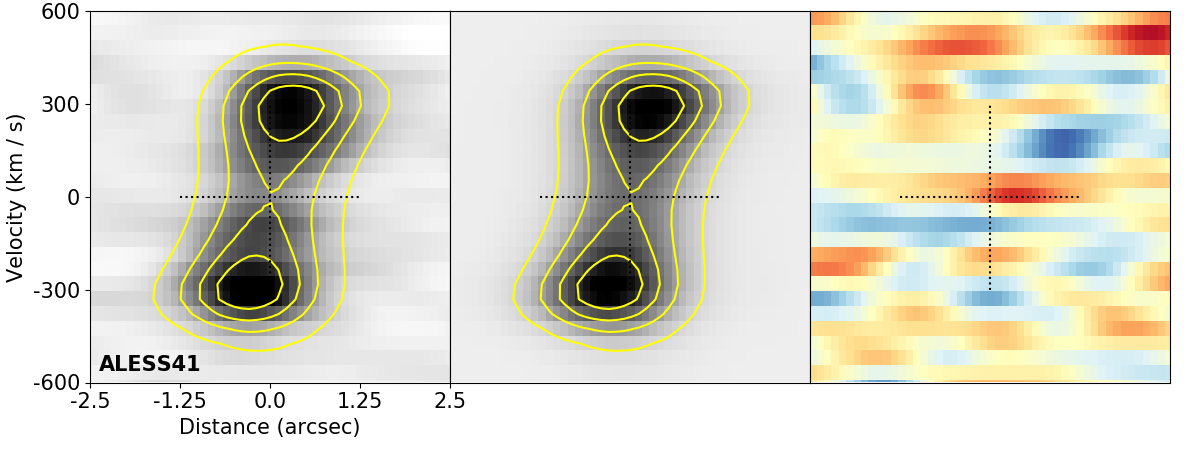} 
& 
\includegraphics[width=0.325\textwidth,height=0.08\textheight]{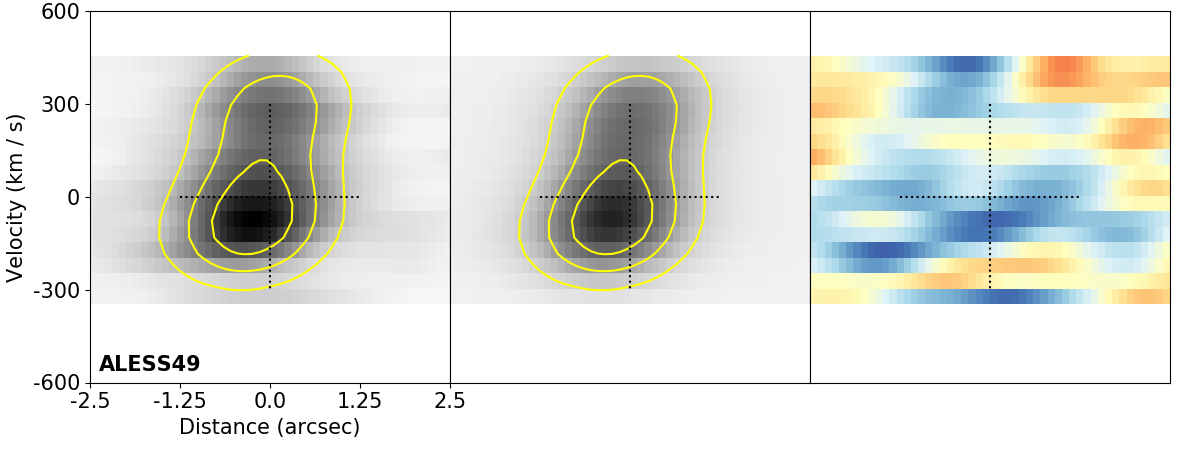}
&
\includegraphics[width=0.325\textwidth,height=0.08\textheight]{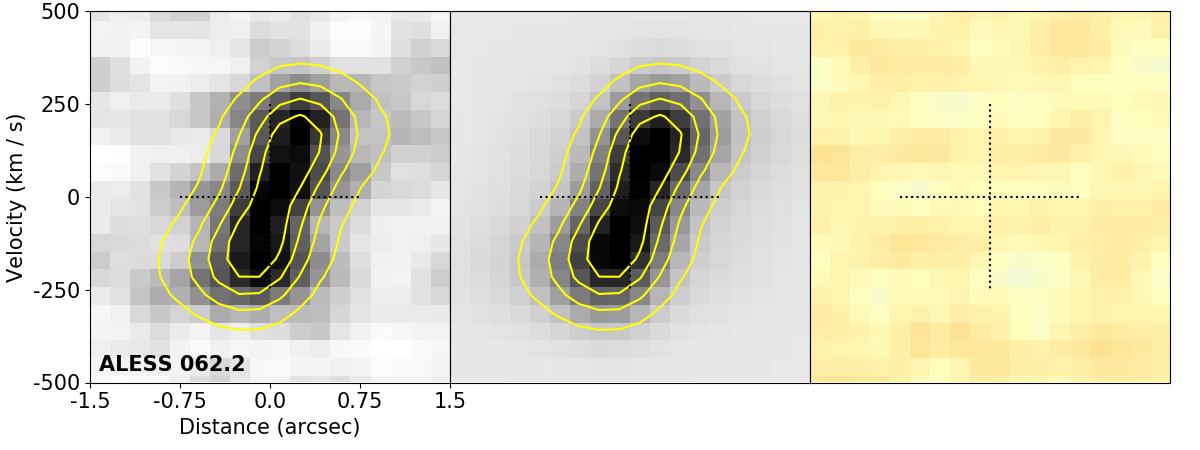}\\[1.0mm]

\includegraphics[width=0.325\textwidth,height=0.08\textheight]{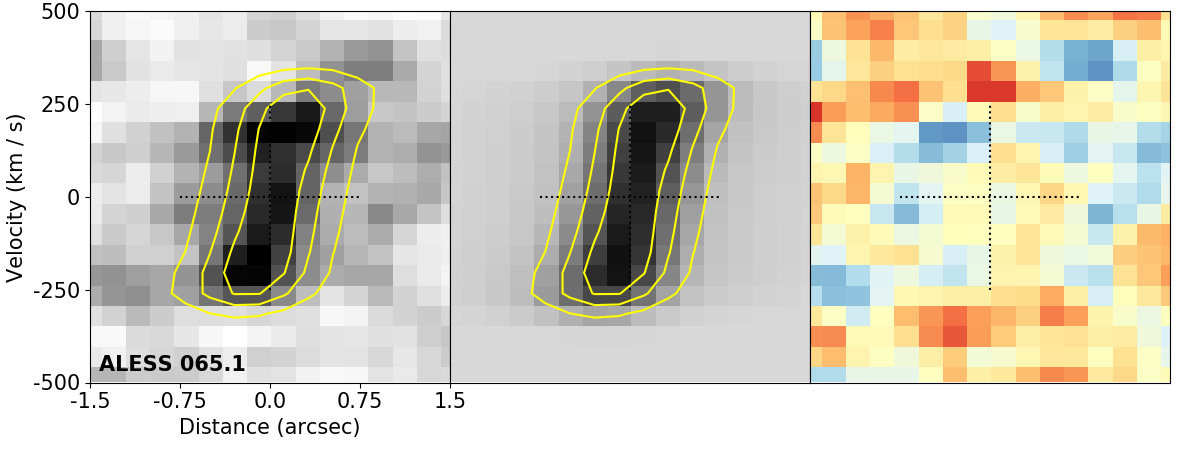} 
& 
\includegraphics[width=0.325\textwidth,height=0.08\textheight]{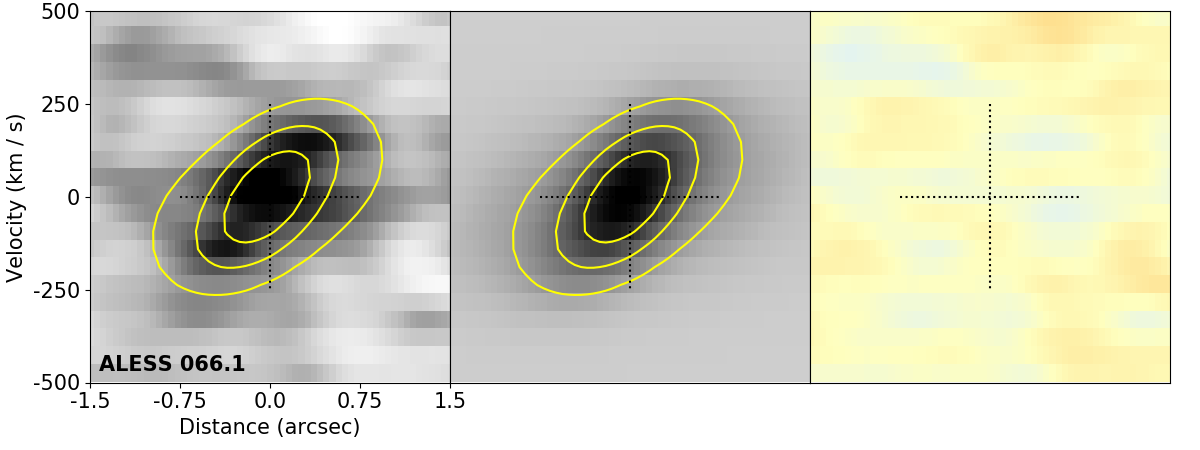}
&
\includegraphics[width=0.325\textwidth,height=0.08\textheight]{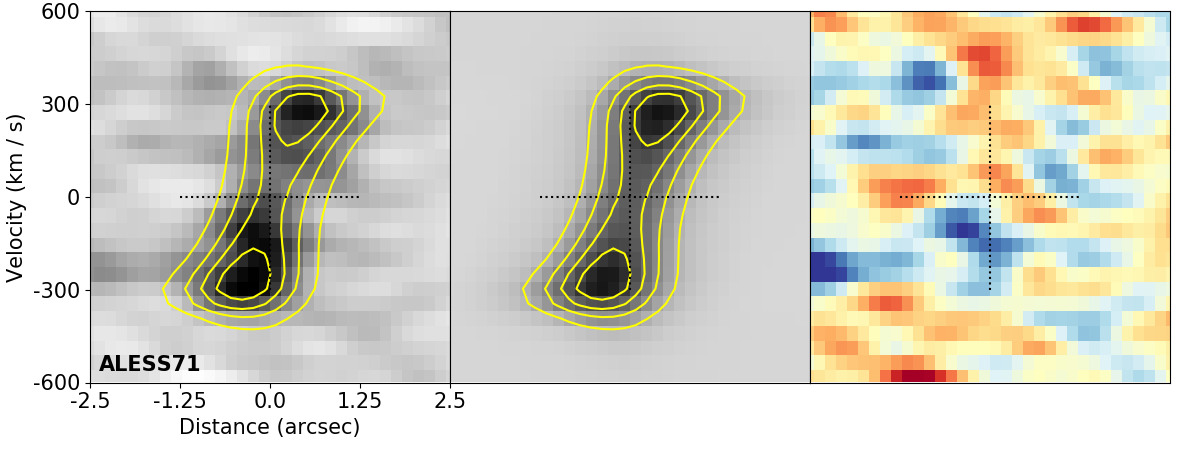}\\[1.0mm]

\includegraphics[width=0.325\textwidth,height=0.08\textheight]{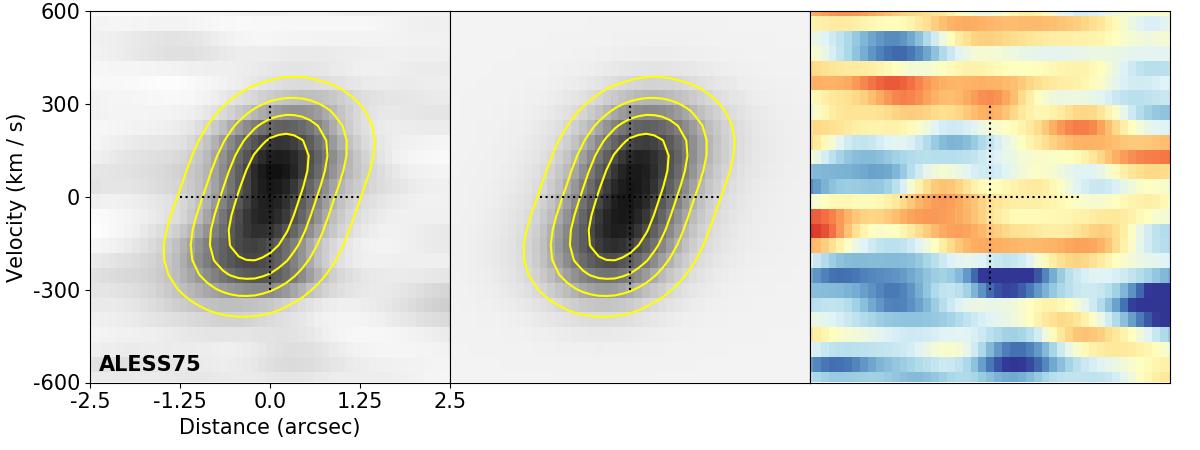}
&
\includegraphics[width=0.325\textwidth,height=0.08\textheight]{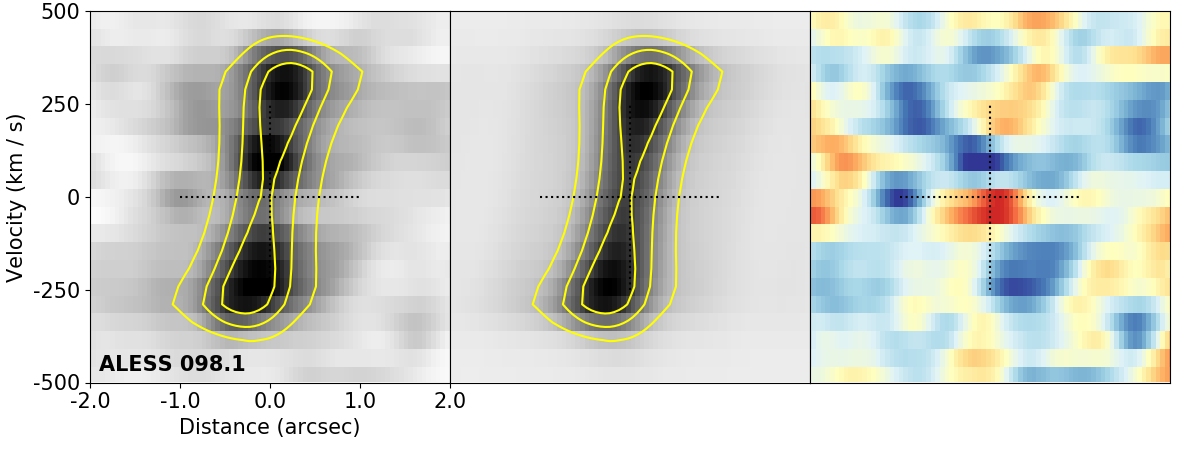}
& 
\includegraphics[width=0.325\textwidth,height=0.08\textheight]{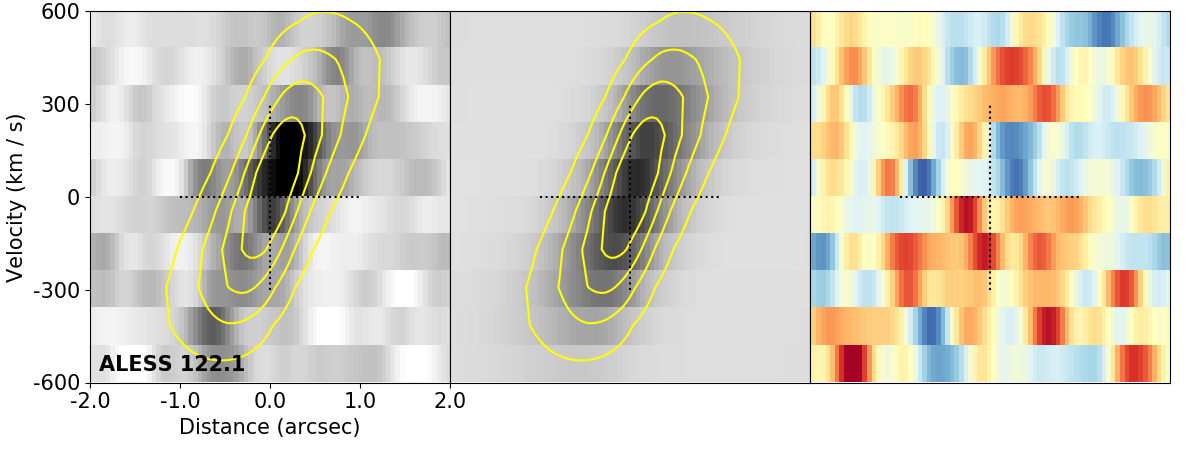}\\[1.0mm]

& \textbf{minor-axis} & \\[1.0mm]

\includegraphics[width=0.325\textwidth,height=0.08\textheight]{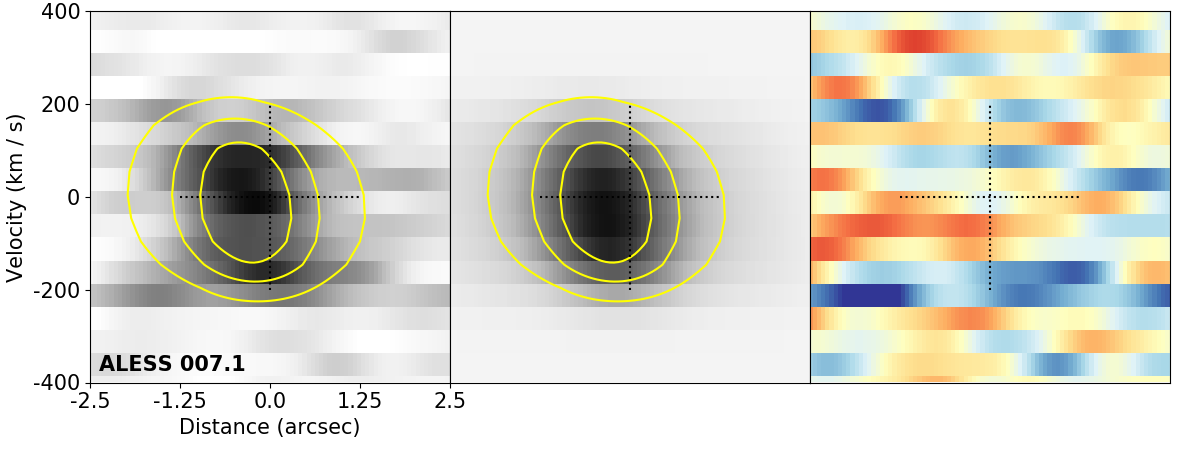}
&
\includegraphics[width=0.325\textwidth,height=0.08\textheight]{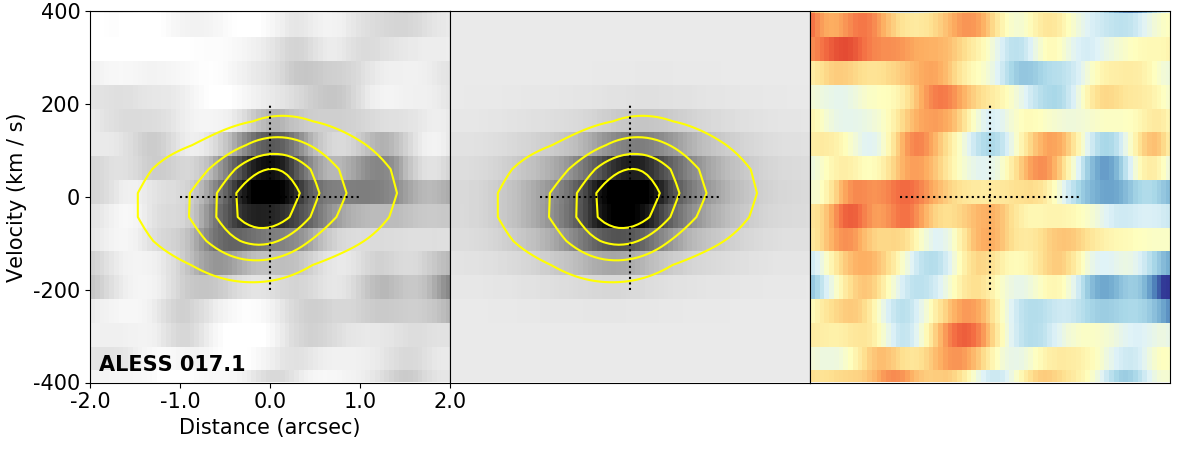}
&
\includegraphics[width=0.325\textwidth,height=0.08\textheight]{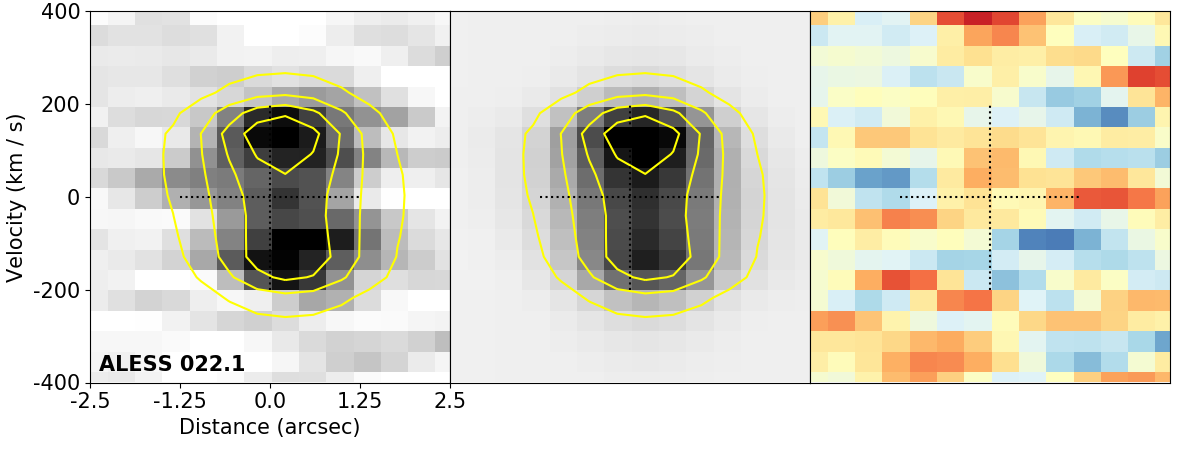}\\[1.0mm]

\includegraphics[width=0.325\textwidth,height=0.08\textheight]{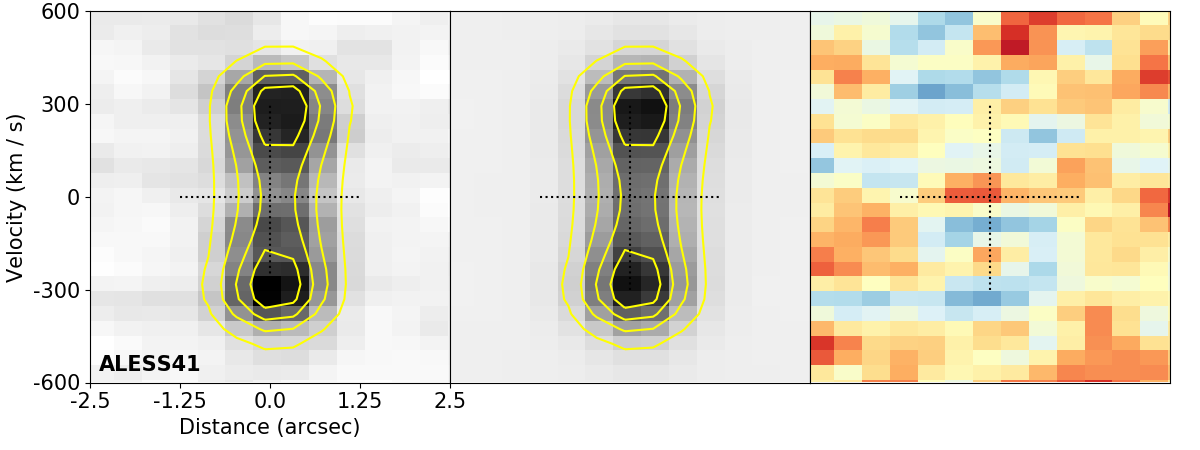} 
& 
\includegraphics[width=0.325\textwidth,height=0.08\textheight]{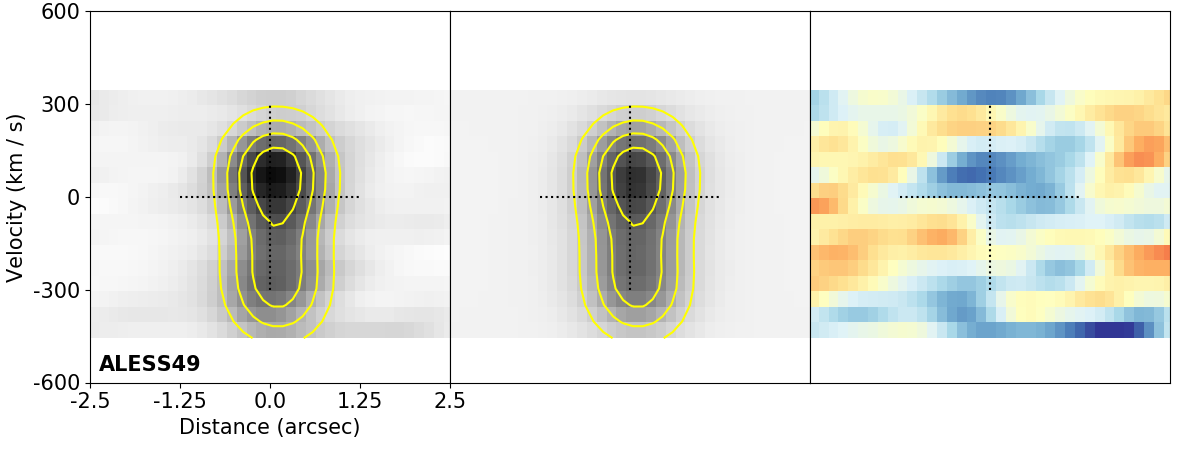}
&
\includegraphics[width=0.325\textwidth,height=0.08\textheight]{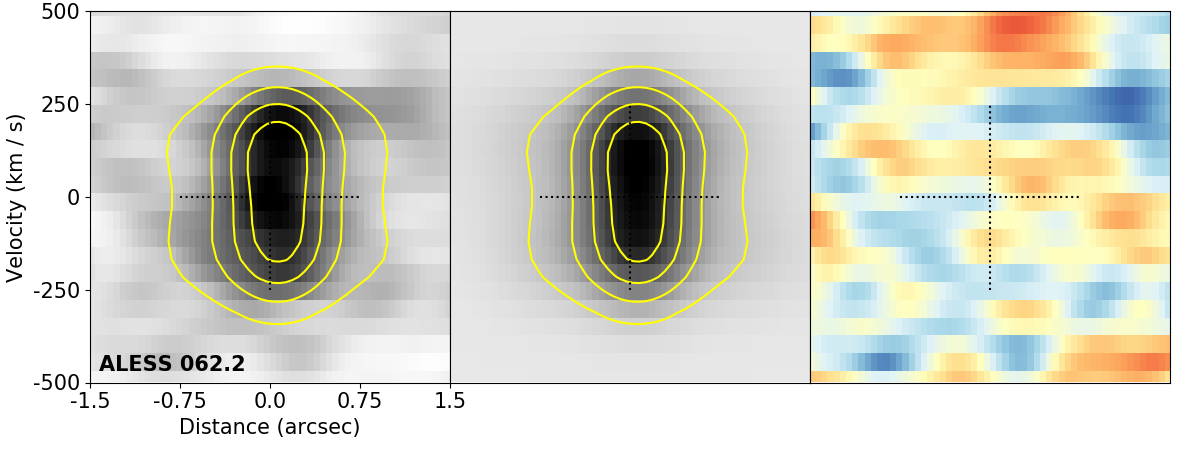}\\[1.0mm]

\includegraphics[width=0.325\textwidth,height=0.08\textheight]{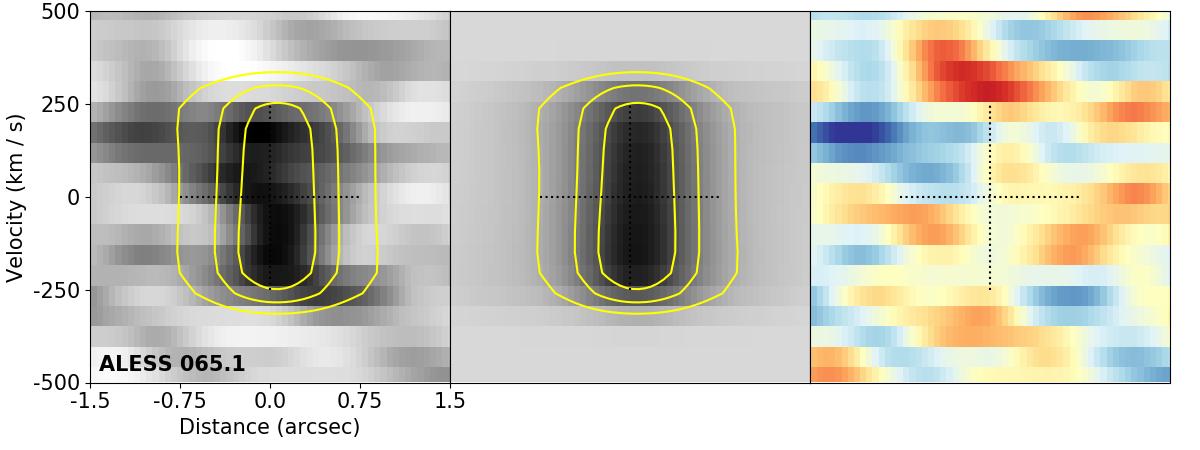} 
& 
\includegraphics[width=0.325\textwidth,height=0.08\textheight]{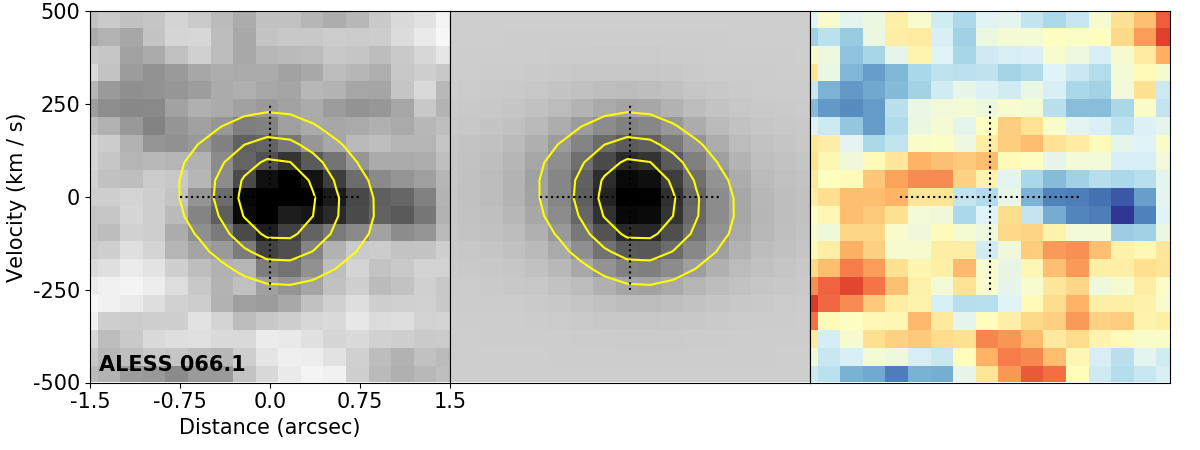}
&
\includegraphics[width=0.325\textwidth,height=0.08\textheight]{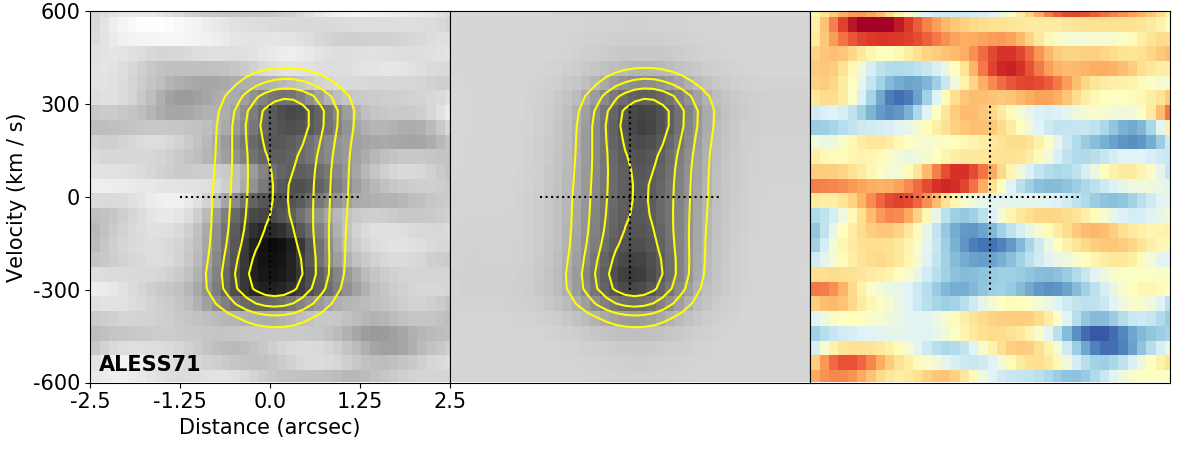}\\[1.0mm]

\includegraphics[width=0.325\textwidth,height=0.08\textheight]{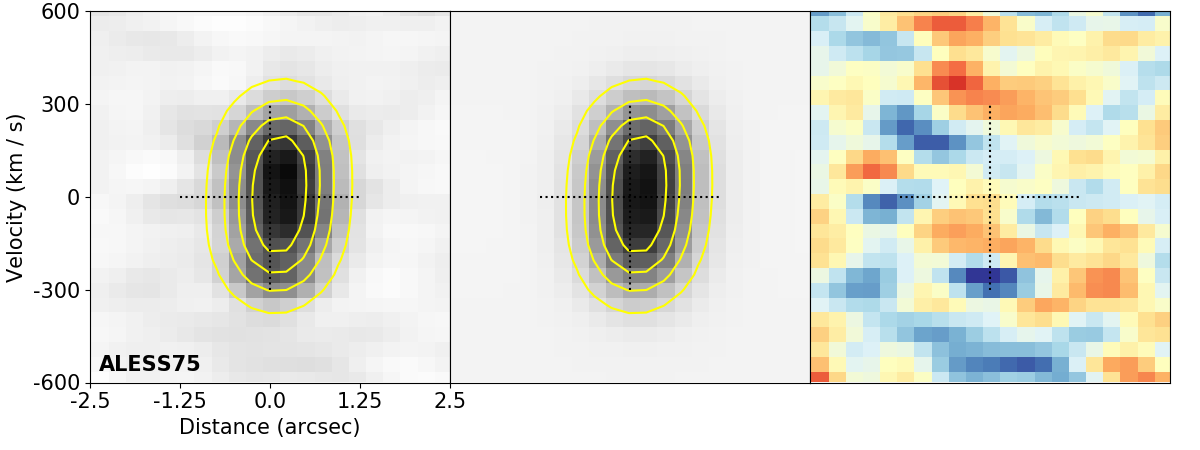}
&
\includegraphics[width=0.325\textwidth,height=0.08\textheight]{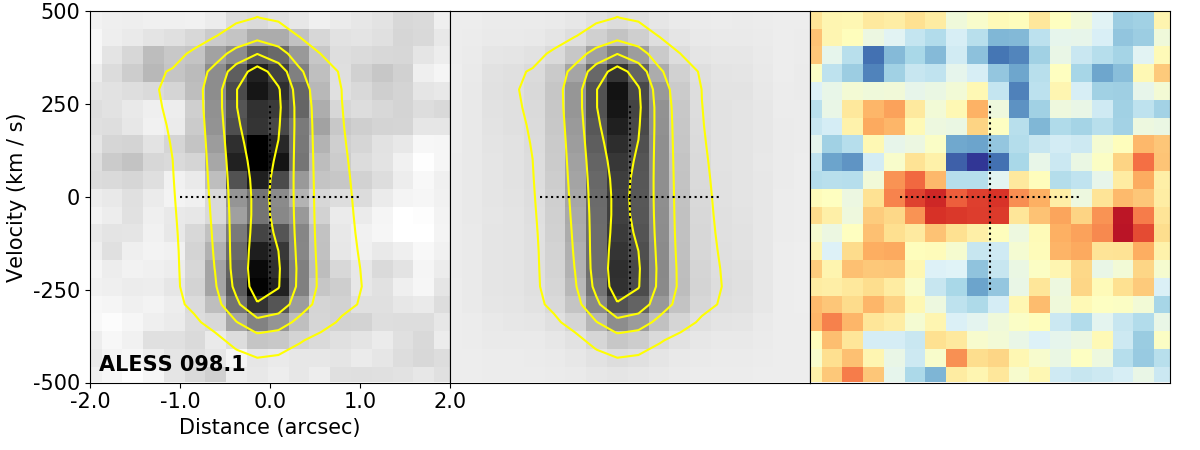}
& 
\includegraphics[width=0.325\textwidth,height=0.08\textheight]{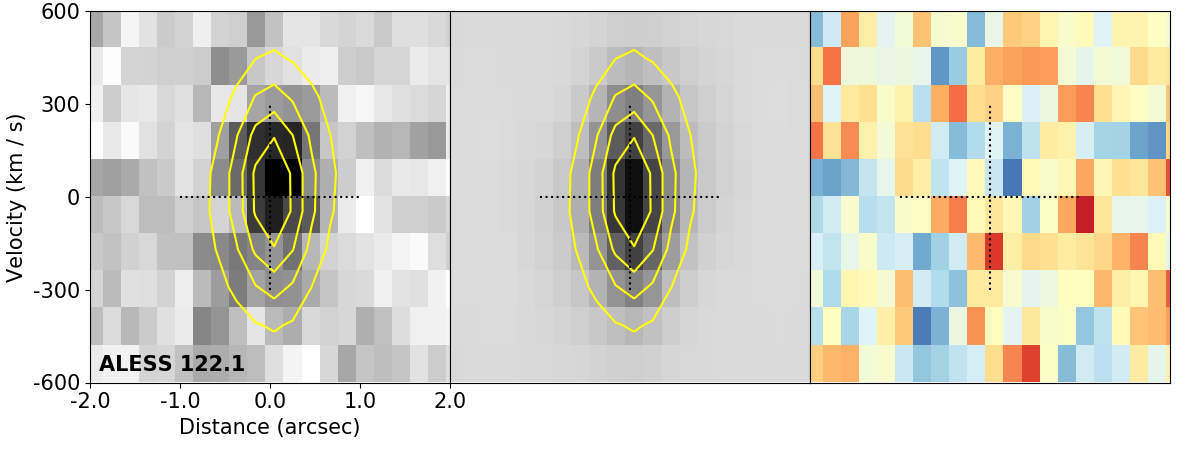}

\end{tabular}
\caption{Position velocity (PV) diagrams along the major/minor axis for the 12 Class \Romannum{1} sources in our sample which are well described by a disk kinematic model. The two left panels in each figure show the PV digrams for the data and model, where yellow contours in both panels are for the data. The right panel in each figure show the residuals.}
\label{fig:figure_Appendix_C}
\end{figure*}

\begin{figure*}
\begin{tabular}{cccc}

\textbf{007.1} & \textbf{017.1} & \textbf{022.1} & \textbf{041.1} \\[-0.5mm]

\includegraphics[width=0.225\textwidth,height=0.125\textheight]{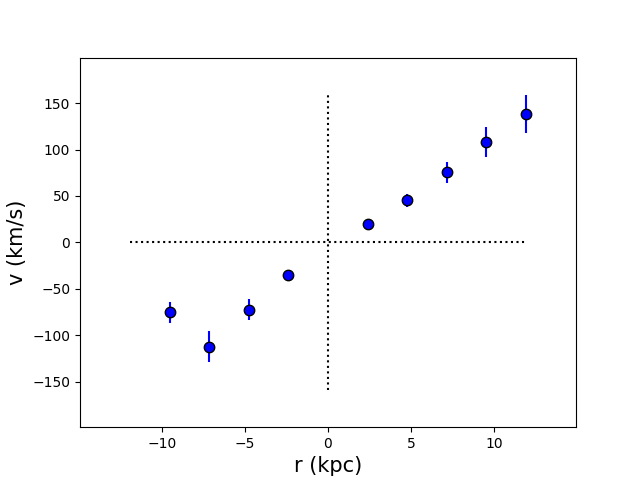}
&
\includegraphics[width=0.225\textwidth,height=0.125\textheight]{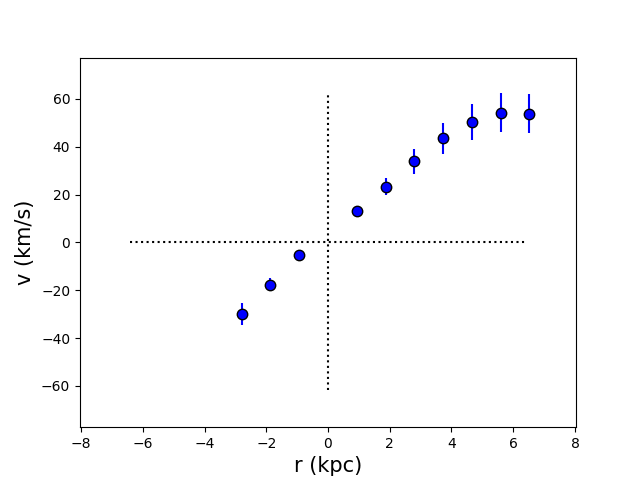}
&
\includegraphics[width=0.225\textwidth,height=0.125\textheight]{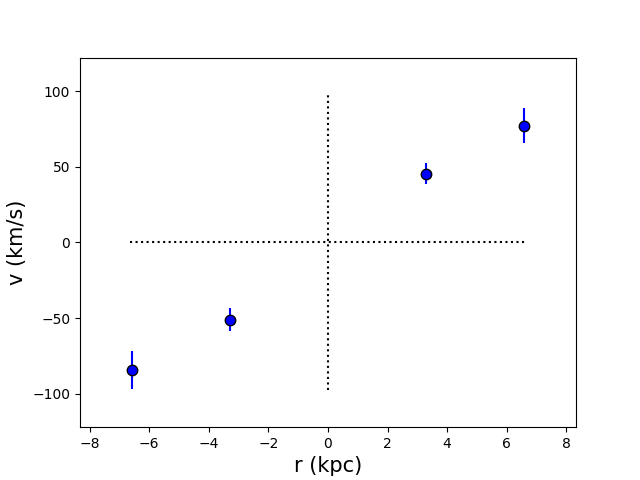} 
& 
\includegraphics[width=0.225\textwidth,height=0.125\textheight]{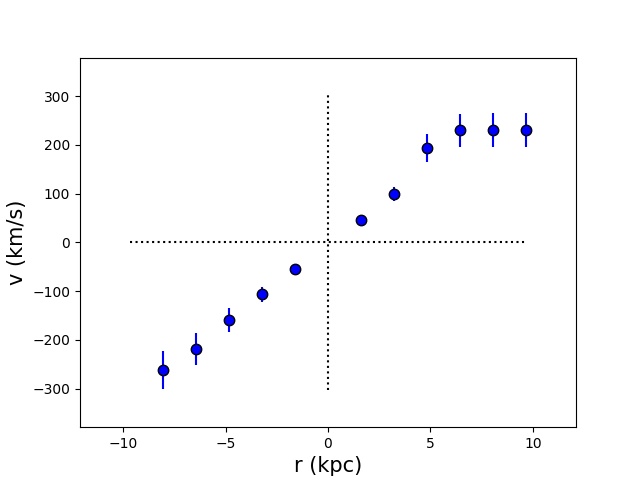}\\[1.0mm]

\textbf{049.1} & \textbf{062.1} & \textbf{065.1} & \textbf{066.1} \\[-0.5mm]

\includegraphics[width=0.225\textwidth,height=0.125\textheight]{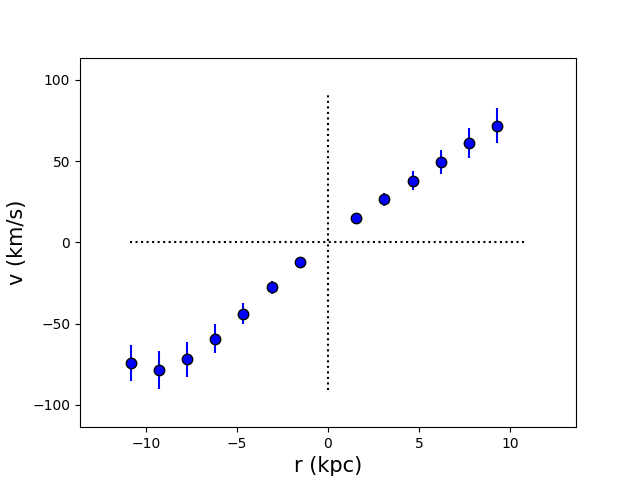}
&
\includegraphics[width=0.225\textwidth,height=0.125\textheight]{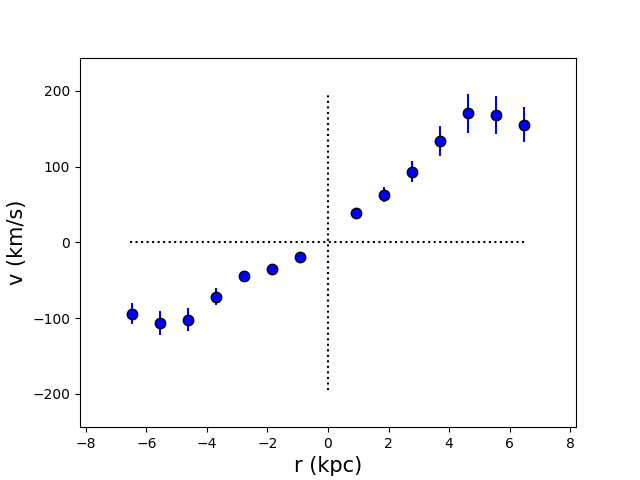}
&
\includegraphics[width=0.225\textwidth,height=0.125\textheight]{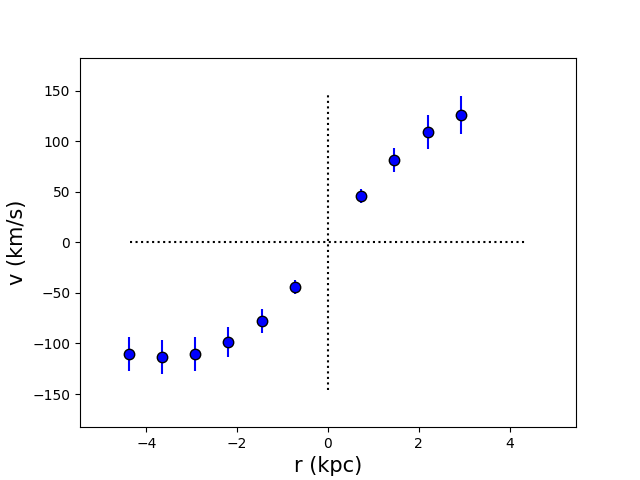} 
& 
\includegraphics[width=0.225\textwidth,height=0.125\textheight]{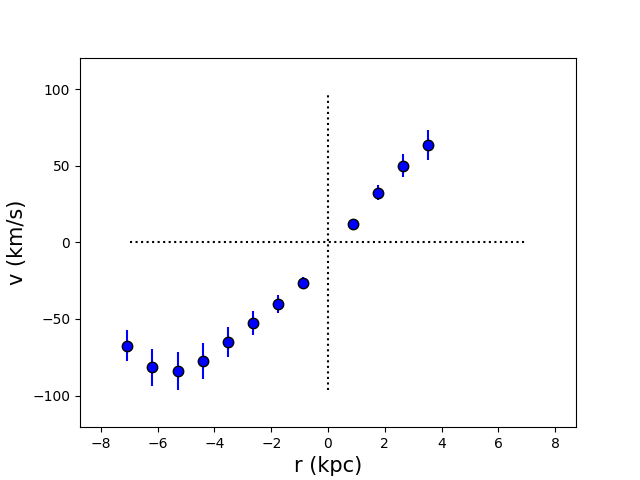}\\[1.0mm]

\textbf{071.1} & \textbf{075.1} & \textbf{098.1} & \textbf{122.1} \\[-0.5mm]

\includegraphics[width=0.225\textwidth,height=0.125\textheight]{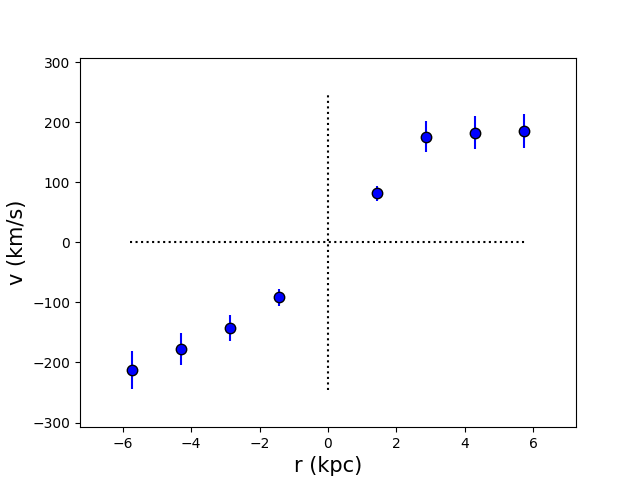}
&
\includegraphics[width=0.225\textwidth,height=0.125\textheight]{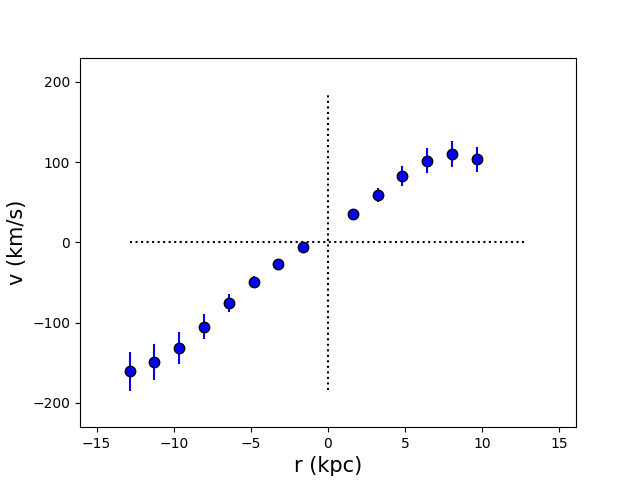}
&
\includegraphics[width=0.225\textwidth,height=0.125\textheight]{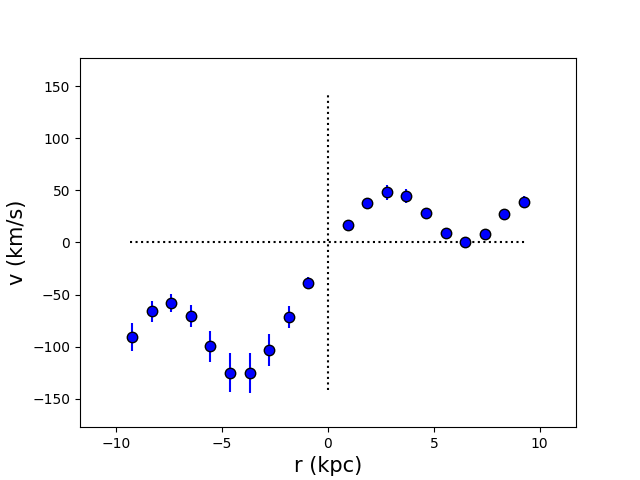} 
& 
\includegraphics[width=0.225\textwidth,height=0.125\textheight]{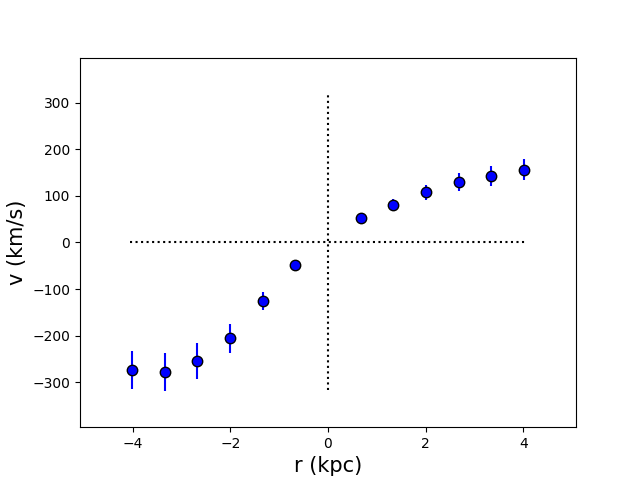}\\[1.0mm]
\end{tabular}
\caption{Rotation curves extracted from velocity maps for the 12 Class \Romannum{1} sources in our sample which are well described by a disk kinematic model.}
\label{fig:figure_Appendix_B}
\end{figure*}
\end{document}